%% file: PhDthesisV2.0.tex
\documentclass[11pt]{book} 
\usepackage{cite}
\usepackage{float}

\usepackage[utf8]{inputenc} 

\usepackage{geometry} 
\usepackage{graphicx} 

\usepackage{booktabs} 
\usepackage{array} 
\usepackage{paralist} 
\usepackage{verbatim} 
\usepackage{subfig} 

\usepackage{fancyhdr} 
\usepackage{sectsty}
\allsectionsfont{\sffamily\mdseries\upshape} 

\usepackage[nottoc,notlot]{tocbibind} 
\usepackage[titles,subfigure]{tocloft} 

\usepackage{amsmath}
\usepackage{amssymb}
\usepackage{amsfonts}
\usepackage{relsize}
\usepackage{stmaryrd}
\usepackage[usenames, dvipsnames]{color}
\usepackage{nicefrac}
\usepackage{slashed}
\usepackage{setspace}
\usepackage{pdfpages}

\DeclareMathOperator{\arcsinh}{arcsinh}

\usepackage[intoc]{nomencl}
\makenomenclature

\usepackage{etoolbox}
\renewcommand\nomgroup[1]{%
	\item[\bfseries
	\ifstrequal{#1}{A}{Abbreviations}{%
	\ifstrequal{#1}{B}{Symbols}{%
	\ifstrequal{#1}{C}{Physical Constants}{}}}%
	]}

\begin{document}

\input{PhDthesisV2.0preamble.tex}
	\input{PhDthesisV2.0chpt01Intro.tex}

	\input{PhDthesisV2.0chpt02MD.tex}

	\input{PhDthesisV2.0chpt03FO.tex}
	\input{PhDthesisV2.0chpt04Constraints.tex}
	\input{PhDthesisV2.0chpt05BID.tex}
	\input{PhDthesisV2.0chpt06HP.tex}
	\input{PhDthesisV2.0chpt07Conclusion.tex}	
	\input{PhDthesisV2.0bib.tex}
	\appendix
	\clearpage

	\makebox[\textwidth]{\centering
		\includegraphics[width=1.05\linewidth,page=1]{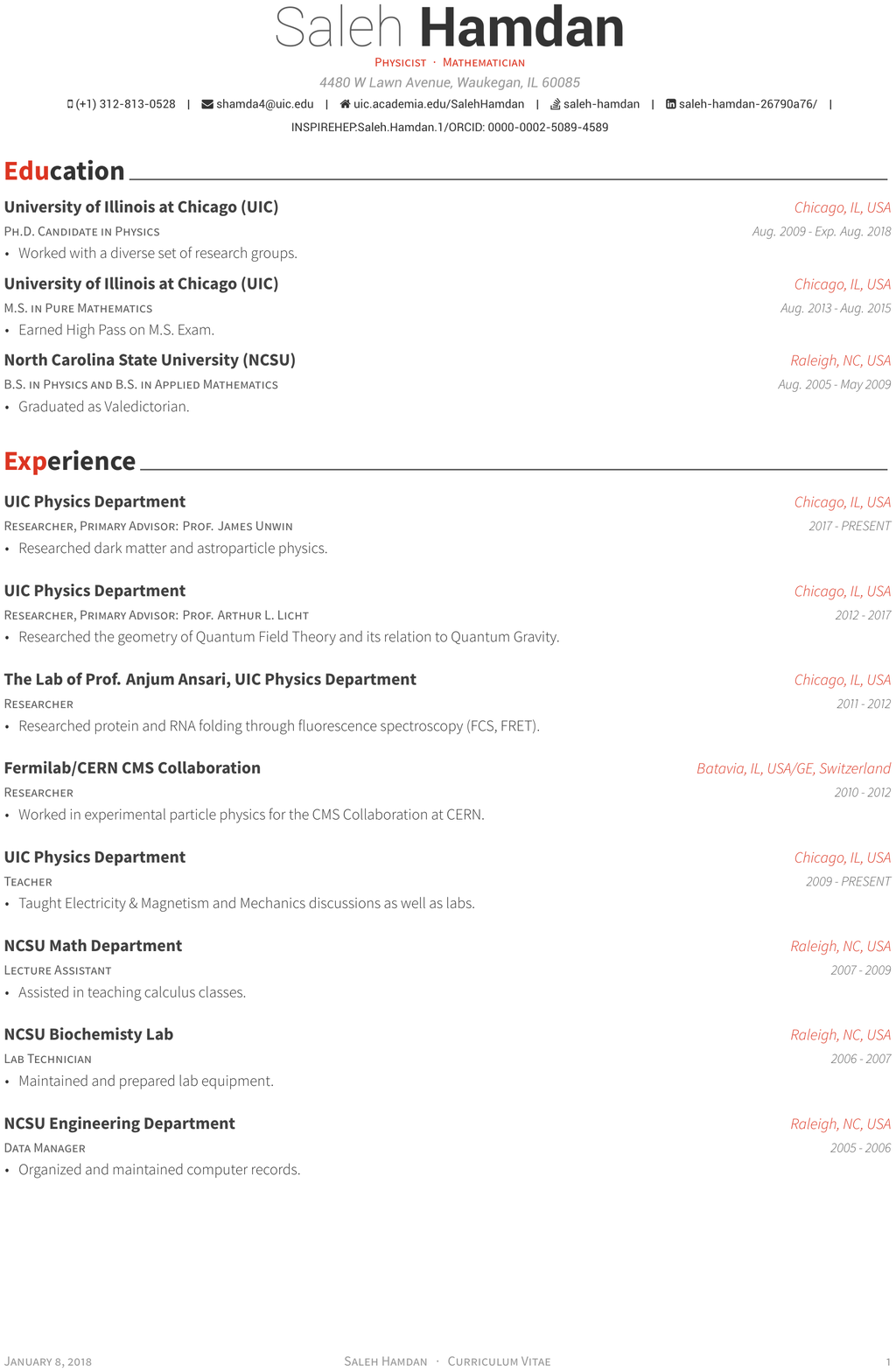}    }
	
	\makebox[\linewidth]{
		\includegraphics[width=1.05\linewidth,page=2]{Figs/Saleh-Hamdan-CV.pdf}    }
	
	\makebox[\linewidth]{
		\includegraphics[width=1.05\linewidth,page=3]{Figs/Saleh-Hamdan-CV.pdf}    }
	
	\addcontentsline{toc}{chapter}{Vita}
	
\end{document}

%% file: PhDthesisV2.0preamble.tex
	\thispagestyle{empty}
	\newgeometry{top=1.1in}
	\begin{center}
	
	{\bf Dark Matter Freeze-out in a Matter Dominated Universe}\\
	\vfill
	BY\\
	\vspace{5mm}
	SALEH HAMDAN\\
	B.S., North Carolina State University, 2009\\
	M.S., University of Illinois at Chicago, Chicago, 2015\\
	\vfill
	THESIS\\
	\vspace{5mm}
	Submitted as partial fulfillment of the requirements\\
	for the degree of Doctor of Philosophy in Physics\\
	in the Graduate College of the\\
	University of Illinois at Chicago, 2018\\
	\vspace{10mm}
	Chicago, Illinois
	\end{center}
	\vspace{5mm}
	
	\noindent Defense Committee:\vspace{5mm}\\
	\hspace*{23mm}Wai-Yee Keung, Chair\\
	\hspace*{23mm}James Unwin, Advisor\\
	\hspace*{23mm}Arthur Licht, Advisor\\
	\hspace*{23mm}Richard Cavanaugh\\
	\hspace*{23mm}Christopher Kolda, University of Notre Dame
	\restoregeometry
	\clearpage
	
	\pagestyle{plain}
	\pagenumbering{roman}
	\setcounter{page}{2}
	\chapter*{Dedication}
		
		{\em I dedicate this work to my dear mother, brother, and remaining family, without whose support this work would have been impossible.}
	
	\chapter*{Acknowledgements}
	\doublespacing
	First and foremost I wish to sincerely thank my advisor James A. Unwin for his guidance and help. I especially want to thank him for suggesting matter-dominated freeze-out as a novel scenario worth researching, which is what grew into this thesis. His guidance was vital throughout my research, writing, publishing, as well as all the other processes that academics must undertake.	
	
	I would also wish to express gratitude to my advisor Arthur (Lew) Licht, who guided me through advanced physics that was essential in completing this thesis, and was helpful and patient with me as I explored research topics. I wish to further express my gratitude toward other members of my thesis and preliminary exam committee: Richard Cavanaugh,  Wai-Yee Keung, Mikhail Stephanov, and Christopher Kolda. 
	
	I am deeply thankful for the teachers I have had throughout my graduate tenure. In particular, Arthur (Lew) Licht taught me Statistical Mechanics, General Relativity, and String Theory; Tom Imbo taught me Quantum Field Theory; Richard Cavanaugh taught me Particle Physics; Wai-Yee Keung taught me Quantum Mechanics; and Mikhail Stephanov taught me Electromagnetism. No words can express the amount of gratitude and respect I hold towards them for passing on their wisdom and knowledge to me.
	
	I would also like to thank the UIC Physics Department and the UIC Mathematics, Statistics, and Computer Science Department for accommodating me as I took classes in their departments and did research. Last, but certainly not least, I thank Fermilab for accommodating me during my early research in particle physics there.

	\chapter*{Statement of Originality}
	
	The ideas contained in chapters of this thesis have appeared in S.~Hamdan, and J.~Unwin \cite{Hamdan:2017psw}, arXiv:1710.03758. Additionally, Chapters \ref{Ch5} and \ref{Ch6} are expected to appear in a separate forthcoming publication. In all of the original work presented in this thesis I am responsible for the major derivations and calculations, while J.~Unwin provided direction, guidance, and checks.

	\newpage	
		
	\chapter*{Summary}
	
	 \begin{center}
	 	\bf{Dark Matter Freeze-out in a Matter Dominated Universe} \\ \textnormal{A thesis submitted for the degree of {\em Doctor of Philosophy}.} \\
	\textnormal{ Saleh Hamdan} \\ \textnormal{University of Illinois at Chicago}
	\end{center}
	
	The universe has evolved through several phases as its various constituents dominated its energy content. Candidate dark matter particles may have undergone freeze-out during any such phase. While the standard freeze-out scenarios have been explored during the radiation-dominated era, and more recently during scalar field decay, this work extends the study of dark matter freeze-out to a potential early period during which the universe is matter-dominated and its evolution adiabatic. Decoupling during an adiabatic matter dominated era changes the freeze-out dynamics, since the Hubble rate is parametrically different for matter and radiation domination. Furthermore, for successful Big Bang Nucleosynthesis the state dominating the early universe energy density must decay, this dilutes (or repopulates) the dark matter. As a result, the masses and couplings required to match the observed dark matter relic density can differ significantly from radiation dominated freeze-out.
	
	\clearpage
	
	\onehalfspace
	\tableofcontents
	\clearpage

	\nomenclature[A]{DM}{Dark matter}
	\nomenclature[A]{RD}{Radiation domination}
	\nomenclature[A]{MD}{Matter domination}
	\nomenclature[A]{BBN}{Big Bang Nucleosynthesis}
	\nomenclature[A]{FLWR}{Friedmann–Lemaitre–Robertson–Walker}
	\nomenclature[A]{MDFO}{Matter dominated freeze-out}
	\nomenclature[A]{RDFO}{Radiation dominated freeze-out}
	\nomenclature[A]{LHC}{Large Hadron Collider}
	\nomenclature[A]{MB}{Maxwell-Boltzmann}
	\nomenclature[A]{SM}{Standard Model}
	\nomenclature[A]{BE}{Boltzmann Equation}
	
	\nomenclature[B]{$T_\star$}{Temperature of thermal bath when decaying species evovles as matter}
	\nomenclature[B]{$\rho_\star$}{Energy density at $T_\star$}
	\nomenclature[B]{$H_\star$}{Hubble parameter at $T_\star$}
	\nomenclature[B]{$a_\star$}{Scale factor at $T_\star$}
	\nomenclature[B]{$t_\star$}{Comoving time at $T_\star$}
	\nomenclature[B]{$T_F$}{Freeze-out temperature}
	\nomenclature[B]{$x_F$}{Dimensionless freeze-out parameter}
	\nomenclature[B]{$r$}{Fraction of energy in radiation at $T_\star$}
	\nomenclature[B]{$g_*$}{Energy relativistic degrees of freedom}
	\nomenclature[B]{$g_{*S}$}{Entropic relativistic degrees of freedom}
	\nomenclature[B]{$T_{\rm MD}$}{Temperature when matter domination ensues}
	\nomenclature[B]{$T_\Gamma$}{Temperature when decay ensues}
	\nomenclature[B]{$T_{\rm RH}$}{Reheat temperature}
	\nomenclature[B]{$\Gamma_\Phi$}{Decay width of decaying species}
	\nomenclature[B]{$Y_\infty$}{Freeze-out abundance}
	\nomenclature[B]{$\zeta$}{Dilution factor}
	\nomenclature[B]{$s_{\rm R}$}{Entropy before dilution}
	\nomenclature[B]{$s_{\rm RH}$}{Entropy after dilution}
	\nomenclature[B]{$H_{\rm MD}$}{Hubble parameter when matter domination ensues}
	\nomenclature[B]{$\mathcal{T}$}{Ratio of $t_\star$ to comoving time}
	\nomenclature[B]{$\mathcal{A}$}{Ratio of $a_\star$ to scale factor}
	\nomenclature[B]{$T_{\rm EV}$}{Temperature when entropy violation becomes considerable}
	\nomenclature[B]{$\mathcal{T}_{\rm EV}$}{$\mathcal{T}$ when entropy violation becomes considerable}
	\nomenclature[B]{$t_{\rm EV}$}{Comoving time when entropy violation becomes considerable}
	\nomenclature[B]{$w$}{Parameter relating pressure and energy density}
	\nomenclature[B]{$\beta$}{Parameter used to define matter domination}
	\nomenclature[B]{$\Omega h^2$}{Fractional critical density with Hubble parameter uncertainty $h$}
	\nomenclature[B]{$T_{\rm BBN}$}{Minimum reheating temperature to fulfill BBN constraint}
	\nomenclature[B]{$\hat{\gamma}$}{Ratio of relativistic degrees of freedom at $T_\star$ to $T_{\rm RH}$}
	\nomenclature[B]{$\alpha$}{Used to parameterize thermally-averaged cross-section}
	
	\printnomenclature
	
	\clearpage
	
	\listoffigures
	
	\clearpage
	
	\newpage\null\newpage

	\doublespace
	\pagenumbering{arabic}
	\pagestyle{fancy}

%% file: PhDthesisV2.0chpt01Intro.tex
	\chapter{Introduction}
	\label{Ch1}
	
	\vspace{-15mm}
	{\em This chapter introduces the main topic of this thesis.}
		
	The current dark matter  density is measured to be $\Omega_{\rm CDM}h^2=0.115$ \cite{Bennett0067-0049-208-2-20}. There has been many proposals for particles which might account for dark matter, ranging from axions, Weakly Interacting Massive Particles (WIMPs), Strongly Interacting Massive Particles (SIMPs), to the variety of Supersymmetric (SUSY) particles and beyond \cite{Bertone:2004pz}. With little known about which, if any, candidate may be dark matter, it is prudent to take a model-independent approach to exploring the parameter space of this elusive constituent of our universe. 
	
	Dark matter may have once been in thermal equilibrium with the photon plasma; such dark matter candidates are termed thermal, and will be the focus of this thesis. An effective method for exploring the validity of a thermal dark matter candidate is to compute its freeze-out abundance. Because an interaction becomes much less likely whenever $\Gamma \lesssim H$, where $\Gamma$ is the interaction rate of a particle and $H$ is the Hubble parameter, dark matter annihilations become much less likely (``Frozen-out''), preserving a relic abundance we may measure today. The goal of such calculations is to investigate the parameter space of a species of particles, such as its mass and interaction strength, consistent with the thermal history of the universe and the abundance we measure of such a particle today. An example of the fruitfulness of such a method is the Lee-Weinberg bound for a stable, cold neutrino species, $m_{\nu}\gtrsim 2$ GeV \cite{LeeWeinbergPhysRevLett.39.165}.
	
	Although freeze-out calculations have  been performed for dark matter candidates, these calculations have typically been carried out assuming freeze-out occurred either during an era of the universe when relativistic particles dominated the energy content of the universe \cite{Kolb:1990vq}, referred to as the radiation domination phase, or while a scalar field undergoes coherent oscillations as it decays with high adiabicity violation (one may also view this latter scenario as the decay of a massive unstable species of particles) \cite{GiudicePhysRevD.64.023508,McDonald:1989jd}. While these scenarios are feasible, there are no violations of known cosmological or physical principles for freeze-out to have occurred during an adiabatic era of the universe when non-relativistic matter dominated its energy content, referred to as the matter domination phase. It is quite possible that dark matter indeed has frozen-out during a matter domination phase of the universe. This is because $\rho_{\rm M}/\rho_{\rm R}\propto a^{-3}/a^{-4}\propto a$, where $\rho_{\rm M}$ is the energy density of decoupled (from thermal bath), non-relativistic matter, $\rho_{\rm R}$ is the energy density of radiation, and $a$ is the scale factor. Hence, the energy density of decoupled matter tends to grow relative to that of radiation. 
	
	The thesis is structured as follows: in the remainder of the introductory chapter we make the case for adopting the dark matter hypothesis and review the methods for searching for dark matter. In Chapter \ref{Ch2} we discuss some pertinent background material, rewrite the Friedmann equation in a manner conducive to studying matter domination and radiation domination limits, and give a non-technical overview of the matter domination freeze-out scenario. Chapter \ref{Ch3} is devoted to a rigorous treatment of freeze-out during matter domination. We then move on to discussing the constraints on the matter-dominated freeze-out scenario in Chapter \ref{Ch4}. Since much of our analysis in Chapters \ref{Ch3} and \ref{Ch4} makes use of the instantaneous decay approximation, in Chapter \ref{Ch5} we demonstrate that this approximation is well justified, and show where it is less reliable. Following the model-independent approach in Chapters \ref{Ch3} and \ref{Ch4}, we apply this analysis to a specific Higgs Portal model  in Chapter \ref{Ch6}, and argue that this portal to dark matter provides  a viable scenario. We summarize and conclude this thesis in Chapter \ref{Ch7}.
	
	\section{The Case for Dark Matter}
	
	\vspace{-4mm}
	
	Until a future discovery sheds light on the precise nature of dark matter, it is more of a concept than a physical object. It is a label physicists use to describe the gap between their suppositions about Nature and what it actually shows them. In the early 19th century Neptune was considered ``dark''; it was just a hypothesis of the astronomers Urbain Le Verrier and John Couch Adams to describe anomalies in the orbit of Uranus \cite{Bertone:2016nfn}. Once the Berlin Observatory used a refractory telescope to observe it, Neptune went from a hypothesis to a planet in fact. The hypothesized planet {\em Vulcan}, which was also predicted by Le Verrier to explain the anomalous precession of the perihelion of Mercury, was never discovered \cite{Bertone:2016nfn}. Instead, the General Theory of Relativity  explained the anomalous precession, and ``Vulcan'' was no more. This example provides an illuminating tale for physicists studying DM. Just as in the past, physicists today confront a major discrepancy between simple suppositions about Nature versus what we factually observe. Nevertheless the existence of a cold (non-relativistic) form of matter that very weakly interacts with baryonic matter is practically a scientific certainty, though its precise nature remains elusive. In this section I will highlight the case for the existence of such matter, commonly referred to as ``dark matter'' (DM).
	
\begin{figure}
		\centering
		\begin{minipage}[t]{0.49\textwidth}
			\centering
			\includegraphics[width=\textwidth,height=5cm]{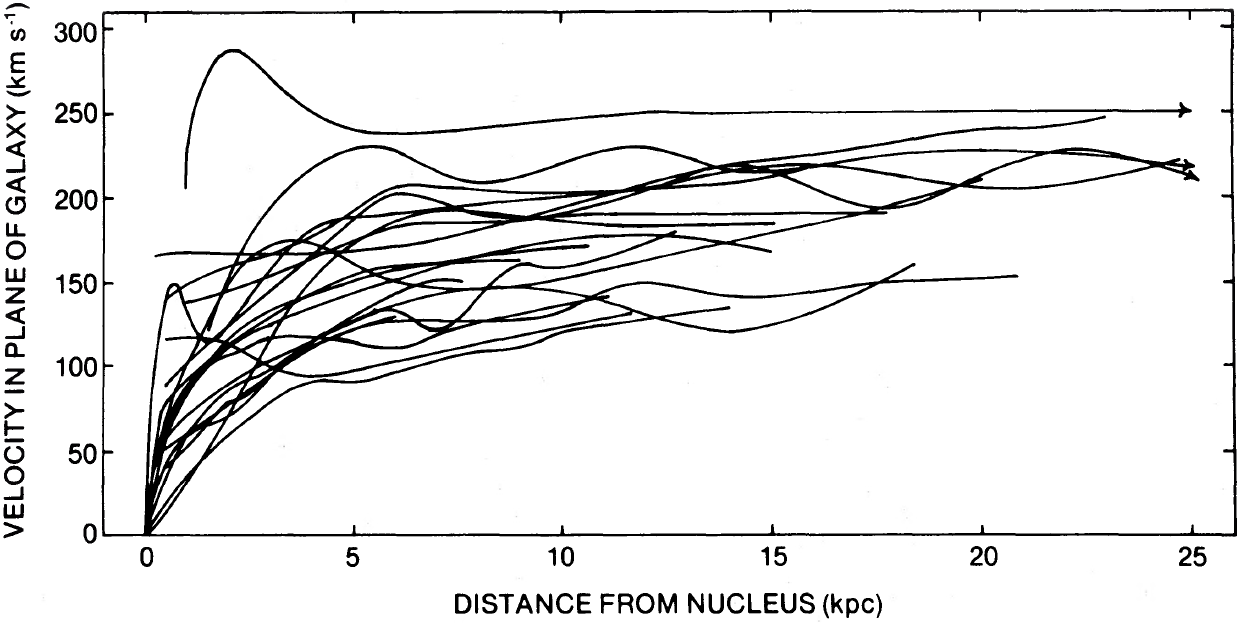}
			\vspace{-0.1mm}
	\caption[Superposition of measured rotational velocities versus radial distance plots for all 21 Sc galaxies, from \cite{Rubin:1980zd}.]{Superposition of measured rotational velocities versus radial distance plots for all 21 Sc galaxies taken from \cite{Rubin:1980zd}. Remarkably, rotational velocity flattens out near disk radius.}
			\label{fig:RubinRotCurve}
		\end{minipage}\hfill
		\begin{minipage}[t]{0.49\textwidth}
			\centering
			\includegraphics[width=\textwidth,height=5cm]{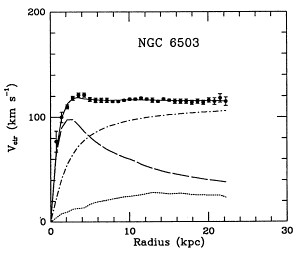}
					\vspace{-0.1mm}
	\caption[Rotation curve fits (solid line) for NGC 6503 taken from \cite{Begeman:1991iy}.]{Rotation curve fits (solid line) for NGC 6503, from \cite{Begeman:1991iy}. Rotation curves for visible matter (dashed), gas (dotted), and dark halo (dashed-dotted) also shown.}
			\label{fig:galaxyrot}
		\end{minipage}
	\end{figure}

	As early as 1933, the astronomer Fritz Zwicky, using the virial theorem, observed that ``the average density in the Coma system would have to be at least 400 times larger than that derived on the grounds of observations of luminous matter,'' and that if that were the case, ``dark matter is present in much greater amount than luminous matter'' \cite{Zwicky:1933gu}. Although Zwicky's argument applied to a cluster of galaxies, the principles apply just as well to any stable gravitational system; for example individual galaxies. Indeed one would expect from Newtonian gravity that the tangential velocity $v_t$ around a spherical mass distribution to behave as
	\begin{equation}\label{eq:vTan}
	v_t(r)=\sqrt{\frac{GM(r)}{r}}_,
	\end{equation}
	where $M(r)$ is the enclosed total mass within the radial distance $r$. Thus if the mass of a galaxy is concentrated in its disk, $M(r)$ should be roughly constant outside the disk, and $v_{t} (r\gtrsim R_{\rm disk})\propto r^{-1/2}$ \cite{Lisanti:2016jxe}. Instead a flattening of the tangential velocity is observed as seen in Figure \ref{fig:RubinRotCurve} taken from Rubin et al. \cite{Rubin:1980zd}. This indicates that $M(r)\propto r$ around the disk, and hence there must be more mass in a galaxy than its luminous matter as seen in Figure \ref{fig:galaxyrot} taken from Begeman et al. \cite{Begeman:1991iy}.
	
	The essence of the problem of the galactic rotational curves is one of a high mass-to-light ratio, $M/L$. More mass is measured in galaxies than what we would expect from the light we observe from them \cite{1985ApJ...295..305V}. But what if that mass came from baryonic objects with low luminosity? Certainly low luminosity objects exist in galaxies, such as planets, dwarf stars, and even black holes. The class of baryonic matter that may make up the missing mass in the galaxy were termed massive astrophysical compact halo objects, or MACHOs. Yet early objections to MACHOs making up the missing mass in galaxies were published \cite{Hegyi:1983bj}, and by the mid-2000s the EROS-2 collaboration had shown using micro-lensing that MACHOs can only make up a maximum of 8\% of the Milky Way halo mass fraction \cite{Bertone:2016nfn,Tisserand:2006zx}. Furthermore, the baryonic budget (the relic fractional density of baryons, $\Omega_b$) which is calculated from primordial nucleosynthesis considerations ($\Omega_b=0.04\pm0.02$), inferred from the cosmic microwave background (CMB) power spectrum ($\Omega_bh^2=0.02273\pm0.00062$), as well as measured from galaxy clusters ($\Omega_b=0.04$) indicates that it is too small to accommodate a MACHO explanation for the galactic rotational curve problem \cite{Bertone:2010zza}. 
	
\begin{figure}[h]
			\centering
			\includegraphics[width=0.8\textwidth]{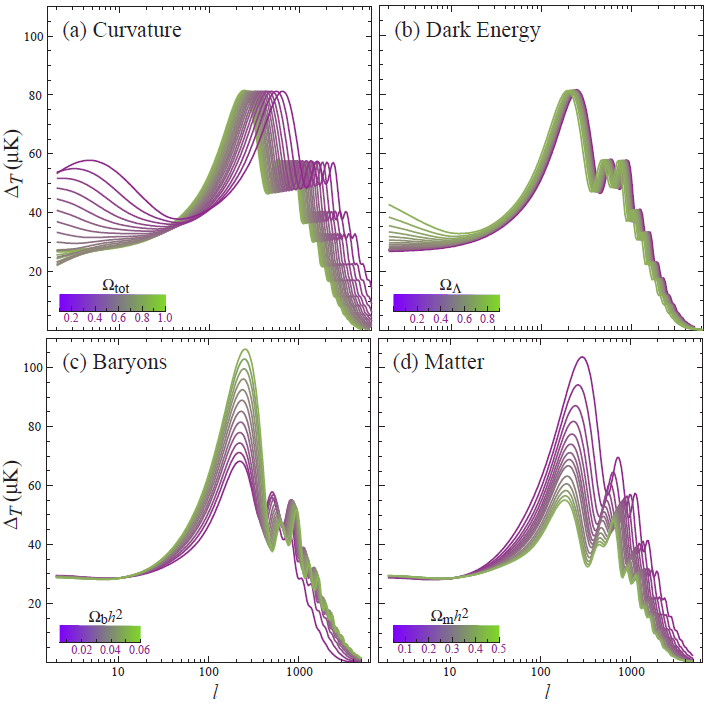}
			\vspace{-3mm}			
			\caption{CMB power spectrum for varying energy content of the universe, from \cite{Hu:2001bc}.}
			\label{fig:AcousticSens}
			\centering
			\includegraphics[width=0.8\textwidth]{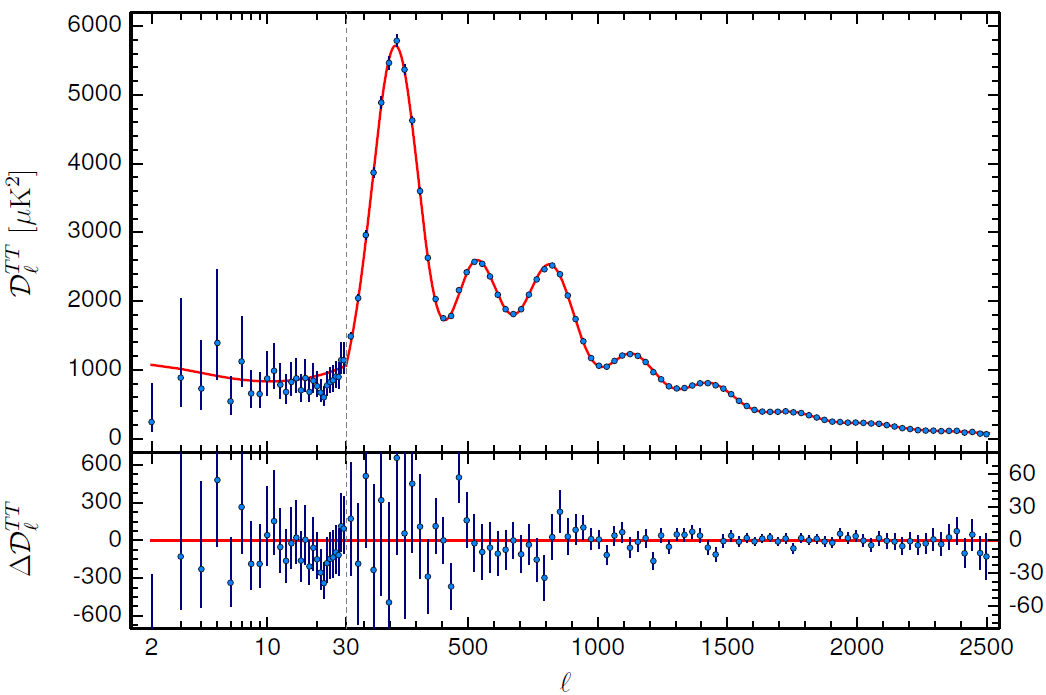}
			\caption[Planck Collaboration CMB temperature power spectrum \cite{Ade:2015xua}.]{Planck Collaboration CMB temperature power spectrum \cite{Ade:2015xua}. Upper panel shows $\Lambda$CDM fit (solid red); lower panel shows residuals with respect to this model.}
			\label{fig:PlanckCMBspec}
		\end{figure}

	These findings suggest that a solution to the galactic rotational curve problem requires a fundamental shift away from known physics. The hypothesis of dark matter provides an elegant solution to this problem, yet that is not its most compelling evidence. What began as a hypothesis to explain the missing mass in galaxies has now become part of a standard cosmological model called $\Lambda$CDM, which assumes that the General Theory of Relativity to be correct, and includes dark energy (represented as the cosmological constant, $\Lambda$, to explain the accelerating expansion of the universe) and cold DM to describe a consistent big bang cosmology. $\Lambda$CDM makes precise predictions about the acoustic power spectrum of the CMB, which is essentially the photons reaching us from the surface of last scattering, or when photons decoupled from the rest of matter and the universe became transparent. The CMB temperature anisotropies give an acoustic power spectrum that is sensitive to the energy content of the universe as can be seen in Figure \ref{fig:AcousticSens}. The measured CMB power spectrum from the Planck Collaboration \cite{Ade:2015xua} fits the $\Lambda$CDM prediction with stunning precision as shown in Figure \ref{fig:PlanckCMBspec}, thus providing compelling evidence for the existence of DM.
	\newpage
	\clearpage
	
	Further strengthening the case for DM is large scale structure formation in the universe. It had been known by the early 80s that the perturbations in the CMB were incompatible with the formation of galaxies and galaxy clusters in a universe composed of only baryonic matter, and that including weakly-interacting massive particles (WIMPs, a candidate DM class) in the energy profile of the universe can aid structure formation \cite{Peebles:1982ff}. Since then computer simulations have demonstrated that $\Lambda$CDM can successfully predict the large scale structure we observe today in the universe \cite{Davis:1985rj,Vogelsberger:2014kha,Schaye:2014tpa}, and have been important in modeling the structure of DM halos \cite{Navarro:1995iw}. Simulations have also played a significant role in ruling out that hot dark matter (which is relativistic at decoupling, such as neutrinos) could be responsible for the DM density \cite{White:1984yj}, thus in this thesis DM shall only refer to cold dark matter.
	
\begin{figure}[h]
		\centering
		\includegraphics[width=0.9\textwidth]{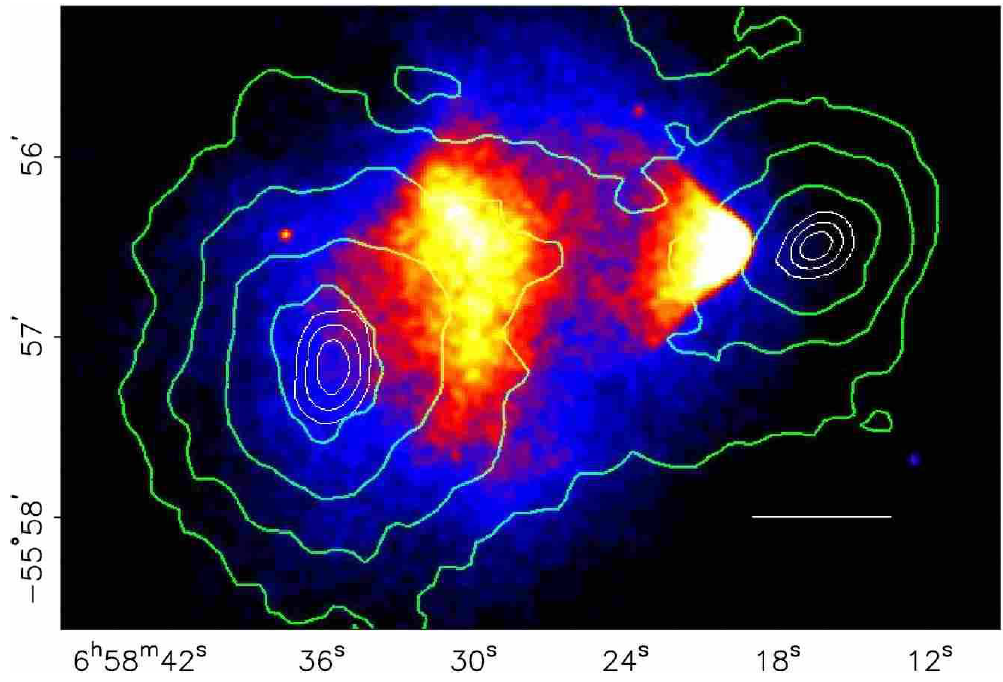}
		\caption[Bullet Cluster image taken from \cite{Clowe:2006eq}.]{Bullet Cluster image taken from \cite{Clowe:2006eq}. The coloring represents X-ray emissions as taken by Chandra. The green lines are reconstructed gravitational weak lensing gradients. The white bar represents 200 kpc at the distance of the cluster.}
		\label{fig:BulletCluster}
	\end{figure}

	Finally, no case for DM would be complete without mentioning a research article boldly titled, ``A direct empirical proof of the existence of dark matter." Clowe et al. \cite{Clowe:2006eq} analyzed the merging cluster 1E0657-558, also known as the Bullet Cluster, which is composed of two galaxy clusters merging together. Individual galaxies in a merging cluster behave as collisionless particles, whereas the intracluster plasma experiences ram pressure and emits X-rays \cite{Clowe:2006eq}. If there was no dark matter, one would expect the gravitational potential to track the spatial configuration of the plasma, since it forms the dominant baryonic mass content of clusters. On the other hand, since dark matter only weakly interacts, it is expected to track the collisionless individual galaxies. Therefore a spatial decoupling of the X-ray emissions and gravitational potential provides compelling evidence that the bulk of mass in galaxy clusters is dark. This spatial decoupling is precisely what Clowe et al. found when they viewed the X-ray emissions from the Bullet Cluster using the Chandra telescope, and used gravitational weak lensing to probe how the gravitational potential was spatially distributed, as shown in Figure \ref{fig:BulletCluster}.
	\clearpage

	Although the precise nature of DM is currently unknown, galactic rotational curves, the baryon budget, the CMB power spectrum, large scale structure formation, and the Bullet Cluster provide compelling evidence that the existence of DM is a near scientific certainty. The evidence for dark matter, not all of which we presented here, has even led some physicists to identify its hypothesis along the lines of a Kuhnian scientific revolution \cite{Tremaine87,Einasto:2011jw}. While other theories have emerged to explain the apparent missing mass in galaxies, most notably Modified Newtonian Dynamics (MOND) \cite{Milgrom:1983ca}, these theories do not provide the same explanatory power that the $\Lambda$CDM model does for the disparate phenomenon we observe. Le Verrie's tale demonstrates that a healthy dose of skepticism is prudent in science, but we believe that a vigorous hunt for DM through theoretical and empirical means is well warranted and exceedingly likely to inaugurate a new particle into physics, beyond the Standard Model (SM).
	\clearpage
	
	\section{Empirical Searches for Dark Matter}
	
	The compelling evidence for the existence of DM has prompted several empirical efforts to detect it. We have already described the analysis of the Bullet Cluster which boasts a direct empirical proof of DM, yet the search is still on for a direct signal of DM that illuminates its precise nature \cite{Arcadi:2017kky}. The major experimental efforts can be classified into three categories: direct detection, indirect detection, and collider searches \cite{PDGroup}. In this section we will describe these three methods and present the current bounds they give on DM parameters.
	
	Direct detection experiments rely on observing recoils from potential DM scatterings off nuclei, in particular elements such as  Xenon, Germanium and Argon, which have high atomic numbers which enhances the DM-nucleon cross-sections ($\sigma_{N}\propto A^2$) for certain interactions. Several factors go into predicting the rate of events that direct detection experiments should expect. Specifically, the differential event rate (with typical units of kg$^{-1}$day$^{-1}$keV$^{-1}$ or `dru' standing for differential rate units) is given by \cite{Bertone:2010zza,Arcadi:2017kky}
	\begin{equation}\label{eq:DRUdd}
	\frac{dR}{dE_{\rm R}}=\frac{\rho_0}{m_Nm_{\rm DM}}\int^{v_{\rm esc}}_{v_{\rm min}}vf(v)\frac{d\sigma_{N}}{dE_{\rm R}}(v,E_{\rm R})dv.
	\end{equation}
	where $\rho_0$ is the local DM energy density, $m_N$ and $m_{\rm DM}$ are the nucleon and DM masses respectively, $\frac{d\sigma_{N}}{dE_{\rm R}}$ is the differential cross-section for elastic DM-nucleon scattering, $f(v)$ is the DM speed distribution in the detector frame, $v_{\rm esc}$ is the local escape speed in the galactic rest frame, and $v_{\rm min}$ is the minimum velocity that can cause recoil energy $E_{\rm R}$. Equation \eqref{eq:DRUdd} is integrated from $E_{\rm T}$, the threshold energy that the detector is capable of measuring to infinity, and it presents the various particle physics and astrophysical inputs involved in direct detection.
	
	\clearpage

	As far as the particle physics input, the DM-nucleon differential cross-section encodes information about the potential interactions that DM may have with quarks. This is done by using an effective Lagrangian approach that may include generic spin-dependent and spin-independent terms \cite{Bertone:2010zza}. In moving from quark-DM interactions to nucleon-DM cross-sections, uncertainties are introduced via the nuclear form factors and hadronic matrix elements \cite{Bertone:2010zza}. So long as dark matter's nature does not stray far from known effective field theoretic principles, this effective Lagrangian approach warrants confidence.
	
	Astrophysical parameters are also involved in modeling direct detection experiment outcomes. In eq.~\eqref{eq:DRUdd} the local DM energy density is calculated by using a model of the Milky Way DM halo (which incorporate fitting of rotation curves), and it is predicted to have the standard value of $\rho_0\simeq0.3$ GeVcm${}^{-3}$, though modeling uncertainities can cause this value to vary by a factor of 2 \cite{Bertone:2010zza}. $f(v)$ is also traditionally taken to be Maxwellian and isotropic, and although modeling efforts have been made to relax these assumptions, it nevertheless introduces modest uncertainties in bridging theory and experiment \cite{Bertone:2010zza}. Finally, the upper limit of integration in eq.~\eqref{eq:DRUdd} is $v_{\rm esc}=\sqrt{2\Theta(R_0)}$, where $\Theta(R_0)$ is the local gravitational potential, and particles of speed greater than $v_{\rm esc}$ are expected not to be gravitationally bound to the Milky Way \cite{Bertone:2010zza}.
	
\begin{figure}[H]
		\centering
		\vspace{2mm}
		\includegraphics[width=0.9\textwidth]{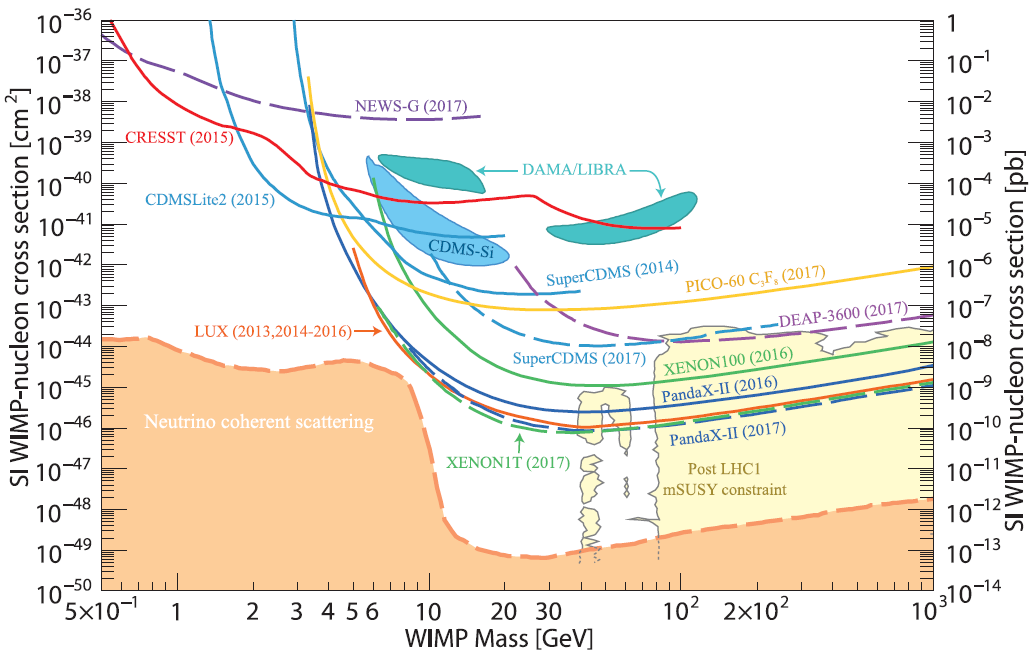}
		\caption[Dark matter-nucleon cross-section exclusion limits from various direct detection experiments from \cite{PDGroup}.]{Dark matter DM-nucleon cross-section exclusion limits from various direct detection experiments taken from \cite{PDGroup}. Black contours provide context for common SUSY models integrating constraints of ATLAS Run 1.}
		\label{fig:TotalDD}
	\end{figure}

	The lower bound of integration, $v_{\rm min}$ depends on the sensitivity of the detector. The typical bounds that direct detection experiments produce have a sharp fall on the lower mass end ($m_{\rm DM}\ll m_N$) of the DM-nucleon cross-section exclusion limits as seen in Figure \ref{fig:TotalDD} taken from \cite{PDGroup}. Below a threshold recoil energy that a particle of speed $v_{\rm min}$ produces \cite{Bertone:2010zza}, the experiment is not sensitive enough to measure this recoil which explains the sharp fall near lower mass in Figure \ref{fig:TotalDD}. On the other hand, Figure \ref{fig:TotalDD} shows that towards higher masses ($m_{\rm DM}\gg m_N$) the exclusionary limits of direct detection experiments also become weaker because the total rate is proportional to the number density of DM, implying that for fixed $\rho_0$, the number density scales as $1/m_{\rm DM}$ \cite{Bertone:2010zza}.
	
	\vspace{-5mm}
	
	Despite the difficulty of separating background events from true DM scatterings, direct detection experiments currently provide among the strongest bounds on DM candidates. Nevertheless, distinguishing DM-nucleon interactions will become much more challenging when the ``neutrino floor'' is reached as indicated by the orange shaded region in Figure \ref{fig:TotalDD}. Beyond the neutrino floor background events caused by neutrino scatterings become difficult to discriminate from DM scatterings. 

	Indirect detection experiments search for byproducts of possible DM annihilations from regions expected to have higher concentrations of DM such as the Sun, dwarf spheroidal galaxies of the Milky Way, or the Milky Way center \cite{Bertone:2010zza,Bertone:2016nfn,Arcadi:2017kky}. The byproduct searches include neutrinos, antimatter, and gamma rays \cite{Bertone:2010zza}. Focusing on gamma rays, the differential flux expected from DM annihilations is given by \cite{Arcadi:2017kky}
	\begin{equation}\label{eq:IND}
	\frac{d\Phi}{d\Omega dE}=\frac{(\sigma v)_{\rm ann}}{8\pi m_{\rm DM}^2}\frac{dN}{dE}\int_{\rm l.o.s}ds\rho_{\rm DM}^2(\vec{r}(s,\Omega)),
	\end{equation}
	where $(\sigma v)_{\rm ann}$ is the DM annihilation cross-section, $\frac{dN}{dE}$ is the energy spectrum of photons from DM annihilations, and the integral is of the DM energy density squared taken over the line of sight. 
	
	Note that eq.~\eqref{eq:IND} depends on the DM energy density profile in the galaxy, which is a source of astrophysical uncertainty. Particle physics uncertainties come from the annihilation cross-section and the expected number of photons produced through various annihilation channels. Definitive signals for DM annihilations would be a pair of gamma rays with the same energy equal to candidate mass $m_{\rm DM}$ \cite{Bertone:2010zza}. Conversely, in the absence of excess signals, exclusionary limits are obtained by observing fluxes from a given source and using eq.~\eqref{eq:IND} to obtain constraints on the DM annihilation cross-section \cite{Arcadi:2017kky}.
	
	The final experimental method for probing DM we will discuss are collider searches. In particle accelerators such as the Large Hadron Collider (LHC), high energy beams of particles are collided and the byproducts of such collisions are analyzed. Collider triggers only directly detect charged particles and photons, thus DM, which is expected to be electrically neutral, cannot be probed directly by these experiments \cite{Kahlhoefer:2017dnp}. Nevertheless a signature of missing energy from mono-X searches   or decay width measurements provide an exciting prospect that DM had been created, and further guide our efforts at directly studying DM.
	
\begin{figure}[t]
		\centering
		\vspace{5mm}
		\includegraphics[width=\textwidth]{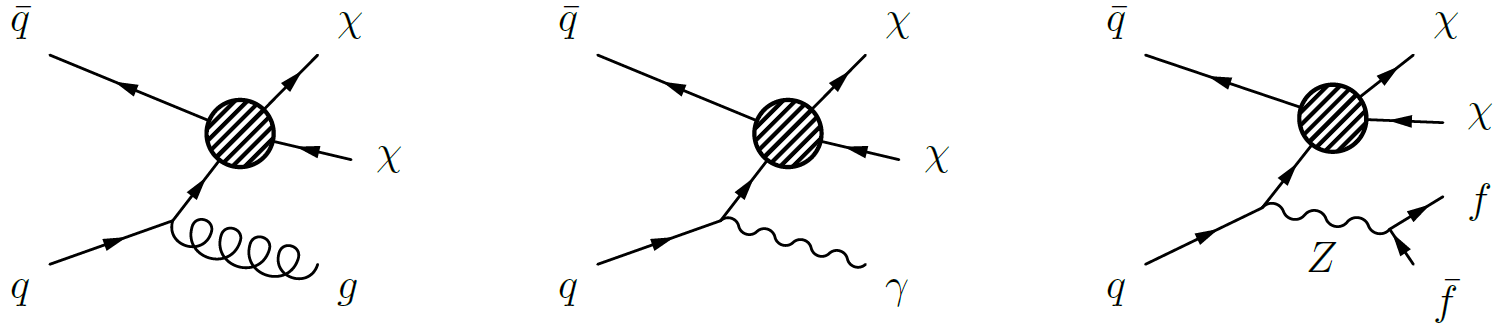}
		\vspace{1mm}
		\caption[Feynman diagrams for mono-X processes taken from \cite{Plehn:2017fdg}.]{Feynmann diagrams for mono-X processes taken from \cite{Plehn:2017fdg}.}
		\label{fig:monoX}
	\end{figure}

	The strategy of mono-X searches are to look for high transverse momentum $p_T$ events alongside missing transverse energy (e.g.~\cite{Fox:2011fx,Rajaraman:2011wf,Fox:2011pm,Haisch:2012kf,Arcadi:2017kky,Kahlhoefer:2017dnp}). The high transverse momentum can be in the form of jets, vector bosons, or even Higgs bosons. The Feynman diagrams for these processes can be seen in Figure \ref{fig:monoX} taken from \cite{Plehn:2017fdg}. An example of a definitive signal from these types of searches would be a mono-photon event where a high-$p_T$ photon is observed without any corresponding leptons \cite{Kahlhoefer:2017dnp}.
	
	Another method that collider searches use to look for missing energy is decay width analysis. By observing decays of a particle, it is possible to discern the fraction of decays going to an invisible sector by comparing the total and observed widths \cite{Arcadi:2017kky,Khachatryan:2016whc}. The Higgs boson decays offer a particularly promising search for DM since the Higgs may directly couple to it. We will study this model called the Higgs Portal in closer detail in Chapter \ref{Ch6}, where we highlight that collider searches are especially effective in constraining DM with mass less than half of that of the Higgs boson. 
	
	Although direct detection, indirect detection, and collider search experiments have thus far produced null results for observing DM, they nevertheless are continuing the search with vigor to confirm the conclusion of Clowe et al. \cite{Clowe:2006eq} of direct empirical proof of the existence of DM. These efforts are largely complimentary as collider searches favor lighter DM masses, indirect detection favors heavier DM masses, and direct detection is optimal for DM masses near heavy nuclear mass ranges. As experimental sensitivities increase and experiments collect more data in the coming years our knowledge of DM will be greatly enhanced, and we may even find a ``smoking gun'' signatures of dark matter. Potentially we are currently witnessing one the most exciting times in science with the hunt for dark matter.
	
	In this thesis we outline a new class of dark matter scenarios, along with a specific `Higgs portal' implementation, which leads to different predictions for the preferred mass range and coupling strength of dark matter. However, notably, in large parts of parameter space the matter dominated freeze-out scenario we highlight here remains discoverable by these standard searches for dark matter.

%% file: PhDthesisV2.0chpt02MD.tex
\chapter{Matter Domination in the Early Universe}
\label{Ch2}

\vspace{-5mm}
{\em This chapter presents original material in Sections \ref{SecIFE} and \ref{SecEarlyScen}, discussed in \cite{Hamdan:2017psw}, and further developed in this thesis.}
\vspace{5mm}

In this chapter we start by deriving useful relationships between time, scale factor, and temperature. Subsequently, we derive an interpolating version of the Friedmann equation which describes both radiation domination and matter domination in different limits. We then discuss the possible scenarios which can arise for dark matter, before studying these via the interpolating Friedmann equation in the next chapter.

\section{Relating Time, Scale Factor, and Temperature}

Consider the Friedmann-Lemaitre-Robertson-Walker (FLRW) cosmology of a perfect fluid with stress-energy tensor $T^{\mu}_{\nu}=\rm{diag}(\rho,-p,-p,-p)$, where $\rho(t)$ is the energy density of the fluid and $p(t)$ is its pressure. Then the $00$ component of the Einstein equation gives the Friedmann equation
\begin{equation}\label{eq:FEback}
\left(\frac{\dot a}{a}\right)^2+\frac{k}{a^2}=\frac{8 \pi G}{3}\rho,
\end{equation}
and $ii$ component
\begin{equation}\label{eq:iiback}
2\frac{\ddot a}{a}+\left(\frac{\dot a}{a}\right)^2+\frac{k}{a^2}=-8\pi Gp,
\end{equation}
where $k$ is the constant parameter of the FLRW metric which determines either a closed ($k=1$), open ($k=-1$), or flat universe ($k=0$), and $G=M_{{\rm pl}}^{-2}$ is the gravitational constant which may be taken as the inverse of planck mass squared. Subtracting equation \eqref{eq:FEback} from \eqref{eq:iiback} gives:
\begin{equation}\label{eq:FEminusiiback}
\frac{\ddot a}{a} = -\frac{4\pi G}{3}(\rho+p)_.
\end{equation}
Moreover, taking the time derivative of equation \eqref{eq:FEback},
\begin{equation}\label{tderFEback}
\frac{\ddot a}{a}=\frac{4\pi G}{3}\Big(\frac{a}{\dot a}\dot\rho + 2\rho\Big)_,
\end{equation}
and equating to equation \eqref{eq:FEminusiiback},
\begin{equation}\label{eq:tarp}
\frac{\dot a}{a}=\frac{\dot\rho}{p}_.
\end{equation}
For the relationship between energy density and pressure, one must solve for the equation of state. For simplicity, we may take the relationship to be defined by:
\begin{equation}\label{eq:pwrho}
p=w\rho,
\end{equation}
where $w$ is a constant that is $w=1/3$ for radiation, $w=0$ for matter, $w=-1$ for vacuum energy, and $w=-1/3$ for curvature. Substituting equation $p=w\rho$ into eq.~\eqref{eq:tarp} then integrating we find,
\begin{equation}\label{eq:rhoarel}
\frac{\rho}{\rho_{0}}=\left(\frac{a}{a_{0}}\right)^{-3(1+w)}_.
\end{equation}
Incidentally one could also derive equation \eqref{eq:rhoarel} by invoking the First Law of Thermodynamics, $d(\rho a^3)=-pd(a^3)$, or equivalently, $\partial_\nu T^{0\nu}=0$.

Substituting equation \eqref{eq:rhoarel} into equation \eqref{eq:FEback} gives:
\begin{equation}\label{eq:adiffeq}
\dot a^2=\frac{8\pi G}{3}\rho_{0}a_{0}^{3(1+w)}a^{-(1+3w)}+k.
\end{equation}
For $k=0$ and $w\neq -1$, integrating equation \eqref{eq:adiffeq} gives:
\begin{equation}\label{eq:avt}
a(t)\propto t^{\frac{2}{3}(1+w)^{-1}}
\end{equation}
and
\begin{equation}\label{eq:Hvt}
H=\frac{\dot a}{a}=\frac{2}{3}(1+w)^{-1}t^{-1}~.
\end{equation}

\vspace{-3mm}

We next derive a useful relation between the FLRW scale factor and temperature, which depends on the equation of state of the universe. Decoupled radiation has the temperature vs scale factor relation $T_{\rm R,decoupled}\propto a^{-1}$, and decoupled matter has the relation, $T_{\rm M,decoupled}\propto a^{-2}$. Here we derive the relation between temperature and scale factor for a thermal bath of both matter and radiation. Suppose, for the sake of simplicity, the universe is filled with an ideal gas at temperature $T$ of relativistic ($T\gg m_{\rm R}$) or non-relativistic particles ($T\ll M_{\rm NR}$) in thermal equilibrium. Further, label the entropy per particle in this universe $\mathcal{S}$, the number density of all particles $n$, and the relativistic and non-relativistic number densities $n_{\rm R}$ and $n_{\rm NR}$, respectively. Then, by the Second Law of thermodynamics, we have,
\begin{equation}\label{eq:dsrhop}
d\mathcal{S}=\frac{d(\varepsilon/n)+pd(1/n)}{T}
\end{equation}
where,
\begin{equation}\label{eq:endens}
\varepsilon=\varepsilon_{\rm R}+\varepsilon_{\rm NR}=3n_{\rm R}T+(3/2)n_{\rm NR}T
\end{equation}
is the energy density, split in relativistic and non-relativistic parts, and
\begin{equation}\label{eq:presdens}
p=n_{\rm R} T+n_{\rm NR}T
\end{equation}
is the pressure. Since $n=n_{\rm R}+n_{\rm NR}$ by assumption, define $y\equiv n_{\rm R}/n$, so that $1-y=n_{\rm NR}/n$. Thus, combining equations \eqref{eq:dsrhop}-\eqref{eq:presdens}, we have,	
\begin{equation}\label{eq:endensimp}
d\mathcal{S}=\left(\frac{3}{2}\right)\frac{d((1+y)T)}{T}+nd(1/n)_.
\end{equation}
Integrating,
\begin{equation}\label{eq:entperpart}
\mathcal{S}=\ln(CT^{(3/2)(1+y)}/n)
\end{equation}
where $C$ is a constant. During equilibrium, $d\mathcal{S}=0$, and $n\propto a^{-3}$, implying $T^{(3/2)(1+y)}a^{3}=const.$, or $T\propto a^{-2/(1+y)}$. Therefore in the limit that the number density of relativistic particles dominate the total number density, $y\approx 1$ and $T\propto a^{-1}$. In the opposite limit of non-relativistic domination of the number density, $y\approx 0$, and $T\propto a^{-2}$. Letting $\eta\equiv n_{\rm NR}/n_{\rm R} = (1-y)/y$, we can rewrite the above relation as $T\propto a^{-2(1+\eta)/(2+\eta)}$. Today, the number density of baryons to photons is of order $10^{-8}$, therefore it is safe to assume that throughout the history of the universe, $\eta\ll 1$, and that the relativistic particles dominate the number density of particles of the universe \cite{weinberg2008cosmology}. We will assume $\eta\ll 1$ throughout this text, hence $T \propto a^{-1}$ during thermal equilibrium.

Additionally, in our universe particles do not stay relativistic, and to a good approximation they become non-relativistic when $T\sim m$. To account for this change, it is common to define a function called the effective entropic degrees of freedom as 
\begin{equation}\label{eq:entrdof}
g_{* S}=\sum_{i=\rm bosons}g_{i}\left(\frac{T_i}{T}\right)^3+\frac{7}{8}\sum_{i=\rm fermions}g_{i}\left(\frac{T_i}{T}\right)^3_,
\end{equation} 
where the sum runs over relativistic particles, and $g_{i}$ is the degrees of freedom for the $i$th species \cite{Kolb:1990vq}. Then the entropy density of the universe can be expressed as follows
\begin{equation}\label{eq:EntropDensUni}
s=\frac{2\pi^2}{45}g_{* S}T^3_.
\end{equation}
Notice that this equation for entropy neglects non-relativistic degrees of freedom, which is a good approximation in our early universe \cite{Kolb:1990vq}. Equation \eqref{eq:EntropDensUni} gives that $g_{* S}a^3T^3=const.$, so that, precisely speaking, $T$ does not exactly evolve as $a^{-1}$ when the number of degrees of freedom change. In our work we will take $g_{* S}\approx 107$ throughout any principal period of interest and $T\propto a^{-1}$.

\section{Interpolating Friedmann Equation}
\label{SecIFE}

Consider a population of states $\Phi$, either a boson or fermion, with energy density $\rho_\Phi$ and suppose at some critical temperature $T_\star $ that $\rho_\Phi$ starts to evolve as matter, i.e.~it scales as $a^{-3}$ in terms of the FLRW scale factor. The simplest example is the case that $\Phi$ is some heavy decoupled species, in which case it become matter-like at $T_\star \sim m_\Phi$ and at temperatures $T>T_\star $ the component of $\Phi$ to the energy density is radiation-like, scaling as $a^{-4}$. It follows that at temperatures below $T_\star $ the Friedmann equation can be expressed as follows\footnote{Note the definition of $r$ differs slightly from that in \cite{Hamdan:2017psw}, but the two definitions coincide with the assumption that $\rho_X$ is negligible, an assumption which will typically hold.}
\begin{equation}\label{eq:FE}
H^2=\frac{8\pi}{3M_{\rm pl}^2}\left[\rho_R+\rho_{\Phi}+\rho_X\right]\simeq\frac{8\pi}{3M_{\rm pl}^2}\left[\rho_\star r\left(\frac{a_\star}{a}\right)^{4}+\rho_\star(1-r)\left(\frac{a_\star}{a}\right)^{3}\right]
\end{equation}
where $\rho_R$ and $\rho_X$ are the energy densities of the Standard Model radiation bath and the DM, respectively and we define, 
\begin{equation}
\rho_\star\equiv\rho_{\rm R}+\rho_{\Phi}|_{T=T_\star }
\label{rhostar}
\end{equation}
and 
\begin{equation}
r\equiv\rho_{\rm{R}}/(\rho_{\rm R}+\rho_{\Phi})|_{T=T_\star }.
\end{equation}
Thus $r$ represents the fraction of the energy in radiation at temperature $T_\star$ and $1-r$ is the fraction of energy in $\Phi$ at $T=T_\star$.
Equation \eqref{eq:FE} may also be written as
\begin{equation}\label{FEhS}
H^2\simeq H_\star ^2\left[r\left(\frac{a_\star}{a}\right)^{4}+(1-r)\left(\frac{a_\star}{a}\right)^{3}\right]_,
\end{equation}
in terms of $H_\star \equiv H|_{T=T_\star}$ which is given by
\begin{equation}\label{eq:FEf}
H_\star ^2\equiv \frac{8\pi}{3M_{{\rm pl}}^2}\rho_\star\equiv\frac{8\pi^3g_{*}(T_\star )}{90M_{{\rm pl}}^2}T_\star ^4,
\end{equation}
	where $g_{*}$ is the effective number of relativistic degrees of freedom
\begin{equation}\label{eq:EffDegFrRho}
g_{*}=\sum_{i=\rm bosons}g_{i}\left(\frac{T_i}{T}\right)^4+\frac{7}{8}\sum_{i=\rm fermions}g_{i}\left(\frac{T_i}{T}\right)^4
\end{equation}
which, like the entropic degrees of freedom, the sum runs over relativistic degrees of freedom. We call \eqref{eq:FE} the Interpolating Friedmann equation because it has the limit of matter domination (MD) when $r\rightarrow 0$, and radiation domination (RD) when $r\rightarrow 1$.

While the DM remains coupled to the thermal bath, it has a thermal distribution and the energy density is given as normal by
\begin{equation}\label{eq:dmrho}
\rho_X=m_Xn_{X}^{EQ}(x)=\left\{
\begin{array}{cc} m_X\frac{g}{(2\pi)^{3/2}}m_X^3x^{-3/2}e^{-x} & \quad T\ll m_X \\[5pt]
\frac{\pi^2}{30}gT^4 & \quad T\gg m_X \end{array}
\right.
\end{equation}
where $m_X$ is the DM mass.
During later times DM will undergo freeze-out, but throughout the parameter range relevant to freeze-out, the energy density of DM will be much smaller than that of $\rho_\star$, as we will show in the Chapter \ref{Ch4}. Thus we neglect the DM energy density in the Friedmann Equation, to aid in calculating freeze-out abundances. 

\section{Early Universe Scenarios with a Matter-like Species}
\label{SecEarlyScen}

Although precise definitions and derivations for terms used in this section will be shown in subsequent chapters, here we give a broad overview of the richer range of scenarios that the inclusion of the matter-like state $\Phi$ offers. Which scenario is realized depends on the ordering of the freeze-out temperature $T_F$, the temperature at which matter dominates the energy density $T_{\rm MD}$,  and the temperature of the bath at the point of $\Phi$ decays $T_{\Gamma}\equiv T(t\sim\Gamma_\Phi^{-1})$:

\begin{enumerate}[i).]
	
	\item {\em Radiation Domination}:  
	If $\Phi$ decays prior to DM freeze-out $T_{\Gamma}\gg T_F$, then DM decouples during radiation domination with $H\propto T^2$ and there is no period of $\Phi$ matter domination. 
	
	\item {\em Radiation Domination with dilution}:
	DM decouples during radiation domination, again $H\propto T^2$,  but $\Phi$ later evolves to dominate the energy density. Decays of $\Phi$ dilute or repopulate the DM freeze-out abundance  \cite{SuperHeavyBramante2017}. In this case DM decouples prior to both matter domination and $\Phi$ decays: $T_F \gg T_{\Gamma}, T_{\rm MD}$. 
	
	\item {\em Matter Domination}:
	DM decouples while $\Phi$ is matter-like and dominates the energy density of the universe, at which point $H\propto T^{\nicefrac{3}{2}}$ assuming adiabicity. This scenario  is realized for $T_{\rm MD} \gg T_F \gg  T_{\Gamma}$.
	
\end{enumerate}

It is also important to compare these quantities to the reheat temperature after $\Phi$ decays\footnote{We use $T_{\rm RH}$ throughout to refer to the heating associated to $\Phi$ decays  (rather than inflationary reheating).} $T_{\rm RH}\simeq\sqrt{M_{\rm Pl}\Gamma_\Phi}$. If one treats $\Phi$ decays as instantaneous, it appears that there is a discontinuous jump in the bath temperature from $T_\Gamma$ to $T_{\rm RH}$. This is an artifact of the sudden decay approximation. Using instead an exponential decay law this jump is absent, rather there is a smooth interpolation and the temperature never rises at any stage  \cite{Scherrer:1984fd}. The effect of $\Phi$ decays is seen as a reduction in the rate of cooling. 

For processes around $T\sim T_{\rm RH}$ the impact of non-instantaneous $\Phi$ decays must be taken into account. Specifically, for DM decoupling with $T_F\sim T_{\rm RH}$, the Boltzmann equations must be modified to include contributions to the number densities due to $\Phi$ decays, similar to \cite{GiudicePhysRevD.64.023508,McDonald:1989jd}. However, for processes active at temperatures away from this period, instantaneous decay is a fine approximation. In Chapter \ref{Ch5} we discuss the case in which there is not a clear separation between the times of freeze-out and reheating due to $\Phi$ decay, i.e.~$T_F\sim T_{\rm RH}$.

While scenarios (i) \& (ii) have been discussed in the literature,  to our knowledge, case (iii) remains largely unstudied. However, matter dominated freeze-out is a very general possibility which readily reproduces the DM relic density and thus we dedicate this thesis to the thorough study of case (iii).

We note here that the prospect of matter dominated freeze-out was remarked upon briefly in other contexts in \cite{Kamionkowski:1990ni,Co:2015pka}. Morever, this scenario is also similar in spirit to \cite{Chung:1998rq,GiudicePhysRevD.64.023508,Gelmini:2006pw} which studied DM freeze-out during inflationary reheating, in which case $H\propto T^{4}$.  Matter-dominated Freeze-out differs from these in several ways, most significantly, these other works consider the case in which the radiation bath is initially negligible. Here, rather, we assume the $\Phi$ states decay at times well after inflationary reheating, when there is a well established thermal bath, and that DM freeze-out occur significantly after inflationary reheating and while $\Phi$ decays are entirely negligible, thus preserving adiabicity.

Interestingly, the impact of an early period of matter domination on DM has been considered from different perspectives in recent papers, e.g.~\cite{Co:2015pka,SuperHeavyBramante2017,Randall:2015xza,Berlin:2016vnh,Mitridate:2017oky}. Also, an interesting variant, not captured in our list (i)-(iii), is if DM freeze-out occurs while the universe is dominated by an energy density redshifting faster than radiation  \cite{DEramo:2017gpl}, one example is kinetic energy (`kination') dominated, e.g.~\cite{Spokoiny:1993kt,Redmond:2017tja}.

%% file: PhDthesisV2.0chpt03FO.tex
	\chapter[Freeze-out of Dark Matter During Matter Domination]{Freeze-out of Dark Matter\\ During Matter Domination}
	\label{Ch3}
	
	\vspace{-5mm}

	{\em This chapter presents some of the main original results of this thesis: eqns.~(\ref{eq:xFO1MD}), (\ref{eq:YinfMD}) \& (\ref{eq:critDensZetaMD})  This has been discussed in  \cite{Hamdan:2017psw}, and is further developed in this thesis.}
	
	\vspace{5mm}

	If DM freeze-out occurs while the universe is dominated by an energy density which redshifts as matter-like, this changes the expansion rate of the universe $H$ and thus will impact the freeze-out dynamics. In this chapter we will study matter dominated freeze-out (MDFO) of DM  in a model independent manner, highlighting the major differences with conventional radiation dominated freeze-out (RDFO). Notably, we observed that the decay of the matter-like field which leads to the early period of matter domination can dilute the DM, thus allowing for smaller annihilation cross section or much heavier DM, while evading experimental searches. In subsequent chapters I will broadly identifying the range of viable parameter space and study this intriguing possibility of in the context of Higgs portal DM.
	
	We begin by explaining the freeze-out process and deriving an interpolating freeze-out temperature formula which has MD and RD limits in the first section. We then move to studying the Boltzmann equation and putting in a from conducive to cosmological studies. This will then allow us to calculate an interpolating abundance of DM in the third section, and the fourth section will use the abundance to present the relic density of DM in light of entropy injections. The final section of this chapter will be used to derive an equation for the dilution factor that quantifies entropy injections in terms of reheating.
	
	\section{Freeze-out Temperature}
	
	First we calculate the freeze-out temperature. Freeze-out is the process that a particle species of the universe undergoes through which the creation and annihilation processes of such a species become decisively rare, thus rendering the number of particles of such a species effectively constant. Consider an interaction of the form, $\Gamma\propto T^\gamma$, and Hubble parameter $-\dot{T}/T=H\propto T^\eta$. Then the number of interactions after comoving time $t$ is:
	\begin{equation}\label{eq:FOmotivate}
		N_{int}=\int_{t}^{\infty}dt'\Gamma(t')=\left(\frac{\Gamma}{H}\right)\bigg|_{T}T^{-(\gamma-\eta)}\int_{0}^{T}dT'(T')^{\gamma-\eta-1}=\left(\frac{\Gamma}{H}\right)\bigg|_{T}\frac{1}{(\gamma-\eta)}_.
	\end{equation}
	Assuming $\gamma-\eta>1$, equation \eqref{eq:FOmotivate} shows that less than one interaction is likely after the temperature drops below the temperature at which $\Gamma=H$. Therefore we define the freeze-out temperature, $T_{F}\equiv m_{X}x_{F}^{-1}$ implicitly using the condition:
	\begin{equation}\label{eq:FOcond}
		\Gamma(x_{F})\equiv H(x_{F})_.
	\end{equation}
	
	We will take the annihilation interaction rate for the dark matter candidate to be, 
	\begin{equation}\label{eq:annintrate}
		\Gamma_{\rm ann}=n_{X}^{\rm EQ}\langle\sigma_{\rm A}|\vec{v}|\rangle,
	\end{equation}
	where,
	\begin{equation}\label{eq:EQnumDens}
		n_{X}^{\rm EQ}(x)=\frac{g}{(2\pi)^{3/2}}m_X^3x^{-3/2}e^{-x}_,
	\end{equation}
	and the thermally average annihilation cross-section is parameterized as,
	\begin{equation}\label{eq:thermAvgPar}
		\langle\sigma_{\rm A}|\vec{v}|\rangle\equiv\sigma_{0}x^{-n}_.
	\end{equation}
	Moreover we define $x_\star \equiv m_{X}/T_\star $. Then, neglecting the small dark matter contribution and using the relation $T/T_\star=a_\star/a$, the Friedmann equation \eqref{FEhS} may be re-written as:
	\begin{equation}\label{eq:FEx}
	H=H_\star \sqrt{1-r}\left(\frac{x_\star }{x}\right)^{3/2}\left[\frac{r}{1-r}\left(\frac{x_\star }{x}\right)+1\right]^{1/2}_.
	\end{equation}
	Using equations \eqref{eq:annintrate}-\eqref{eq:FEx} in definition \eqref{eq:FOcond}, we find $x_F$ to be
	\begin{equation}\label{eq:xFO}
	x_F=\ln{\left[\frac{gm_X^3\sigma_{0}}{(2\pi)^{3/2}H_\star x_\star ^{3/2}}(1-r)^{-1/2}\left(x_F^{2n}+\frac{rx_\star }{1-r}x_F^{-1+2n}\right)^{-1/2}\right]}_.
	\end{equation}
	Equation \eqref{eq:xFO} can be solved numerically, but is highly insensitive to the parameter space we consider. Therefore we may approximate it iteratively, by initially guessing $x_F^{(0)}=1$ on the RHS, and getting that the LHS is
	\begin{equation}\label{eq:xFO1}
	x_F^{(1)}\approx\ln{\left[\frac{gm_X^3\sigma_{0}}{(2\pi)^{3/2}H_\star x_\star ^{3/2}}(1-r)^{-1/2}\left(1+\frac{rx_\star }{1-r}\right)^{-1/2}\right]}_.
	\end{equation}
	
	By taking the $r\rightarrow 1$ and $r\rightarrow 0$ limits of equation \eqref{eq:xFO1} we obtain the RD and MD limits, respectively:
	\begin{equation}\label{eq:xFO1RD}
	(x_F^{\rm RD})^{(1)}\equiv\lim_{r\rightarrow 1}x_F^{(1)}=\ln{\left[\frac{gm_{X}^3\sigma_{0}}{(2\pi)^{3/2}H_\star x_\star ^2}\right]}=
	\ln{\left[\frac{3}{4\pi^3}\sqrt{\frac{5}{2}}\frac{g}{\sqrt{g_{*}(T_\star )}}m_XM_{\rm pl}\sigma_{0}\right]}_,
	\end{equation}
	and
	\begin{equation}\label{eq:xFO1MD}
	(x_F^{\rm MD})^{(1)}\equiv\lim_{r\rightarrow 0}x_F^{(1)}=\ln{\left[\frac{gm_{X}^3\sigma_{0}}{(2\pi)^{3/2}H_\star x_\star ^{3/2}}\right]}=
	\ln{\left[\frac{3}{4\pi^3}\sqrt{\frac{5}{2}}\frac{g}{\sqrt{g_{*}(T_\star )}}\frac{m_X^{3/2}M_{\rm pl}\sigma_{0}}{\sqrt{T_\star }}\right]}_.
	\end{equation}
	\newpage
	The numerical solution to equation \eqref{eq:xFO}, using $n=0$ (the only case we consider in this work), with $m_X=T_\star =\sigma_{0}^{-1/2}=10^2$ GeV and $r\approx 0$ (MD), is $x_F=34.4$. Using the weak scale again, but with $r\approx 1$ (RD) yields $x_F=36.2$. Using the same equation, but with $n=0$, $m_X=T_\star =\sigma_{0}^{-1/2}=10^9$GeV and $r\approx 0$ gives $x_F=18.3$. Using the $10^9$GeV scale but with $r\approx 1$ gives $x_F=19.81$. 
	
	Had we used the approximations in eqns.~\eqref{eq:xFO1RD} and \eqref{eq:xFO1MD}, the only significant difference would be for the case of RD, with $(x_F^{\rm RD})^{(1)}\approx34.4$ rather than the $36.2$ above in the weak scale, and $(x_F^{\rm RD})^{(1)}\approx18.3$ rather than $19.8$ above in the $10^9$GeV scale. In summary, the approximation eqns.~\eqref{eq:xFO1RD} and \eqref{eq:xFO1MD} are in excellent agreement with the exact results in the MD case, and only differs by $\sim5\%$ in the RD case. Moreover, using the approximation eqns.~\eqref{eq:xFO1RD} and \eqref{eq:xFO1MD} yields very little change between the RD and MD cases (they agree exactly up to the significant digits we used), while the MD and RD case differ by the same $\sim5\%$ in the exact numerical solution of eq.~\eqref{eq:xFO}.

	Because there is little change in the values of $x_F$ throughout the wide range of parametrics we consider, we will take the characteristic $x_F$ of the Weak scale to be $x^{\rm char, weak}_F\simeq 35$, and in the $10^9$GeV scale to be,  $x^{\rm char, 10^9GeV}_F\simeq 19$. Moreover, to ensure a finite freeze-out temperature when using approximation \eqref{eq:xFO1MD}, in the case of MD we need that $x_F^{\rm MD}>0$
	or, equivalently, 
	\begin{equation}\label{eq:finitexFO1MDb}
	T_\star \lesssim\frac{45}{32\pi^6}\frac{g^2}{g_{*}(T_\star )}m_X^3M_{\rm pl}^2\sigma_{0}^2.
	\end{equation}

	For an upper bound on $x_\star $ in the case of MD we must ensure that 	$x_\star \leq x_F$ so that the freeze-out temperature is lower than the critical temperature. Although this condition cannot be written analytically, solving it iteratively gives a rough upper bound to $x_\star $ as
	\begin{equation}\label{eq:RoughUpperBoundxs}
	x_\star \lesssim \ln{\left[\frac{3}{4\pi^3}\sqrt{\frac{5}{2}}\frac{g}{\sqrt{g_{*}(T_\star )}}m_XM_{\rm pl}\sigma_{0}\right]}_.
	\end{equation}
	More stringent constraints will be derived in the Constraints segment of this work.
	
	\section{The Boltzmann Equation}
	
	\vspace{-5mm}
	
	In cosmology the Boltzmann Equation is used to describe the change in the phase space distribution, and consequently the number density, of a species in scenarios beyond thermal equilibrium. The relativistic Boltzmann Equation is
	\begin{equation}\label{eq:BEorig}
		\boldsymbol{\hat{\rm{L}}}[f]=\boldsymbol{\rm C}[f]_,
	\end{equation}
	where $f$ is the phase space distribution function, $\boldsymbol{C}$ is the collision operator, and $\boldsymbol{\hat L}$ is the relativistic Liouville operator:
	\begin{equation}\label{eq:LopOrig}
		\boldsymbol{\hat{\rm{L}}}=p^{\alpha}\partial_{\alpha}-\Gamma^{\alpha}_{\beta\gamma}p^{\beta}p^{\gamma}\frac{\partial}{\partial p^{\alpha}}_.
	\end{equation}
	Since the FLRW cosmology is homogeneous and isotropic, $f=f(E,t)$ if follow that
	\begin{equation}\label{eq:Lhom}
		\boldsymbol{\hat{\rm{L}}}[f(E,t)]=E\frac{\partial f}{\partial t}-\frac{\dot a}{a}|\vec{p}|^{2}\frac{\partial f}{\partial E}=\boldsymbol{\rm C}[f]_.
	\end{equation}
	Multiplying the LHS of eq.~\eqref{eq:Lhom} by $g/((2\pi)^3E)$ and integrating gives
	\begin{equation}\label{eq:Lsimpa}
		\frac{g}{(2\pi)^3}\int \frac{d^{3}p}{E}\boldsymbol{\hat{\rm{L}}}[f]  =\frac{g}{(2\pi)^3}\int d^{3}p\frac{\partial f}{\partial t}-\frac{g}{(2\pi)^3}\int \frac{d^{3}p}{E}\frac{\dot a}{a}|\vec{p}|^{2}\frac{\partial f}{\partial E}_.
	\end{equation}
	The first term on the RHS of equation \eqref{eq:Lsimpa} is just $\frac{\partial}{\partial t}\frac{g}{(2\pi)^3}\int d^3pf=\frac{\partial n}{\partial t}$, where $n$ is the number density. For the second term on the RHS of equation \eqref{eq:Lsimpa} we will integrate by parts and use the relation $EdE=pdp$:
	\begin{equation}\label{eq:Lsimpb}
		-\frac{g}{(2\pi)^3}\int \frac{d^{3}p}{E}\frac{\dot a}{a}|\vec{p}|^{2}\frac{\partial f}{\partial E}=\frac{-Hg(4\pi)}{(2\pi)^3}\int \frac{dpp}{E}p^3\frac{\partial f}{\partial E}=\frac{Hg(4\pi)}{(2\pi)^3}\int dE\frac{\partial p^3}{\partial E}f.
	\end{equation}
	Taking the derivative of $p^3$ and using $EdE=pdp$ again:
	\begin{equation}\label{eq:Lsimpc}
		\frac{3Hg(4\pi)}{(2\pi)^3}\int dEEpf=3H\frac{g}{(2\pi)^3}\int d^3pf=3Hn.
	\end{equation}
	Therefore the Boltzmann equation becomes
	\begin{equation}\label{eq:BEcoll}
		\dot n+3Hn=\frac{g}{(2\pi)^3}\int \frac{d^3p}{E}\boldsymbol{\rm C}[f]_.
	\end{equation}
	
	Focusing on the collision term, and considering the Boltzmann equation for a species $X$, the collision term for the most general process $X+a+b+\cdots\longleftrightarrow i+j+\cdots$ may be written as
	\begin{multline}\label{eq:CollTerm}
		\frac{g}{(2\pi)^3}\int \frac{d^3p}{E}\boldsymbol{\rm C}[f]=-\int d\Pi_{X}d\Pi_{a}d\Pi_{b}\cdots d\Pi_{i}d\Pi_{j}\cdots\\
		\times(2\pi)^4\delta^{4}(p_X+p_a+p_b\cdots-p_i-p_j\cdots)\\
		\times[|\mathcal{M}|^{2}_{X+a+b+\cdots\rightarrow i+j+\cdots}f_{a}f_{b}\cdots f_X(1\pm f_i)(1\pm f_j)\cdots]\\
		-|\mathcal{M}|^2_{i+j+\cdots\rightarrow X+a+b+\cdots}f_if_j\cdots(1\pm f_a)(1\pm f_b)\cdots(1\pm f_X)]
	\end{multline}
	where the various $f_s$ are phase space densities for their respective species $s$; $(+)$ or $(-)$ in the $(1\pm f_s)$ are used for bosons and fermions, respectively; finally, for $g$ internal degrees of freedom:
	\begin{equation}\label{eq:CollVol}
		d\Pi\equiv g\frac{1}{(2\pi)^3}\frac{d^3p}{2E}_.
	\end{equation}
	
	Following \cite{Kolb:1990vq} we will make a series of reasonable assumptions to simplify the collision part of the Boltzmann equation. The first is $T$-symmetry in any interactions. The second is the use of Maxwell-Boltzmann (MB) statistics, thus we ignore the emission factors $1\pm f_{s}\approx 1$, where for MB statistics, $f_{s}=\exp{[-(E_s-\mu_s)/T]}$. The third assumption is that any dark matter candidate we consider will be stable relative to the age of the universe, so that decaying dark matter will not be considered. The fourth assumption is $C$-symmetry, so that we treat dark matter particles and anti-particles on the same footing, and with the same abundances. The fifth assumption is that the chemical potentials of the annihilation products are zero, as well as the initial chemical potential of the dark matter species (though the DM chemical potential changes when the temperature of the bath drops well below its mass \cite{Bernstein:1985th}). The sixth assumption is that the dark matter candidate is in \emph{kinetic}\footnote{It is possible to maintain a MB thermal distribution for DM at the temperature of the bath, termed kinetic equilibrium, while not maintaining an equilibrium balance of chemical potentials, termed chemical equilibrium \cite{Bernstein:1985th}.} equilibrium with the rest of the thermal bath, sharing the temperature of photons and keeping a MB thermal distribution. Making all these assumptions, the Boltzmann equation can be written as
	\begin{equation}\label{eq:BEn}
		\dot n +3Hn=-\langle\sigma_{\rm A}|\vec{v}|\rangle[n^2-(n^{\rm EQ})^2]_,
	\end{equation}
	where $\langle\sigma_{\rm A}|\vec{v}|\rangle$ is the thermally-averaged annihilation cross-section.
	
	To mod out the change in number density due to expansion we define the dimensionless $Y\equiv n/s$, termed abundance. Moreover, define $x\equiv m_X/T$ so that we may consider only dimensionless parameters. When the entropy per comoving volume is conserved it follows that
	\begin{equation}\label{eq:BEmid}
		\dot n +3Hn=s\dot Y=s\frac{dY}{dx}\frac{dx}{dt}_.
	\end{equation}
	We thus finally arrive at the standard BE used for cosmological studies
	\begin{equation}\label{eq:BEy}
		\frac{dY}{dx}=-\left(\frac{dt}{dx}\right)\frac{\langle\sigma_{\rm A}|\vec{v}|\rangle}{s}[n^2-(n^{\rm EQ})^2]=-\left(\frac{dt}{dx}\right)\langle\sigma_{\rm A}|\vec{v}|\rangle s[Y^2-Y_{\rm EQ}^2]_.
	\end{equation}
	
	\section{Freeze-out Abundance}
	\label{SecAbundance}
	
	Next we wish to calculate the freeze-out abundance. Let $\Delta\equiv Y-Y_{EQ}$, then equation \eqref{eq:BEy}, using equations \eqref{eq:EntropDensUni} and \eqref{eq:thermAvgPar}, may be written:
	\begin{equation}\label{eq:delBEgen}
		\Delta'=-Y_{\rm EQ}'-\left(\frac{dt}{dx}\right)\sigma_{0}x^{-n} \left(\frac{2\pi^2}{45}g_{* S}m_X^3x^{-3}\right)\Delta[2Y_{\rm EQ}+\Delta]_.
	\end{equation}
	Furthermore, for the $\frac{dt}{dx}$ relation, we use that $a\propto T^{-1}$ (hence $H=-\dot T/T$) to arrive at
	\begin{equation}\label{eq:dxdt}
		\frac{dx}{dt}=\frac{m_X}{T}\left(\frac{-\dot T}{T}\right)=xH.
	\end{equation}
	From equations \eqref{eq:delBEgen} and \eqref{eq:dxdt} we have
	\begin{equation}\label{eq:delBE}
		\Delta'=-Y_{\rm EQ}'-\left[\frac{2\pi^2g_{*S}m_X^3\sigma_{0}}{45H_\star x_\star ^{3/2}}\right](1-r)^{-1/2}\left(1+\frac{r}{(1-r)}\frac{x_\star }{x}\right)^{-1/2}x^{-5/2-n}\Delta[2Y_{\rm EQ}+\Delta]_,
	\end{equation}
	where primes stand for derivative with respect to $x$. Define,
	\begin{equation}\label{eq:lambdaBE}
		\lambda\equiv\frac{2\pi^2g_{*S}m_X^3\sigma_{0}}{45H_\star x_\star ^{3/2}}=\frac{\sqrt{\pi}}{3\sqrt{5}}\frac{g_{*S}}{\sqrt{g_{*}(T_\star )}}\frac{M_{{\rm pl}}m_X^{3/2}\sigma_{0}}{\sqrt{T_\star }}_.
	\end{equation}
	Since we are interested the freeze-out abundance we consider primarily the domain $x\gg x_F$. In this regime we have $Y\gg Y_{\rm EQ}$, so that $\Delta\approx Y$. We may also neglect $Y_{\rm EQ}'$ in the late time regime. Then equation \eqref{eq:delBE} approximates to
	\begin{equation}\label{eq:delDiffEq}
		\Delta'\approx -\lambda (1-r)^{-1/2}\left(1+\frac{r}{(1-r)}\frac{x_\star }{x}\right)^{-1/2}x^{-5/2-n}
		\Delta^2
	\end{equation}
	Integrating equation \eqref{eq:delDiffEq}, we arrive at $Y(x\rightarrow\infty)\equiv Y_{\infty}$
	\begin{equation}\label{eq:Yinf}
		Y_{\infty}\approx\Delta_{\infty}\approx\left[\lambda (1-r)^{-1/2}\int_{x_F}^{\infty}dx\left(1+\frac{r}{(1-r)}\frac{x_\star }{x}\right)^{-1/2}x^{-5/2-n}
		\right]^{-1}_.
	\end{equation}
	Note that we used $Y_{\infty}\ll Y_{x_F}$, in arriving at equation \eqref{eq:Yinf}. The integral on the RHS of equation \eqref{eq:Yinf} is not particularly useful in its present form, with Mathematica finding,
	\begin{equation}\label{eq:intYinf}
		\int_{x_F}^{\infty}dx\left(1+\frac{r}{(1-r)}\frac{x_\star }{x}\right)^{-1/2}x^{-5/2-n}=\left(\frac{-rx_\star }{1-r}\right)^{-3/2-n}B(-rx_\star /(1-r)x_F; 3/2+n, 1/2),
	\end{equation}
	where $B(\cdot;\cdot,\cdot)$ is the Incomplete Beta function. A more enlightening form comes when $n=0$,
	\begin{multline}\label{eq:intYinf0}
		\int_{x_F}^{\infty}dx\left(1+\frac{r}{(1-r)}\frac{x_\star }{x}\right)^{-1/2}x^{-5/2}\\=\frac{(1-r)}{rx_\star x_F}\left(\sqrt{x_F+rx_\star /(1-r)}-x_F\sqrt{(1-r)/rx_\star }\arcsinh{\sqrt{rx_\star /(1-r)x_F}}\right)_.
	\end{multline}
	
		To see the behavior of $Y_{\infty}$ in the limits of RD and MD, consider first the standard radiation dominated case (as in \cite{Scherrer:1985zt})
	\begin{equation}\label{eq:YinfRD}
		Y^{\rm RD}_{\infty}\equiv\lim_{r\rightarrow 1}Y_{\infty}=\frac{\sqrt{x_\star }}{\lambda}(n+1)x_F^{n+1}
		=3\sqrt{\frac{5}{\pi}}\frac{\sqrt{g_{*}(T_\star )}}{g_{*S}(T_F)}\frac{(n+1)x_F^{n+1}}{M_{{\rm pl}}m_X\sigma_{0}}_.
	\end{equation}
	Finally for the freeze-out abundance in the MD case we have
	\begin{equation}\label{eq:YinfMD}
		Y^{\rm MD}_{\infty}\equiv\lim_{r\rightarrow 0}Y_{\infty}=\frac{1}{\lambda}(n+3/2)x_F^{n+3/2}=
		3\sqrt{\frac{5}{\pi}}\frac{\sqrt{g_{*}(T_\star )}}{g_{*S}(T_F)}\frac{(n+3/2)x_F^{n+3/2}}{M_{{\rm pl}}m_X\sigma_{0}\sqrt{x_\star }}_.
	\end{equation}
	One immediate difference is that $Y^{\rm MD}_{\infty}\propto (x_F^{\rm MD})^{n+\nicefrac{3}{2}}$, whereas in the radiation dominated case $Y^{\rm RD}_{\infty}\propto (x_F^{\rm RD})^{n+1}$.  Another notable difference is  that  $Y^{\rm MD}_{\infty}$ has a new parameter dependence, namely varying with the quantity $x_\star$.
	
	\section{Dark Matter Relic Density}

	Even if freeze-out occurred during MD, to recover the successful Big Bang Nucleosynthesis (BBN) paradigm RD must be restored at its onset. This is possible through the decay of a massive particle species or coherent oscillations of a scalar field, thus injecting a large amount of entropy dominated by light degrees of freedom \cite{SuperHeavyBramante2017}.
	
	Not only do entropy injections ensure RD at the inception of BBN, but they may also considerably increase the parameter space. Suppose there is an instant decay of a massive particle species (or an instantaneously decaying coherently oscillating scalar field), such that the products of this decay are relativistic and do not repopulate the candidate DM species (the exact parametrics for ensuring this scenario will be discussed in the Constraints segment of this work). Then the number density of candidate DM stays fixed, while the entropy density increases by the inverse of
	\begin{equation}\label{eq:zetaBegin}
		\zeta\equiv s_{\rm before}/s_{\rm after}.
	\end{equation}
	Then, the relic abundance relative to the computed freeze-out abundance above (label it $Y^F_{\infty}$ for this section) is
	\begin{equation}\label{eq:YrelicZeta}
		Y^{\rm relic}_{\infty}=n_X/s_{\rm after}=\zeta n_X/s_{\rm before}=\zeta Y^F_{\infty}.
	\end{equation}
	
	\vspace{-4mm}	
	The DM abundance $Y$ can be re-expressed in terms of 
	\begin{equation}\label{eq:standardCritDens}
		\Omega_Xh^2=s_0m_XY^{\rm relic}_\infty/\rho_{\rm critical},
	\end{equation}
	by scaling with the critical density $\rho_{\rm critical} \simeq8.13\times10^{-47}$GeV$^4$ 
	and the entropy density today $s_0 \simeq2.29\times10^{-38}$GeV$^3$. Consequently, this changes the relic density by a factor $\zeta$,
	\begin{equation}\label{eq:critDensZeta}
		\Omega_X^{\rm relic}h^2=\zeta\Omega_X^Fh^2.
	\end{equation}
	
\begin{figure}[h]
		\centering
		\vspace*{2cm}
		\includegraphics[width=0.95\textwidth]{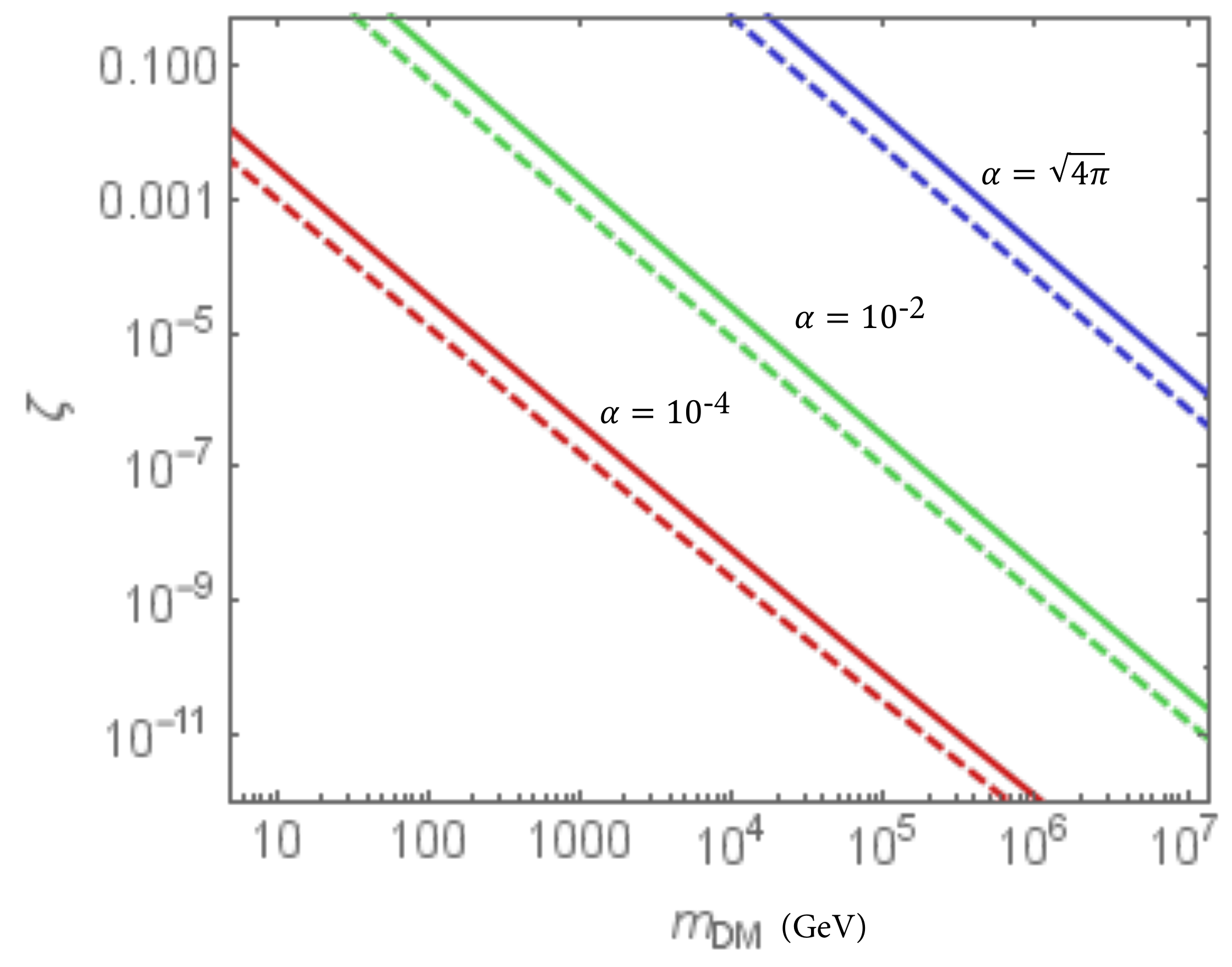}
		\caption[Plot of dark matter mass $m_{X}$ vs dilution $\zeta$  in matter dominated freeze-out.]{$m_{X}$ vs $\zeta$ plot in MD FO case for $\alpha=\sqrt{4\pi}$ (blue); $\alpha=10^{-2}$ (green); and $\alpha=10^{-4}$ (red). The dashed lines are for $x_\star =1$ and solid lines, $x_\star =10$. The dilution factor $\zeta$ allows for much heavier DM, given the above coupling strengths, in order to match the CDM relic density observed today.}
		\label{fig:zetaVSmDM}
	\end{figure}

\begin{figure}[h]
		\centering
		\vspace*{4cm}
		\includegraphics[width=0.95\textwidth]{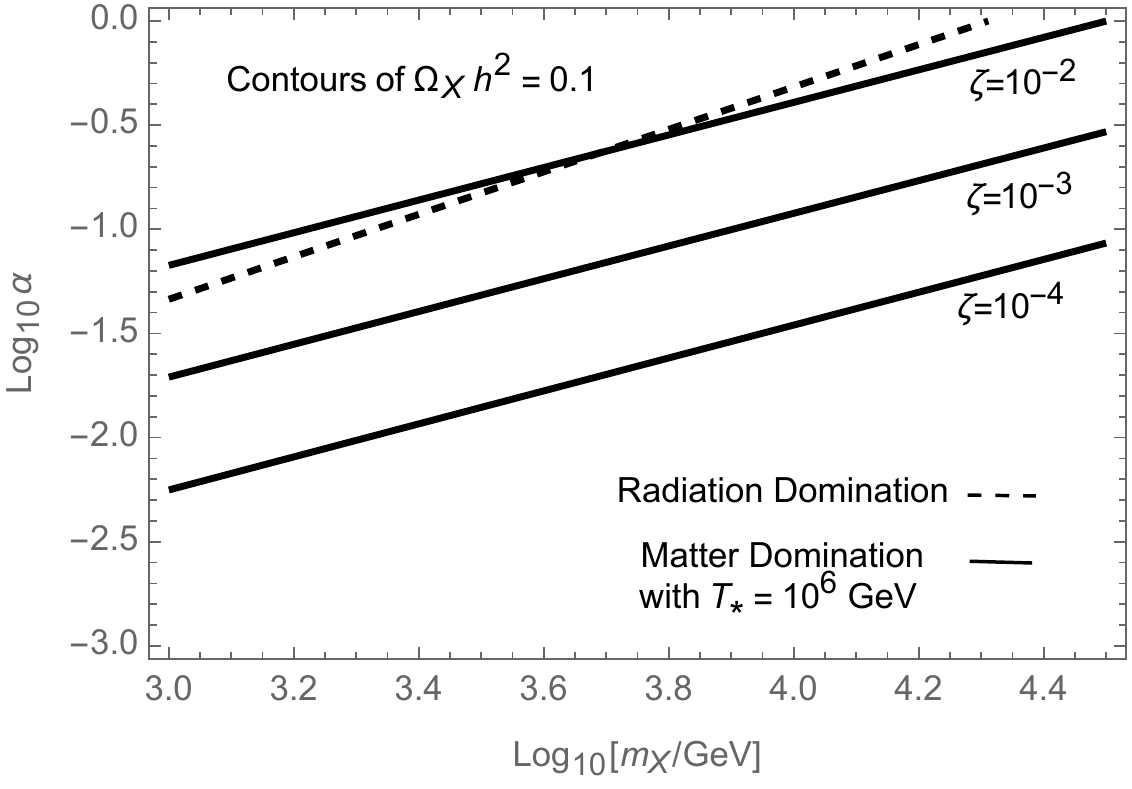}
		\caption[Contours of $\Omega_X h^2=0.1$ for matter and radiation dominated freeze-out]{Contours of $\Omega_X h^2=0.1$ for matter (solid) and radiation (dashed) dominated freeze-out scenarios as we vary DM mass and coupling $\alpha$. Matter domination has two extra parameter freedoms, we fix $T_\star=10^6$ GeV and vary $\zeta$.}
		\label{fig:3}
	\end{figure}

	Thus with an entropy injection following radiation or matter dominated freeze-out, the final abundance of dark matter can be drastically altered via the dilution factor
	\begin{equation}\label{eq:critDensZetaRD}
		(\Omega_X^{\rm RD})^{\rm relic}h^2\sim 0.1\cdot\left(\frac{m_X}{10^6{\rm{GeV}}}\right)^{2}\left(\frac{0.1}{\alpha}\right)^{2}\left(\frac{\zeta}{10^{-5}}\right)~,
	\end{equation}
	\begin{equation}\label{eq:critDensZetaMD}
		(\Omega_X^{\rm MD})^{\rm relic}h^2\sim 0.1\cdot\left(\frac{m_X}{10^6{\rm{GeV}}}\right)^{2}\left(\frac{0.1}{\alpha}\right)^{2}\left(\frac{10^2}{x_\star}\right)^{1/2}\left(\frac{\zeta}{10^{-5}}\right)~.
	\end{equation}
	where we parameterized $\sigma_0=\alpha^2/m_X^2$, set $g_*\sim g_{*S}\simeq107$, $n=0$, and fixed $x_F$ to a characteristic value of 27. Figures \ref{fig:zetaVSmDM} and \ref{fig:3} show the MD parametrics.

	\clearpage
	\clearpage
	
	\section{The Dilution Factor}
	
	We next derive specific relations for the $\zeta$ parameter discussed above. To find an expressions for $\zeta$ we  derive expressions for $T_{\Gamma}\equiv T(H=\Gamma)$ and  the reheat temperature $T_{\rm RH}$, being the temperature of the decay products of $\Phi$ assuming instantaneous reheating.
	
	Decay of the $\Phi$ matter occurs when $H\equiv \Gamma$. Subscripts of $\Gamma$ represent the value of the corresponding parameter at the time of decay. The scale factor at the time of decay may be obtained by solving the Friedmann equation when $H=\Gamma$. To obtain a solution, change variables in the Friedmann equation using the transformation,
	\begin{equation}\label{eq:conjfracM}
		\bar{f}_{\Phi}\equiv\frac{(1-r)}{r}\left(\frac{a}{a_\star}\right)_,
	\end{equation}
	so that the Friedman equation can be written as,
	\begin{equation}\label{eq:friedpoly}
		\xi\bar{f}_{\Phi}^{4}-\bar{f}_{\Phi}-1=0,
	\end{equation}
	where 
	\begin{equation}\label{eq:eta}
		\xi\equiv\left(\frac{\Gamma}{H_\star}\right)^2\frac{r^3}{(1-r)^4}_,
	\end{equation}
	The only real and positive solution is, in exact form,
	\begin{equation}\label{eq:Agam}
		\left(\frac{a_\star}{a}\right)^{-1}_{\Gamma}=\frac{r}{1-r}\frac{1}{2(6)^{1/3}\sqrt{\xi}q^{1/6}}\left[\sqrt{p}+\left(8(3)^{1/3}\xi-2^{1/3}q^{2/3}+12\sqrt{\frac{\xi q}{p}}\right)^{1/2}\right]
	\end{equation}
	where,
	\begin{equation}
		\begin{aligned}
			q&\equiv9\xi+\sqrt{3}\xi\sqrt{27+256\xi}_,\\
			p&\equiv-8(3)^{1/3}\xi+2^{1/3}q^{2/3}_.
		\end{aligned}
	\end{equation}
	Equation (\ref{eq:Agam}) has simpler expressions in the extreme limits of $r$, but also when the Friedmann equation is dominated by the matter component. The latter condition places the constraint that $\Gamma$ is sufficiently small relative to $H$ so that matter will come to dominate before decay. Thus, we may expand equation (\ref{eq:Agam}) in powers of $\Gamma/H_\star $ around zero. Doing so we obtain the approximation,
	\begin{equation}\label{eq:AgamTaylor}
		\left(\frac{a_\star}{a}\right)_\Gamma\approx\frac{1}{(1-r)^{1/3}}\left(\frac{\Gamma}{H_\star }\right)^{2/3}-\frac{1}{3}\frac{r}{(1-r)^{5/3}}\left(\frac{\Gamma}{H_\star }\right)^{4/3}+\mathcal{O}\left(\left(\frac{\Gamma}{H_\star }\right)^{8/3}\right)
	\end{equation}
	Moreover, when $r\ll 1$, we may expand equation \eqref{eq:Agam} in powers of $r$ thusly
	\begin{equation}\label{eq:AgamTaylorR}
		\left(\frac{a_\star}{a}\right)_\Gamma\approx\left(\frac{\Gamma}{H_\star }\right)^{2/3}+\frac{1}{3}r\left(\frac{\Gamma}{H_\star }\right)^{2/3}+\mathcal{O}(r^{2}).
	\end{equation}
	Notice that $(1-r)^{-1/3}\approx 1+r/3+\mathcal{O}(r^2)$, so that both expansions above approximately agree to first order terms. Finally, since $T/T_\star=a_\star /a$, the temperature of the thermal bath at decay is,
	\begin{equation}\label{eq:Tgam}
		T_{\Gamma}=T_\star \left(\frac{a_\star}{a}\right)_{\Gamma.}
	\end{equation}

	We will assume the $\Phi$ matter decays to radiation instantaneously, then the temperature of the decay products is
	\begin{equation}\label{eq:TD}
		T_{\rm RH}\equiv\left(\frac{30}{\pi^2g_{*}(T_{\rm RH})}\rho_{\Phi}|_{t\sim\Gamma^{-1}}\right)^{1/4}=\left(\frac{30}{\pi^2g_{*}(T_{\rm RH})}(1-r)\rho_{\star}\left(\frac{a_\star}{a}\right)_{\Gamma}^3\right)^{1/4}_.
	\end{equation}
	This equation can be further simplified,
	\begin{equation}
		T_{\rm RH}=\left((1-r)\hat{\gamma}\left(\frac{a_\star}{a}\right)_{\Gamma}^3\right)^{1/4}T_\star ,
		\label{76}
	\end{equation}
	where,
	\begin{equation}\label{eq:gamhat}
		\hat{\gamma}\equiv\frac{g_{*}(T_\star)}{g_{*}(T_{\rm RH})}_.
	\end{equation}
	This radiation energy subsequently thermalizes instantaneously with the radiation present right before decay. We will generally be interested in the case in which radiation from decay is much larger than the radiation present before decay.
	
	Given the expressions for $T_{\rm RH}$ and $T_{\Gamma}$ the dilution parameter $\zeta$ can be expressed 
	\begin{equation}\label{eq:zeta1}
		\zeta\equiv\frac{s_{\rm before}}{s_{\rm after}}=\frac{[s_{R}]_{\rm before}}{[s_{\rm RH}]_{\rm after}}=\frac{\frac{2\pi^2}{45}g_{*S}(T_\Gamma)T_\Gamma^3}{\frac{2\pi^2}{45}g_{*S}(T_{\rm RH})T_{\rm RH}^3}_,
	\end{equation}
	where,
	\begin{equation}\label{eq:srh}
		s_{\rm RH}\equiv\frac{2\pi^2}{45}g_{*S}(T_{\rm RH})T_{\rm RH}^3,
	\end{equation}
	is the entropy in the Reheat radiation after decay and thermalization. Simplifying equation (\ref{eq:zeta1}) we obtain,
	\begin{equation}\label{eq:zeta2}
		\zeta=\left(\frac{g_{*S}(T_\Gamma)}{g_{*S}(T_{\rm RH})}\right)\left(\frac{T_{\Gamma}}{T_{\rm RH}}\right)^{3}_.
	\end{equation}
	Finally, for $\Gamma/H_\star \ll1$  so that we use only the first term of approximation \eqref{eq:AgamTaylor}, we have,
	\begin{equation}\label{eq:zetaapprox}
		\zeta\approx\frac{g_{*}(T_\Gamma)/g_{*}(T_{\star})}{(1-r)}\frac{T_{\rm RH}}{T_\star }_,
	\end{equation}
	where we have used $g_*\approx g_{*S}$ and the following relationship that is a consequence of $\Gamma/H_\star \ll1$:
	\begin{equation}\label{eq:GammaTrh}
		\Gamma\approx\frac{H_\star }{\sqrt{\hat{\gamma}}T_\star ^2}T_{\rm RH}^2.
	\end{equation}
	
	It follows that $\zeta$ is parametrically
	\begin{equation}
	\zeta  \sim 10^{-5}\times \left(\frac{ 0.1}{1-r} \right) \left(\frac{ T_{\rm RH}}{1~{\rm GeV}} \right) \left(\frac{10^8~{\rm GeV}}{T_\star} \right).
	\label{zeta}
	\end{equation}
	Thus although  the parameter $\zeta$, critical for setting the DM relic density, can take a wide range of values, it is not unbounded. Indeed, there are several consistency conditions and constraints which must be satisfied for matter dominated DM freeze-out to be viable. Note that in all cases of interest we have that $T_\star\gg  T_{\rm RH}, T_{\rm FO}, T_{\Gamma}$. In the next chapter we will explore the wider range of theoretical and experimental constraints on this scenario.

%% file: PhDthesisV2.0chpt04Constraints.tex
\chapter{Constraints on Matter Dominated Freeze-out}
\label{Ch4}

\vspace{-5mm}

{\em This chapter presents original results which  have been discussed briefly in  \cite{Hamdan:2017psw}, but is significantly expanded upon in this thesis.}

\vspace{5mm}

In this chapter we enumerate the constraints on the Matter Dominated Freeze-out scenario. The following constraints provide the strongest model independent bounds on the parameter space: 
\begin{enumerate}
	\item Freeze-out of dark matter before decay of $\Phi$ matter. 
	\vspace{-1mm}
	\item Freeze-out of dark matter during matter domination. 
	\vspace{-1mm}
	\item Sufficiently high reheat temperature from $\Phi$ Decays.
	\vspace{-1mm}
\end{enumerate} 
In addition to the above, as we discuss in Section \ref{sec-ord}, the following requirements follow from our assumption on the ordering of scales that $T_{\rm MD} \gg T_{\rm FO} \gg  T_{\Gamma}$, namely 
\begin{enumerate}
	\setcounter{enumi}{3}
	\item[4a.] Decay of $\Phi$ matter occurs below $T_\star$ and $T_{\rm MD}$.
	\vspace{-1mm}
	\item[4b.] Freeze-out temperature of dark matter is finite and less than $T_\star$  and $T_{\rm MD}$.
	\vspace{-1mm}
	\item[4c.] The reheat temperature does not exceed $T_\star$  and $T_{\rm MD}$.
\end{enumerate} 
We consider each of these constraints in the subsections which follow and in Section \ref{4.7} we provide a summary of the leading constraints and an illustrative example of the viable parameter space for such models. We also require that decays of $\Phi$ are negligible during freeze-out, essentially that $T_{\rm RH}\ll T_{\rm FO}$, but we reserve a careful discussion of this requirement for the next chapter. 

\section{Freeze-out of dark matter during matter domination}
\label{c1}

For this section we need a precise definition of what we mean by ``matter domination''. In \cite{SuperHeavyBramante2017} the authors define matter domination as the state of the universe when matter constitutes at least half of the energy density of the universe. Here we take a more general approach that is able to recover the definition of the authors of \cite{SuperHeavyBramante2017}. The definition we take of ``matter domination'' is as follows: Matter domination is the state of a universe constituted primarily of matter and radiation when the Hubble parameter is sufficiently small relative to a characteristic Hubble parameter scale such that any matter component of the universe constitutes the principal, and to a \emph{fair approximation} the only, contributor to the dynamics of the Friedmann Equation. Mathematically, we have from the Friedmann equation that when $H/H_\star \ll 1$,
\begin{equation}\label{eq:approxmd}
	\left(\frac{a_\star}{a}\right)\approx\frac{1}{(1-r)^{1/3}}\left(\frac{H}{H_\star }\right)^{2/3}-\frac{1}{3}\frac{r}{(1-r)^{5/3}}\left(\frac{H}{H_\star }\right)^{4/3}+\mathcal{O}\left(\left(\frac{H}{H_\star }\right)^{8/3}\right)_.
\end{equation} 
To reasonable approximation matter is the only contributor to the dynamics of the Friedmann equation when the second term in equation (\ref{eq:approxmd}) is much smaller than the first. The exact relation between the first two terms may be parameterized by using a factor $\beta$, implicitly defined along with $H_{\rm MD}$ as,
\begin{equation}\label{eq:beta}
	\frac{\beta}{(1-r)^{1/3}}\left(\frac{H_{\rm MD}}{H_\star }\right)^{2/3}\equiv\frac{1}{3}\frac{r}{(1-r)^{5/3}}\left(\frac{H_{\rm MD}}{H_\star }\right)^{4/3}
\end{equation}
Thus, $H_{\rm MD}$ is the Hubble parameter below which the regime of matter domination ensues, and it is such that the second term of equation (\ref{eq:approxmd}) is a factor $\beta$ smaller than the first. Equation (\ref{eq:approxmd}) may be solved explicitly for $H_{\rm MD}$ as,
\begin{equation}\label{eq:hmd}
	\frac{H_{\rm MD}}{H_\star }\equiv(1-r)^{2}\left(\frac{3\beta}{r}\right)^{3/2}_.
\end{equation}
Moreover, since $H_{\rm MD}$ is by definition the Hubble parameter such that the first term is the only term one must keep from expansion (\ref{eq:approxmd}), we define,
\begin{equation}\label{eq:amd}
	\left(\frac{a_\star}{a}\right)_{\rm MD}\equiv\frac{1}{(1-r)^{1/3}}\left(\frac{H_{\rm MD}}{H_\star }\right)^{2/3}=3\beta\frac{(1-r)}{r}_.
\end{equation}
We define the fraction of energy in $\Phi$ at the point matter domination ensues as
\begin{equation}\label{eq:fmd}
	f_{\Phi, \rm{MD}}\equiv\frac{1}{\frac{r}{(1-r)}\left(\frac{a_\star}{a}\right)_{\rm MD}+1}=\frac{1}{3\beta+1}_.
\end{equation}

For $\beta=1/3$ the second term in expansion (\ref{eq:approxmd}) is $1/3$ the magnitude of the first, then matter domination ensues when the $\Phi$ matter accounts for half of the energy density of the universe: $f_{\Phi, \rm{MD}}=1/2$, just as matter domination was defined in \cite{SuperHeavyBramante2017}.

Then the constraint that freeze-out occurs during matter domination may be expressed as,
\begin{equation}\label{eq:c1}
	T_F\lesssim T_{\rm MD}\equiv T_\star\left(\frac{a_\star}{a}\right)_{\rm MD}=3\beta\frac{(1-r)}{r}T_\star,
\end{equation}
which may also be written as,
\begin{equation}\label{eq:c1b}
	T_\star\gtrsim \frac{1}{3\beta}\frac{r}{(1-r)}\frac{m_{X}}{x_F}_.
\end{equation}

It should be emphasized that what we mean by the Hubble parameter being ``sufficiently small'' is intimately linked to the $r$ parameter. As we showed in expansion \eqref{eq:AgamTaylorR} when $r$ is small, the terms up to linear order in $r$ give approximately the first term of expansion \eqref{eq:AgamTaylor}. We will derive below an $r_{\star,{\rm MD}}$ which will represent the upper bound on $r$ such that matter domination ensues immediately when $\Gamma=H_\star $, to demonstrate that it is not necessary for $\Gamma\ll H_\star $ in order to be in a matter domination scenario.

\section{Freeze-out of dark matter before decay of $\Phi$ matter}
\label{c2}

Relative to the temperature $T_\star$ (at which the energy density of $\Phi$  begins to evolve as a matter-like contribution) the temperature of the thermal bath which corresponds to the $\Phi$ lifetime is given by
\begin{equation}
T_{\Gamma}=T_\star \left(\frac{a_\star}{a}\right)_{\Gamma}\simeq \frac{T_\star }{(1-r)^{1/3}}\left(\frac{\Gamma}{H_\star }\right)^{2/3}_,
\end{equation}
Since, in this work we study the effects of entropy injections on a frozen-out population of dark matter, we require that dark matter freezes-out prior to $\Phi$ matter decay.	 This condition can be expressed as
$T_F\gtrsim T_{\Gamma}$
and this can be written in terms of the reheat temperature using eq.~\eqref{76} 
\begin{equation}\label{eq:c3b}
	T_F\equiv	\frac{m_X}{x_F}\gtrsim\frac{T_{\rm RH}^{4/3}}{[(1-r)\hat{\gamma}T_\star ]^{1/3}}
\end{equation}

Additionally, we should require that $\Phi$ dominates the energy density at the time of decay and thus $\rho_{X}\ll\rho_\Phi$. Indeed, this has been implicitly assumed in Section (\ref{SecIFE}) when neglected the dark matter contribute to the interpolating Friedmann equation and we check this assumption now. This can be shown by comparing  the dark matter freeze-out abundance to the $\Phi$ matter energy density at the point of freeze-out. For the sake of simplicity, we use the freeze-out abundance, $Y_{\infty}=Y^{\rm MD}_{\infty}$, the freeze-out abundance that would occur in the limit that $r\rightarrow 0$. Since we will be mainly concerned in the $\Gamma\ll H_\star$ or small $r$ regime, and we are only looking for an order of magnitude estimate, this should not affect the accuracy of this derivation (in fact, we arrive at generally more stringent constraints than needed since the MD abundances before dilution are greater than those of RD). Notice that after freeze-out is complete,
\begin{equation}\label{eq:rhoxs}
	\rho_{X}=m_XY^{\rm MD}_{\infty}s=m_X\left(3\sqrt{\frac{5}{\pi}}\frac{\sqrt{g_{*}(T_\star )}}{g_{* S}(T_F)}\frac{(n+3/2)x_F^{n+3/2}}{M_{{\rm pl}}m_X\sigma_{0}\sqrt{x_\star }}\right)\left(\frac{2\pi^2}{45}g_{* S}(T_F)T^3\right)_.
\end{equation}
Furthermore,
\begin{equation}\label{eq:TtoPhi}
	T^{3}=T_\star ^3\left(\frac{a_\star}{a}\right)^{3}=T_\star ^3\frac{\rho_{\Phi}}{(1-r)\rho_\star}_.
\end{equation}
Plugging eq.~\eqref{eq:TtoPhi} into eq.~\eqref{eq:rhoxs}, taking $n=0$, and parameterizing the cross section as $\sigma_{0}=\alpha^2/m_X^2$, we have,
\begin{equation}\label{eq:c10b}
	\frac{\rho_X}{\rho_{\Phi}}=\left[6\sqrt{\frac{5}{\pi}}\frac{\sqrt{x_\star }x_F^{3/2}}{(1-r)\sqrt{g_*(T_\star )}\alpha^2}\right]\left(\frac{m_X}{M_{\rm Pl}}\right)_.
\end{equation}
Thus this fraction is parametrically of order
\begin{equation}
	\frac{\rho_X}{\rho_{\Phi}}\sim3\times10^{-13}\times\left(\frac{10^6~{\rm GeV}}{T_\star}\right)^{1/2}
	\left(\frac{10^{-2}}{\alpha}\right)^2\left(\frac{m_X}{1~{\rm TeV}}\right)^{3/2}_. 
\end{equation}
Thus unless the annihilation rate of the dark matter is small (in which case it will likely freeze-out whilst relativistic) or the mass is very high then $\rho_X/\rho_{\Phi}\ll1$. Thus assuming that the dark matter undergoes a period of freeze-out it is typically safe to assume that $\rho_{\rm DM}\ll \rho_\Phi$ for $T<m_{\rm DM}$. This implies that provided $\rho_\Phi> \rho_R$ then $\Phi$ dominates the energy density when it decays and, moreover,  radiation domination is restored after $\Phi$ decay, as necessary for successful BBN. We discuss the requirements for reproducing the successes of BBN further in the next section.

\section{Constraints on the reheating temperature}
\label{c3}

Big Bang Nucleosynthesis (BBN) is a remarkably successful theory in predicting the current cosmic abundances of light elements \cite{Taoso:2007qk,Kolb:1990vq}. Thus, any theory of DM should respect its constraints as a benchmark for success \cite{Taoso:2007qk}. The BBN period begins when the temperature of the bath $T\gg1$ MeV, and at this temperature scale the proton to neutron ratio is roughly unity \cite{Kolb:1990vq}. As the temperature drops to $T\simeq1$ MeV the ratio of protons to neutrons changes beginning a series of events that eventually lead to the light element abundances we observe today \cite{Kolb:1990vq}. So long as DM decoupling and entropy injections happen well before BBN onset, they will not interfere with BBN \cite{Taoso:2007qk}.
 
We thus require that DM be decoupled before the beginning of BBN and that the minimum temperature at its inception be $T_{\rm BBN}\simeq 10$ MeV. Thus, the reheat temperature should be at least $T_{\rm RH}\gtrsim T_{\rm BBN}$, setting up the initial conditions for BBN. However, the reheat temperature is intrinsically tied to the relic density of dark matter in this scenario because of the dilution due to the entropy injection. Specifically the relic density of dark matter scales with dilution factor $\zeta$ as follows
\begin{equation}\label{eq:zetasol}
	(\Omega_X^{\rm MD})^{\rm relic}h^2\sim 0.1\cdot\left(\frac{m_X}{10^6~{\rm{GeV}}}\right)^{5/2}\left(\frac{0.1}{\alpha}\right)^{2}\left(\frac{10^7~{\rm GeV} }{T_\star }\right)^{1/2}   \left(\frac{\zeta}{10^{-4}}\right)_.
\end{equation}
Since $\zeta$ is related to $T_{\rm RH}$ via eq.~(\ref{eq:zetaapprox}): $\zeta\approx\frac{g_{*}(T_\Gamma)/g_{*}(T_{\star})}{(1-r)}\frac{T_{\rm RH}}{T_\star }$, it follows that
\begin{equation}\label{eq:zetasolrh}
	T_{\rm RH}\sim 10^3~{\rm{GeV}}\cdot\left(\frac{10^6~{\rm{GeV}}}{m_X}\right)^{5/2}\left(\frac{\alpha}{0.1}\right)^{2}\left(\frac{T_\star }{10^7~{\rm GeV} }\right)^{3/2}  \gtrsim T_{\rm BBN}.
\end{equation}
Thus the requirement of reproducing BBN observables bounds the parameter space, we will give an example shortly which demonstrates how these bounds manifest. 

Furthermore, if the dark matter is produced via $\Phi$ decays, then it can have two components, an initial thermal component which freezes-out during matter domination and a component associated with the direct decay products of $\Phi$. To restrict to the case of matter dominated freeze-out we assume that $\Phi$ does not decay non-thermally to dark matter. However, since the dark matter couples to the thermal bath, if the $\Phi$ decay implies a reheat temperature comparable to the freeze-out temperature, then interactions in the  bath can lead to dark matter production. Thus to focus on the clean regime of freeze-out during matter domination followed by dilution we require $T_{\rm RH}\ll T_F$.  Comparing eqs.~\eqref{eq:xFO1MD} \&~\eqref{eq:zetasolrh} both $T_{\rm RH}$ and $T_F$ are controlled by $m_X$, $\alpha$ and $T_\star$ thus this constraint also significantly constrains the parameter space, as we illustrate shortly. We will discuss  in more detail  the case of $T_{\rm RH}\sim T_F$, which is a consistent and interesting alternate scenario for dark matter freeze-out, in the next Chapter.

\section{Ordering of scales}
\label{sec-ord}

As discussed in Section \ref{SecEarlyScen} the matter dominated freeze-out scenario relies on an assumed ordering of scales 
\begin{equation}
	T_\star\geq T_{\rm MD} \gg T_F \gg  T_{\Gamma}
\end{equation}
This implies that a period of matter domination occurs, then subsequent dark matter freezes out and after which the matter-like state $\Phi$ decays. Thus in this section we briefly comment on the implications this ordering  applies on the relative size of different parameters. 

Specifically, here we address the requirements 
\begin{enumerate}
	\setcounter{enumi}{4}
	\item[4a.] Decay of $\Phi$ matter occurs below $T_\star$ and $T_{\rm MD}$.
	\item[4b.] Freeze-out temperature of dark matter is less than $T_\star$  and $T_{\rm MD}$.
	\item[4c.] The reheat temperature does not exceed $T_\star$  and $T_{\rm MD}$.
\end{enumerate} 
Note that the lifetime of $\Phi$ is controlled by an independent coupling and thus $T_{\Gamma}$ (the temperature $T$ of the bath at $H\sim\Gamma$) can be set independently of the other parameters in the model.

\newpage

If $\Phi$ matter decays prior to matter domination, not only will the entropy injection be small, but freeze-out would not occur prior to decay when satisfying the constraint of Section \ref{c1}. Therefore we require that,
\begin{equation}\label{eq:c2}
	\Gamma\lesssim H_{\rm MD}\equiv (1-r)^{2}\left(\frac{3\beta}{r}\right)^{3/2}H_{\star},
\end{equation}
which by using approximation \eqref{eq:GammaTrh} may be written as,
\begin{equation}\label{eq:c2b}
	T_{\rm RH}\lesssim \hat{\gamma}^{1/4}(1-r)\left(\frac{3\beta}{r}\right)^{3/4}T_\star .
\end{equation}

It is required that the decay of $\Phi$ matter when the bath is at temperature $T_\Gamma$ occurs below $T_\star$, the relative orderings of these scales can be checked by inspection of the Friedmann equation. In the Friedmann equation \eqref{eq:FE} there is an implicit $\Theta$-function, such that above temperature $T_\star $ all energy is radiation, and below which the matter component appears. Moreover, matter domination may ensue immediately at $T_\star $. For example, for $H_{\rm MD}=H_\star$ this would imply
\begin{equation}\label{eq:c4ex}
	r_{\star,{\rm MD}}\equiv r\big|_{H_{\rm MD}=H_\star}=\frac{1}{(3\beta)^{-1}+1}
\end{equation}
with $\beta=1/3$, this corresponds to a value $r=0.5$. Thus, if matter constitutes greater than 50\% of energy at $T_\star $, matter domination begins immediately at $T_\star $. 
Requiring that $\Phi$ decay does not occur above $T_\star $ implies the constraint $\Gamma\lesssim H_\star$, or in terms of temperatures
\begin{equation}\label{eq:c4b}
	T_{\rm RH}\lesssim \hat{\gamma}^{1/4}T_\star.
\end{equation}
Furthermore, since our analysis of the Friedman equations fixes an initial condition at $T=T_\star$ it follows that for our analysis to be valid all temperature scales which we are interested should be below or equal to $T_\star$, including the freeze-out temperature of dark matter: $m_X/x_F=T_F\lesssim T_\star.$

\section{Summary}
\label{4.7}	

\vspace{-3mm}
Above we have explored the various constraints,  before closing this section we summarize the leading constraints and show how they apply for a particular example. We assume the ordering of scales $T_\star\gg  T_{\rm RH}, T_F, T_{\Gamma}$, as discussed above.

\begin{figure}[h]
	\centering
	\vspace{25mm}
	\includegraphics[width=0.8\textwidth]{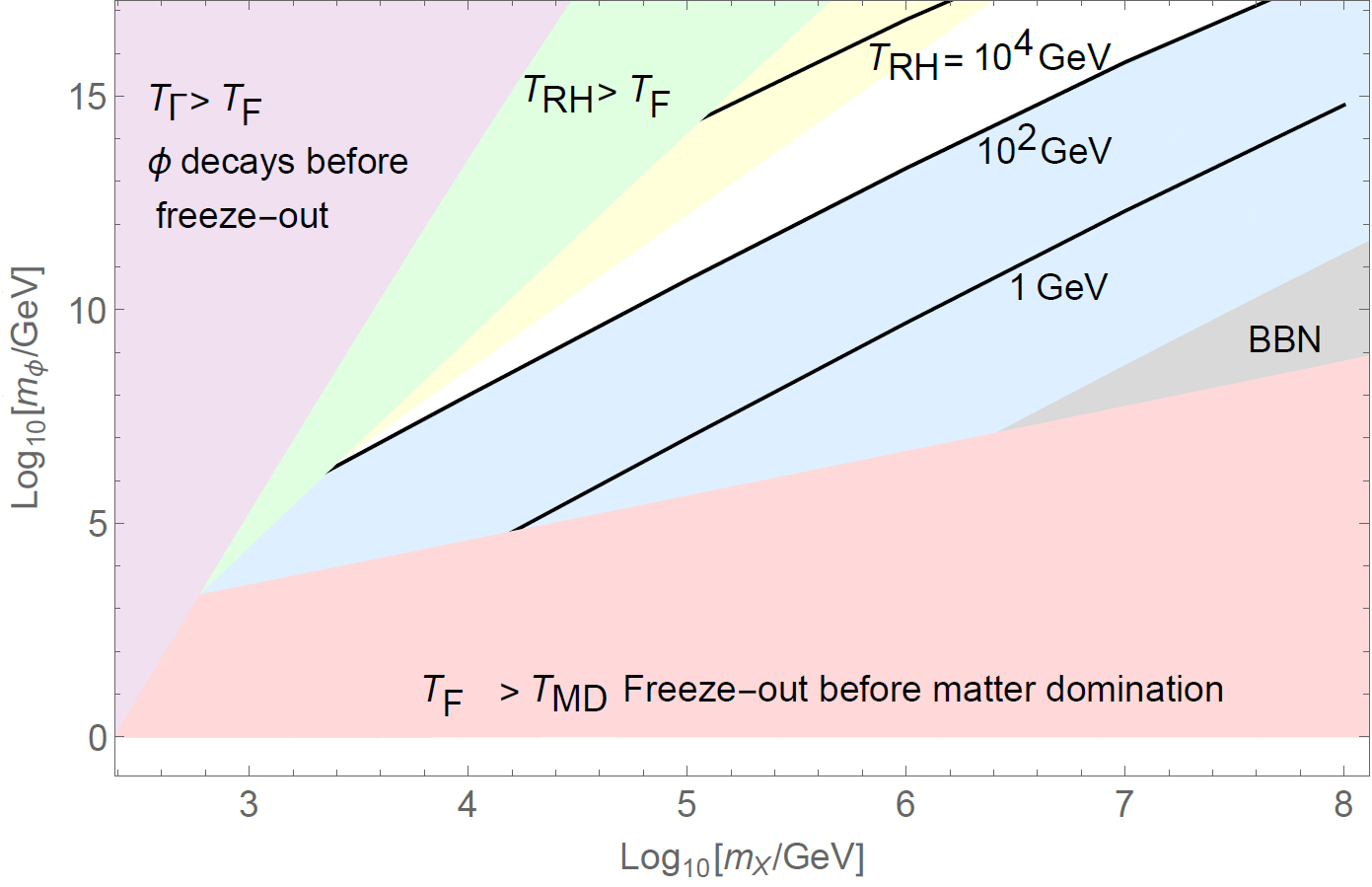}
	\caption[Parameter space of matter dominated dark matter freeze-out]{Parameter space of matter dominated DM freeze-out assuming an $s$-wave annihilation with $\sigma_{0}\equiv\alpha^2/m_X^2$ for $\alpha=0.1$ with $T_\star\simeq m_\Phi$, and $r\approx0.99$. 
		We show contours of $T_{\rm RH}$ which reproduce the dark matter relic density. The main constraints are that the dark matter decouples prior to $\Phi$ decays (purple), but after the universe is $\Phi$ matter dominated (red). We also require $T_F>T_{\rm RH}$ (green), in the yellow region the instantaneous decay approximation is less reliable. We shade regions in which the reheat temperature is low: for $T_{\rm RH}\lesssim10$ MeV BBN observables are disrupted (grey) and for $T_{\rm RH}\lesssim100$ GeV baryogenesis is challenging~(blue).}
	\label{fig:2}
\end{figure}

As discussed in Section \ref{c1}  the dark matter should decouple only after the universe is matter dominated: $T_{\rm MD}\gtrsim T_F^{\rm RD}$ (excluded region shaded red in Fig.~\ref{fig:2}). Thus we should compare the radiation dominated freeze-out temperature to  $T_{\rm MD}$ derived in eq.~(\ref{eq:c1}). This also implies that $\Phi$ decays only after it comes to dominate the energy density of the universe. From eq.~(\ref{eq:Tgam}) \& (\ref{eq:c1}) one can verify that $T_{\rm MD} \gg T_{\Gamma}$ provided that $T_\star\gg  T_{\rm RH}$. Furthermore, as in Section \ref{c2}, the dark matter should decouple prior to the period of $\Phi$ decays: $T_{\Gamma}\lesssim T_F^{\rm MD}$ (purple in Fig.~\ref{fig:2}).

In Section \ref{c3} we argued that to avoid conflict with BBN we require $T_{\rm RH}\gtrsim10$ MeV (grey in Fig.~\ref{fig:2}). Both $T_{\rm RH}$ and the dark matter relic density are controlled by $\zeta$ and $T_\star$, but these requirements can generally be simultaneously satisfied. Higher reheat temperatures are often desirable for baryogenesis mechanisms, e.g.~electroweak baryogenesis \cite{Morrissey:2012db} typically requires $T_{\rm RH}\gtrsim100$ GeV  (blue in Fig.~\ref{fig:2}). Moreover, to remain in the matter dominated freeze-out regime we require  $T_F\gg T_{\rm RH}$ (green in Fig.~\ref{fig:2}). The yellow region indicates where we expect the instantaneous decay approximation to be unreliable, which we discuss more in the next Chapter. Additionally, to restore radiation domination after $\Phi$ decays, it is necessary that the the dark matter energy density remains small relative to that of $\Phi$  immediately prior to $\Phi$ decays. Since the the dark matter number density is Boltzmann suppressed, $\rho_\Phi\gg \rho_{X}$ at $T=T_F$ and this requirement is typically readily satisfied.

Finally, the observed the dark matter relic density must be reproduced ($\Omega_{\rm DM}h^2\sim0.1$), as determined by eq.~(\ref{eq:critDensZetaMD}). In Figure \ref{fig:2} we show how the parameter space is constrained by the requirements above for  $s$-wave annihilations with $\alpha=0.1$, where we take $T_\star\simeq m_\Phi$ (thus $r\approx0.99$), and we neglect changes in $g_*$.

Observe that the leading constraints are due to the requirement that the dark matter decouples after the universe is $\Phi$ matter dominated (shaded red) and $T_F\gg T_{\rm RH}$ (green/yellow). Requiring a reheat temperature above the electroweak  phase transition $T_{\rm RH}\gtrsim100$ GeV (blue) further constrains the parameter space. For the parameter choices made in Figure \ref{fig:2}, the model prefers heavier  dark matter, and can accommodate TeV scale dark matter. 

\clearpage

%% file: PhDthesisV2.0chpt05BID.tex
\chapter{Beyond Instantaneous Decay}
\label{Ch5}

\vspace{-12mm}

{\em This chapter presents original results which have not been  previously presented.}

\vspace{2mm}

In the previous chapter we assumed an instantaneous decay and thermalization at $t\sim\Gamma^{-1}$. In this chapter we take a closer look at the equations governing the decay of a massive species or oscillating scalar field. We also derive constraints that assure accuracy in assuming instantaneous decay in the previous chapter.

\vspace{-5mm}

\section{Decay Equations}

\vspace{-5mm}

The equation governing the decay of a coherently oscillating field or massive particle species $\Phi$ is as follows
\begin{equation}\label{eq:PhiDecay}
\dot{\rho_{\Phi}}+3H\rho_{\Phi}=-\Gamma\rho_{\Phi},
\end{equation}
which may be written in a more suggestive form as
\begin{equation}\label{eq:PhiDecay2}
\frac{d}{dt}\left(\rho_{\Phi}a^{3}\right)=-\Gamma\left(\rho_{\Phi}a^{3}\right).
\end{equation}
Integrating from time $t_\star$, the time when coherent oscillations of a scalar field begin or a decaying particle species begins to evolve as matter\footnote{This time will roughly be $t_\star\sim H_\star ^{-1}$ but here we keep the analysis general.}, to time $t$ we have
\begin{equation}\label{eq:PhiSol}
\rho_{\Phi}=\rho_{\Phi \star}\left(\frac{a_\star}{a}\right)^{3}e^{-\tau},
\end{equation}
where $\rho_{\Phi \star}$ is the energy density of the $\Phi$ matter at $t_\star$, and $\tau\equiv\Gamma t$. Expanding \eqref{eq:PhiSol} in powers of $\tau$ 
\begin{equation}\label{eq:PhiSolApprox}
\rho_{\Phi}=\rho_{\Phi \star}\left(\frac{a_\star}{a}\right)^{3}(1+\mathcal{O}(\tau^{1})).
\end{equation}
From relation \eqref{eq:avt} we also have $a(t)\propto t^{\frac{2}{3}(1+w)^{-1}}$ (recall that $w=0$ for MD and $w=1/3$ for RD), hence
\begin{equation}\label{eq:avt2}
\frac{a_\star}{a}=\left(\frac{t_\star}{t}\right)^{\frac{2}{3}(1+w)^{-1}}_.
\end{equation}
Thus, collecting only the $\mathcal{O}(\tau^{0})$ term of expansion \eqref{eq:PhiSolApprox} and using eq.~\eqref{eq:avt2} we have
\begin{equation}\label{eq:PhiSolApprox2}
\rho_{\Phi}\approx\rho_{\Phi \star}\left(\frac{t_\star}{t}\right)^{2(1+w)^{-1}}_.
\end{equation}
Assuming that $\Phi$ matter decays solely to the radiation sector, the equation governing the radiation energy density is
\begin{equation}\label{eq:RDecay}
\dot{\rho_r}+4H\rho_r=\Gamma\rho_{\Phi},
\end{equation}
which may be rewritten as
\begin{equation}\label{eq:RDecay2}
\frac{d}{dt}\left(\rho_{r}a^{4}\right)=\Gamma\rho_{\Phi}a^{4}.
\end{equation}
Using approximation \eqref{eq:PhiSolApprox2} and relation \eqref{eq:avt2}, eq.~\eqref{eq:RDecay2} may be approximated as
\begin{equation}\label{eq:RDecayApprox}
\frac{d}{dt}\left(\rho_{r}a^{4}\right)\simeq \Gamma\rho_{\Phi \star}\left(\frac{t_\star}{t}\right)^{2(1+w)^{-1}}a^{4}\simeq\Gamma\rho_{\Phi \star}a_\star^{4}\left(\frac{t_\star}{t}\right)^{-(2/3)(1+w)^{-1}}
\end{equation}
Integrating equation \eqref{eq:RDecayApprox} from $t_\star$ to $t$ gives
\begin{equation}\label{eq:RDecaySol}
\rho_r\simeq\rho_{r\star}\left(\frac{a_\star}{a}\right)^4+\rho_{\Phi\star}\left(\frac{a_\star}{a}\right)^4\frac{1}{(\nicefrac{2}{3})(1+w)^{-1}+1}\left(\tau\left(\frac{\tau_\star}{\tau}\right)^{-(2/3)(1+w)^{-1}}-\tau_\star\right)_,
\end{equation}
where $\tau_\star=\Gamma t_\star$. In \cite{Scherrer:1984fd} they refer to the first term on the RHS of equation \eqref{eq:RDecaySol} as the ``old'' radiation and the latter term as the ``new'' radiation. We will adopt the same naming conventions here. Writing equation \eqref{eq:RDecaySol} solely in terms of $\tau$ gives
\begin{equation}\label{eq:RDecaySol2}
\rho_r\simeq\rho_{r\star}\left(\frac{\tau_\star}{\tau}\right)^{(8/3)(1+w)^{-1}}+\frac{\rho_{\Phi\star}}{(\nicefrac{2}{3})(1+w)^{-1}+1}\left(\tau\left(\frac{\tau_\star}{\tau}\right)^{2(1+w)^{-1}}-\tau_\star\left(\frac{\tau_\star}{\tau}\right)^{(8/3)(1+w)^{-1}}\right)_.
\end{equation} 	
Notice that when $\tau_\star\ll 1$ the last term on the RHS of equation \eqref{eq:RDecaySol2} may be neglected (as they have done in \cite{Scherrer:1984fd}), but here we keep it for completeness. Defining 
\begin{equation}
\mathcal{T}\equiv\tau_\star/\tau=t_\star/t
\end{equation} 
and using $\rho_{r\star}=r\rho_\star$ and $\rho_{\Phi\star}=(1-r)\rho_\star$, we write eq.~\eqref{eq:RDecaySol2} as
\begin{equation}\label{eq:RDecaySolT}
\frac{\rho_r}{\rho_\star}\simeq r\mathcal{T}^{(8/3)(1+w)^{-1}}+\frac{(1-r)\tau_\star}{(\nicefrac{2}{3})(1+w)^{-1}+1}\left(\mathcal{T}^{2(1+w)^{-1}-1}-\mathcal{T}^{(8/3)(1+w)^{-1}}\right)_.
\end{equation}

\section{Breakdown of Entropy Conservation in the Bath}

In \cite{Scherrer:1984fd} they find that the entropy remains roughly constant until the first and second terms (or old and new radiation terms, respectively) on the RHS of equation \eqref{eq:RDecaySolT} become comparable. We call this point at which the assumption of entropy conservation is violated $\mathcal{T}_{\rm EV}$ and thus  define $\mathcal{T}_{\rm EV}$ implicitly by
\begin{equation}\label{eq:TentDef}
r\mathcal{T}_{\rm EV}^{(8/3)(1+w)^{-1}}\equiv\frac{(1-r)\tau_\star}{(\nicefrac{2}{3})(1+w)^{-1}+1}\left(\mathcal{T}_{\rm EV}^{2(1+w)^{-1}-1}-\mathcal{T}_{\rm EV}^{(8/3)(1+w)^{-1}}\right)_.
\end{equation}
Solving \eqref{eq:TentDef} for $\mathcal{T}_{\rm EV}$, and letting $v\equiv(\nicefrac{2}{3})(1+w)^{-1}+1$ for concision, we have
\begin{equation}\label{eq:TEV}
\mathcal{T}_{\rm EV}=\left(1+\left(\frac{r}{1-r}\right)\frac{v}{\tau_\star}\right)^{-1/v}
\end{equation}

Our freeze-out and abundance calculations in Chapter \ref{Ch3}, and constraints in Chapter \ref{Ch4}  relied on the assumption of constant entropy until an instantaneous decay occurs at $t\sim\Gamma^{-1}$. Instead considering an exponential decay law (as in \cite{Scherrer:1984fd}), there is still a considerable parameter space to allow freeze out to occur prior to non-negligible contributions of entropy. The precise time when this occurs is given by eq.~\eqref{eq:TEV}.

\begin{figure}[h!]
	\includegraphics[width=0.8\textwidth]{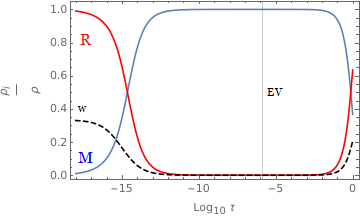}\\
	\vspace{3mm}
	\centering
	\includegraphics[width=0.8\textwidth]{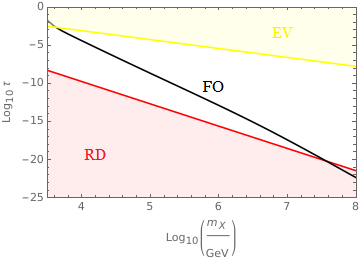}
	\caption[Relevant events for matter dominated freeze-out without the instantaneous decay approximation.]{Logarithmic time scales and orderings of relevant events for MD freeze-out scenario from $\tau=\tau_\star$ till $\tau=1$ ($t\sim\Gamma^{-1}$) for $T_\star=10^7$ GeV and $r=0.99$. Top: fraction of energy density in matter (blue) and radiation (red), for $T_{\rm RH}=10$ MeV and $w$ (dashed) parameterized between RD ($w=1/3$) and MD ($w=0$). Vertical line represents moment when entropy violation is considerable, and first intersection of blue and red line signifies MD onset. Bottom: logarithmic time of freeze-out versus DM mass (black) for $T_{\rm RH}$ required to produce observed DM relic density. RD era shaded in red and times of considerable entropic violations shaded in yellow; adiabatic MD freeze-out occurs between shaded regions.}
	\label{fig:EV}
\end{figure}

We fixed $T_\star=10^7$ GeV, $r=0.99$, $T_{\rm RH}=10$ MeV, and $w$ is taken to be an affine interpolation between RD ($w=1/3$) and MD ($w=0$) values and plot the evolution of the system in the top panel of Figure \ref{fig:EV}. The $w$ interpolation is weighted linearly according to the relative fraction of energy that radiation and matter constitute, and is plotted as the dashed line in Figure \ref{fig:EV}. The the vertical black line shows the point at which $T=T_{\rm EV}$. Notice that despite the $\Phi$ matter (whose fraction of the total energy is the blue line, and radiation's fraction is the red line) constituting only 1\% of the energy of the universe at $\tau_\star$, it comes to dominate relatively early on during decays, and only precipitously falls near $\tau=1$ (or $t\sim \Gamma^{-1}$).

Figure \ref{fig:EV} also shows that it is possible to have a large amount of time between when MD ensues (when matter constitutes half the energy, hence when the red and blue line intersect) till when entropy violations become considerable. The new aspect of dark matter which this thesis (and the associated paper \cite{Hamdan:2017psw}) presents is the possibility of freeze-out in the matter dominated regime which had been previously unstudied. This difference between our work and previous related scenarios (such as \cite{GiudicePhysRevD.64.023508}) is most clearly illustrated in the lower panel of Figure \ref{fig:EV} which shows that the time of freeze-out (black curve) can readily fall between when matter domination begins (red shaded region is radiation dominated) and when entropy violations are large (yellow shaded region). For the bottom of Figure \ref{fig:EV} $T_{\rm RH}$ was not fixed as in the top (the rest of the parameters were fixed the same); instead \eqref{eq:zetaapprox} was used to ensure that the observed DM relic density was matched.
\newpage

\clearpage

Let $t_{\rm EV}\equiv t_\star/\mathcal{T}_{\rm EV}$, which from relation \eqref{eq:avt2} corresponds to scale factor 
\begin{equation}
(a_\star/a)_{\rm EV}\equiv (t_\star/t_{\rm EV})^{v-1}
\end{equation}
Then, because entropy remains roughly constant until $a=a_{\rm EV}$, we may use the $T\propto a^{-1}$ relation during the period while entropy remains constant to obtain the temperature which corresponds to $T(a_{\rm EV})$. Notably, $T\propto a^{-1}$ is not valid once the approximation of  entropy conservation is strongly perturbed. To obtain an expression for $T_{\rm EV}$ we consider the evolution from $T=T_\star$
\begin{equation}\label{eq:TempEnt2}
T_{\rm EV}\equiv T_\star (a_\star/a)_{\rm EV}\simeq T_\star (t_\star/t_{\rm EV})^{v-1}
\end{equation}
Using the definition $ \mathcal{T}_{\rm EV}\equiv (t_\star/t_{\rm EV})$ it follows
\begin{equation}\label{eq:TempEnt}
T_{\rm EV}\simeq T_\star \mathcal{T}_{\rm EV}^{v-1}\simeq T_\star \left(1+\left(\frac{r}{1-r}\right)\frac{v}{\tau_\star}\right)^{(1-v)/v}.
\end{equation}
Therefore the requirement of constant entropy during freeze-out implies an additional constraint in the event of non-instantaneous decay:
\begin{equation}
T_F\gtrsim T_{\rm EV}.
\end{equation}

For a more precise derivation  of the temperature of the thermal bath at any given moment we should track changes in the degrees of freedom, we may  use equation \eqref{eq:RDecaySolT} since
\begin{equation}\label{eq:TempOfThermBathAlt}
\frac{T}{T_\star }=\left(\frac{g_*(T_\star )}{g_*(T)}\frac{\rho_r}{\rho_\star}\right)^{1/4}_.
\end{equation}
Then, by eq.~\eqref{eq:TEV} in \eqref{eq:RDecaySolT} we find that $T_{\rm EV}$ may also be defined as
\begin{equation}\label{eq:TentAltDef}
T_{\rm EV}\equiv T_\star \left(\frac{g_*(T_\star )}{g_*(T)}\frac{\rho_r}{\rho_\star}\right)^{1/4}\bigg|_{\mathcal{T}=\mathcal{T_{\rm EV}}}.
\end{equation}
Writing out \eqref{eq:TentAltDef} explicitly, with the aid of eq.~ \eqref{eq:RDecaySolT} and eq.~\eqref{eq:TEV}
\begin{equation}\label{eq:TentAltDef2}
T_{\rm EV}\simeq T_\star \left(\frac{g_*(T_\star )}{g_*(T_{\rm EV})}\right)^{1/4}\left[ r\mathcal{T}_{\rm EV}^{4(v-1)}+\frac{(1-r)\tau_\star}{v}\left(\mathcal{T}_{\rm EV}^{3v-4}-\mathcal{T}_{\rm EV}^{4(v-1)}\right)\right]^{1/4}.
\end{equation}
But by definition of $\mathcal{T}_{\rm EV}$, the two terms in the square bracket must be equal, so that the value within the square bracket is $2r\mathcal{T}_{\rm EV}^{4(v-1)}$, and thus $T_{\rm EV}$ may be written as
\begin{equation}\label{eq:TentAltDef3}
T_{\rm EV}\simeq  T_\star (2r)^{1/4}\left(\frac{g_*(T_\star )}{g_*(T_{\rm EV})}\right)^{1/4}\left(1+\left(\frac{r}{1-r}\right)\frac{v}{\tau_\star}\right)^{(1-v)/v}
\end{equation}
Unless there are significant numbers of new states at high temperatures we expect $g_*(T_\star )/g_*(T_{\rm EV})\sim\mathcal{O}(1)$ and thus this gives a similar result to eq.~(\ref{eq:TempEnt}) above. 

\vspace{-3mm}
\section{Hubble Parameter During Decays}
\vspace{-3mm}

When entropy produced by $\Phi$ decays becomes comparable to the entropy due to the old radiation the assumption of constant entropy per comoving volume ceases to hold. Moreover, even if freeze-out occurred during the time prior to significant entropy injection, the changing dynamics of the universe can conceivably cause a frozen-out species to recouple solely due to the changing relationship between the Hubble parameter and temperature, thus altering the freeze-out temperature. This is because the Hubble parameter in the RD and MD case have relation $H_{\rm RD}\propto T^2$ and $H_{\rm MD}\propto T^{3/2}$, but if entropy injection becomes significant $H_{\rm EV}\propto T^4$. 

Freeze-out temperature is defined by the condition $H=\Gamma_{\rm ann}$, thus a change in the relationship between the Hubble parameter and temperature changes the freeze-out temperature also. Below we derive a general freeze-out temperature that encompasses various scenarios and derive a constraint to ensure that dark matter is not repopulated.

First we derive a relation between temperature and scale factor during decays, following \cite{GiudicePhysRevD.64.023508}. From eq.~\eqref{eq:avt2}, and defining $\mathcal{A}\equiv a_\star/a$, we have that,
\begin{equation}\label{eq:TvA}
\mathcal{T}=\mathcal{A}^{1/(v-1)}=\mathcal{A}^{(3/2)(1+w)}.
\end{equation}
The decay equations prior to $t\sim \Gamma^{-1}$ may also be written in terms of $\mathcal{A}$ thusly
\begin{equation}\label{eq:RDecaySolA}
\frac{\rho_r}{\rho_\star}\simeq r\mathcal{A}^{4}+\frac{(1-r)\tau_\star}{v}\left(\mathcal{A}^{4\bar{v}}-\mathcal{A}^{4}\right)_,
\hspace{15mm}
\frac{\rho_\Phi}{\rho_\star}\simeq (1-r)\mathcal{A}^3,
\end{equation}
where $\bar{v}\equiv \nicefrac{((\nicefrac{3}{4})v-1)}{(v-1)}$. Since $\bar{v}<1$ (strictly this is true only for $v>0$, which is our concern), the term in \eqref{eq:RDecaySolA} proportional to $\mathcal{A}^{4\bar{v}}$ will eventually come to dominate prior to $t\sim\Gamma^{-1}$ when $r$ is sufficiently small; the precise conditions were derived above with $\mathcal{T}_{\rm EV}$. For the temperature versus scale factor relation, we use \eqref{eq:TempOfThermBathAlt} and \eqref{eq:RDecaySolA} to obtain
\begin{equation}\label{eq:TempvA}
T\simeq T_\star \left(\frac{g_*(T_\star )}{g_*(T)}\right)^{1/4}\left[r\mathcal{A}^{4}+\frac{(1-r)\tau_\star}{v}\left(\mathcal{A}^{4\bar{v}}-\mathcal{A}^{4}\right)\right]^{1/4}_.
\end{equation}
If the $\mathcal{A}^{4\bar{v}}$ comes to dominate, and $\bar{v}=3/8$, then we have the relation $T\propto a^{-3/8}$ found in \cite{GiudicePhysRevD.64.023508} and \cite{Scherrer:1984fd}. Furthermore,  the Hubble parameter during decay is given by
\begin{equation}\label{eq:HDecay}
H\simeq H_\star \left[\frac{\rho_r}{\rho_\star}+\frac{\rho_\Phi}{\rho_\star}\right]^{1/2}\simeq H_\star \left[r\mathcal{A}^{4}+\frac{(1-r)\tau_\star}{v}\left(\mathcal{A}^{4\bar{v}}-\mathcal{A}^{4}\right)+(1-r)\mathcal{A}^3\right]^{1/2}_.
\end{equation}

\vspace{-3mm}

The requirement that DM is not repopulated during decays can be expressed using the results above. For MD freeze-out, we require that $H(T_F)=\Gamma_{\rm ann}$ occur while $T\propto a^{-1}$ and the matter component dominates the energy density. This is achieved with $T_F\gtrsim T_{\rm EV}$, as derived above. A more general freeze-out temperature may be obtained by using eq.~\eqref{eq:HDecay} as a function of temperature (by inverting eq.~\eqref{eq:TempvA}), and setting this equal to the annihilation rate. Requiring DM is frozen out (i.e.~$H(T)\gtrsim\Gamma_{\rm ann}(T)$) before decays are important implies
$\tau_\star\lesssim\mathcal{T}(T_F)\lesssim \mathcal{T}_{\rm EV}$
or equivalently 
\begin{equation}
T_\star\lesssim T_F\lesssim {T}_{\rm EV}
\end{equation}

Let us now make the following assumptions that will make the above analysis simpler. First, assume that $t_\star=H_\star ^{-1}$, implying 
\begin{equation}
\tau_\star=\frac{\Gamma}{H_\star} \approx\hat{\gamma}^{-1/2}\left(\frac{T_{\rm RH}}{T_\star}\right )^{2},
\end{equation}
 where the approximation comes from \eqref{eq:GammaTrh}. Second, assume $\tau_\star\ll 1$. Third, assume that $r$ is sufficiently small such that immediately at $T_\star $ matter domination ensues. This can be achieved by taking $r\lesssim0.5$ when $\beta=1/3$. This will allow us to take $w=0$, or equivalently $v=5/3$ or $\bar{v}=3/8$, from time $t=H_\star ^{-1}$ till time $t=\Gamma^{-1}$. This simplifies our analysis since there is no time in the domain we are interested in where radiation is dominant\footnote{This assumption is not necessary, if an interpolation of $w$ is used and with matter constituting only 1\% of the energy density at $\tau_\star$, MD ensues much earlier than entropy violation as shown in Figure \ref{fig:EV}}. Our fourth assumption will be that temperature is always falling, and never rises. From \eqref{eq:TCalmax}, which is derived in the next section, this requirement can be expressed as $\tau_\star<8r/3(1-r)$.

With these simplifying assumptions we can approximate $T_{\rm EV}$ as follows
\begin{equation}\label{TentMD}
T_{\rm EV}\approx T_\star \left(\frac{3}{5}\right)^{2/5}\left(\frac{1-r}{r}\right)^{2/5}\hat{\gamma}^{1/5}\left(\frac{T_{\rm RH}}{T_\star }\right)^{4/5}_.
\end{equation}
For $T\gtrsim T_{\rm EV}$ we assume $T\propto a^{-1}$, and for $T\lesssim T_{\rm EV}$ then $T\propto a^{-3/8}$. We also assume that the energy density continues to be dominated by matter until $t=\Gamma^{-1}$.  Thus to evaluate the requirement that $T_F>T_{\rm EV}$ one can compare eq.~(\ref{TentMD}) to eq.~(\ref{eq:xFO1MD})

Returning to Figure \ref{fig:2}, here  we illustrated the parameter space for a characteristic model and the yellow shaded region shown here indicates the region in which entropy conservation is no longer a good approximation. Specifically, the yellow shading indicates the parameter region in which $T_F<T_{\rm EV}$, and thus the dynamics of freeze-out are no longer matter dominated freeze-out but rather decoupling during a period of entropy injection as described above. This latter scenario is reminiscent to that studied in \cite{GiudicePhysRevD.64.023508}.

\section{Temperature Maxima During Decay}

\vspace{-3mm}

Although the decay of a massive species or coherent oscillations of a scalar field are said to ``reheat'' the universe, the temperature need not rise during this reheating \cite{Scherrer:1984fd}. In \cite{GiudicePhysRevD.64.023508} it is shown that after an initial rise in temperature to what is termed $T_{\rm MAX}$, the temperature actually \emph{falls} to $T_{\rm RH}$. This is why, in \cite{GiudicePhysRevD.64.023508},  Giudice, Kolb and Riotto define $T_{\rm RH}$ as the highest temperature of the radiation domination era. Thus after the temperature of the bath falls below $T_{\rm RH}$, radiation comprises the dominant component of the universe's energy. In \cite{GiudicePhysRevD.64.023508} they consider decay in the absence of ``old'' radiation (i.e. $\rho_{r\star}\approx0$). We next comment on the case omitted in \cite{GiudicePhysRevD.64.023508},  which includes an old radiation component  and we derive a more general $T_{\rm MAX}$ than \cite{GiudicePhysRevD.64.023508}.

To find the time at which the temperature is at an extremum, we find the critical point of eq.~\eqref{eq:RDecaySolT} with respect of $\mathcal{T}$, which gives
\begin{equation}\label{eq:TCalmax}
\mathcal{T_{\rm MAX}}=\bar{v}^{1/v}\left(1-\left(\frac{r}{1-r}\right)\frac{v}{\tau_\star}\right)^{-1/v}_.
\end{equation} 
Thus for a sensible analysis, we require $\tau_\star<\mathcal{T_{\rm MAX}}<1$ (otherwise the time that the maximum temperature occurs after 
$t\sim\Gamma^{-1}$ or prior to $t_\star$). For the $\tau_\star$ lower bound and $v>0$, these give $\tau_\star^v-(r/(1-r))v\tau_\star^{v-1}<\bar{v}$. The $\mathcal{T_{\rm MAX}}<1$ upper bound gives $\tau_\star>4(v-1)r/(1-r)$. Whenever any of these bounds are violated, which is readily achievable as shown in the previous section, there will be no physical extremum of temperature. Thus the temperature smoothly continues dropping as the universe expands, and the global maximum of temperature would be $T_\star $. To obtain the temperature when a local maximum is present, we insert \eqref{eq:TCalmax} into \eqref{eq:TempOfThermBathAlt} and \eqref{eq:RDecaySolT} to obtain
\begin{equation}\label{eq:TempMax}
T_{\rm MAX}\simeq T_\star \left(\frac{g_*(T_\star )}{g_*(T_{\rm MAX})}\right)^{1/4}\left[ r\mathcal{T}_{\rm MAX}^{4(v-1)}+\frac{(1-r)\tau_\star}{v}\left(\mathcal{T}_{\rm MAX}^{3v-4}-\mathcal{T}_{\rm MAX}^{4(v-1)}\right)\right]^{1/4}_,
\end{equation}
This generalizes the result of \cite{GiudicePhysRevD.64.023508} to include an ``old'' radiation component and, furthermore, for $r=0$ we recover the result of  \cite{GiudicePhysRevD.64.023508}. 

%% file: PhDthesisV2.0chpt06HP.tex
\chapter[Matter Dominated Freeze-out via the Higgs Portal]{Matter Dominated Freeze-out\\ via the Higgs Portal}
\label{Ch6}

\vspace{-5mm}

{\em This chapter presents original results which have not been  previously presented.}

\vspace{5mm}

In previous chapters we studied a model-independent, matter-dominated freeze-out scenario for dark matter. We found that there was a considerable amount of viable parameter space for a wide range of dark matter mass and annihilation cross-section values. Here we take our previous work on matter-dominated freeze-out and apply it to the specific model of dark matter annihilations through the Higgs Portal.

Very little is known about the interactions of dark matter, but possible interactions with the Higgs boson provide a promising means to link dark matter to the Standard Model \cite{Silveira:1985rk,McDonald:1993ex,Burgess:2000yq,Patt:2006fw}. The precise relations of the Higgs Portal depend on the nature of the dark matter quantum numbers, in this thesis we consider only the case of a real scalar dark matter particle $\varphi$, whose Lagrangian terms are $\mathbb{Z}_2$ invariant
\begin{equation}\label{eq:scalarLagrange}
\mathcal{L}=\mathcal{L}_{\rm SM}+\frac{1}{2}\partial_\mu\varphi\partial^\mu\varphi-\frac{1}{2}\mu^2\varphi^2-\frac{1}{4!}\lambda\varphi^4-\frac{1}{2}\kappa\varphi^2 H^\dagger H
\end{equation}
where $H$ is the Standard Model Higgs and $\mathcal{L}_{\rm SM}$ is the Standard Model Lagrangian. 

Thus this extension of  $\mathcal{L}_{\rm SM}$ corresponds to one real scalar DM particle with a coupling $\kappa$ to the Higgs boson. Although $\mu$ and  $\lambda$ are often reserved for the  Higgs quadratic  and quartic couplings, since we won't make reference to these quantities, we use these instead for the $\varphi$ couplings; importantly they are unrelated to the Higgs coupling. 

Once the Higgs boson acquires vacuum expectation value $\langle H\rangle=v_0$ after electroweak symmetry breaking \cite{Glashow:1961,Weinberg:1967tq,Salam}, and using the Unitary Gauge so that $H^\dagger=(0,(v_0+h)/\sqrt{2})$, the Lagrangian may be expanded around this value giving
\begin{equation}\label{eq:scalarLagrangeExpand}
\mathcal{L}=\mathcal{L}_{\rm SM}+\frac{1}{2}\partial_\mu\varphi\partial^\mu\varphi-\frac{1}{2}\mu^2\varphi^2-\frac{1}{4!}\lambda\varphi^4-\frac{1}{4}\kappa\varphi^2 h^2-\frac{1}{4}\kappa\varphi^2 v_0^2-\frac{1}{2}\kappa\varphi^2 v_0h.
\end{equation}
Thus, the mass of the scalar DM particle is given by
\begin{equation}\label{eq:DMmass}
m_\varphi^2=\mu^2+\frac{1}{2}\kappa v_0^2,
\end{equation}
and the Higgs decay width to $\varphi$ is
\begin{equation}\label{eq:hDecayPhi}
\Gamma(h\rightarrow\varphi\varphi)=\frac{\kappa^2v_0^2}{32\pi m_h}\sqrt{1-\frac{4m_\varphi^2}{m_h^2}},
\end{equation}
where $m_h$ is the Higgs boson mass. Note that the unbroken $\mathbb{Z}_2$ symmetry which stabilises the dark matter also forbids $\varphi$ from obtaining a non-zero vacuum expectation value.

\section{Dark Matter Annihilation to Standard Model States}

For the model-independent approach taken above, we parameterized the thermally-averaged cross-section of dark-matter annihilation to SM particles by parameterizing the cross section as $\sigma_{0}=\alpha^2/m_X^2$. In this section we derive the Higgs Portal amplitudes for various DM annihilations to the SM particles, concluding with the total center-of-mass frame, thermally-averaged annihilation cross-section.

The three main annihilation routes for scalar DM  via the Higgs portal are
\begin{itemize}
	\item Dark matter to Standard Model Fermions, $\varphi\varphi\rightarrow f\bar{f}$,
	\item Dark matter to Standard Model Vector Bosons, $\varphi\varphi\rightarrow V\bar{V}$
	\item Dark matter to Higgs Bosons, $\varphi\varphi\rightarrow hh$
\end{itemize}
which we explore in order below.

\subsubsection{Dark matter to Standard Model Fermions, $\varphi\varphi\rightarrow f\bar{f}$}

\begin{figure}[b!]
	\centering
	\includegraphics[width=0.4\textwidth]{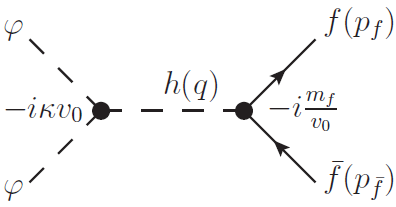}
	\caption{Dark Matter to Standard Model Fermions, $\varphi\varphi\rightarrow f\bar{f}$}
	\label{fig:toff}
\end{figure}

Figure \ref{fig:toff} shows the DM annihilation to two fermions, 
the amplitude for which is  ,
\begin{equation}\label{eq:phiTofAmp}
i\mathcal{M}_{\varphi\varphi\rightarrow f\bar{f}}=(1)(-i\kappa v_0)(1)\frac{i}{q^2-m_h^2+im_h\Gamma_h}\bar{u}^r(p_f)\left(-i\frac{m_f}{v_0}\right)v^{s}(p_{\bar{f}})_.
\end{equation}
where $m_f$ is the mass of the fermion and $\Gamma_h$ is the total width of the Higgs boson.
The cross-section for this process, in the center-of-mass (CM) frame is then
\begin{equation}\label{eq:phiTofCS}
\sigma_{\varphi\varphi\rightarrow f\bar{f}}=\frac{1}{16\pi s}\frac{\sqrt{1-4m_f^2/s}}{\sqrt{1-4m_\varphi^2/s}}\sum_{\rm{spin, color}}|\mathcal{M}_{\varphi\varphi\rightarrow f\bar{f}}|^2_,
\end{equation}
where $s$ is the Mandelstam variable. Carrying out the sum over spin and color gives
\begin{multline}\label{eq:AmpSpinSum}
\begin{split}
&\sum_{\rm{color,spin}}|\mathcal{M}_{\varphi\varphi\rightarrow f\bar{f}}|^2=\frac{N_c\kappa^2m_f^2}{(s-m_h^2)^2+m_h^2\Gamma_h^2}\sum_{r,s}\bar{u}^{r}(p_f)v^{s}(p_{\bar{f}})(\bar{u}^{r}(p_f)v^{s}(p_{\bar{f}}))^*\\
&=\frac{N_c\kappa^2m_f^2}{(s-m_h^2)^2+m_h^2\Gamma_h^2}\mathrm{tr}\left[\left(\slashed{p}_f+m_f\right)\left(\slashed{p}_{\bar{f}}-m_f\right)\right]\\
&=\frac{N_c\kappa^2m_f^2}{(s-m_h^2)^2+m_h^2\Gamma_h^2}\left(4p_fp_{\bar{f}}-4m_f^2\right)=\frac{2N_c\kappa^2m_f^2(s-4m_f^2)}{(s-m_h^2)^2+m_h^2\Gamma_h^2}_,
\end{split}
\end{multline}
where $N_c$ is the number of colors, which we take to be equal to 3 for quarks and 1 for leptons. The cross-section is then
\begin{equation}\label{eq:phiTofCS2}
\sigma_{\varphi\varphi\rightarrow f\bar{f}}=\frac{N_c\kappa^2m_f^2}{8\pi s}\frac{\sqrt{1-4m_f^2/s}}{\sqrt{1-4m_\varphi^2/s}}\frac{s-4m_f^2}{(s-m_h^2)^2+m_h^2\Gamma_h^2}_.
\end{equation}
Letting $r_i\equiv m_i/2m_\varphi$, $r_\Gamma\equiv m_h\Gamma_h/4m_\varphi^2$, and $v$ be the relative velocity between the two incoming particles (in the CM frame, $v\approx2v_1$ in the non-relativistic limit, where $v_1$ is the velocity of the incoming particles), we expand the cross-section multiplied by $v$ in powers of relative velocity thusly
\begin{equation}\label{eq:phiTofCSexp}
\sigma_{\varphi\varphi\rightarrow f\bar{f}}v=\sigma_{\varphi\varphi\rightarrow f\bar{f}}^{(0)}+\sigma_{\varphi\varphi\rightarrow f\bar{f}}^{(2)}v^2+\mathcal{O}(v^4)_,
\end{equation}
where
\begin{equation}\label{eq:phiTofCSexp0}
\sigma_{\varphi\varphi\rightarrow f\bar{f}}^{(0)}\equiv\frac{\kappa^2}{16\pi m_\varphi^2}\frac{r_f^2(1-4r_f^2)^{3/2}}{(1-r_h^2)^2+r_\Gamma^2}
\end{equation}
and
\begin{equation}\label{eq:phiTofCSexp2}
\sigma_{\varphi\varphi\rightarrow f\bar{f}}^{(2)}\equiv\frac{\kappa^2}{32\pi m_\varphi^2}\frac{r_f^2(1-4r_f^2)^{1/2}(1-7r_f^2+r_h^2(1-10r_f^2)+3r_f^2(r_h^4+r_\Gamma^2))}{((1-r_h^2)^2+r_\Gamma^2)^2}_.
\end{equation}

\subsubsection{Dark matter to Standard Model Vector Bosons, $\varphi\varphi\rightarrow V\bar{V}$}
\begin{figure}[b!]
	\centering
	\includegraphics[width=0.45\textwidth]{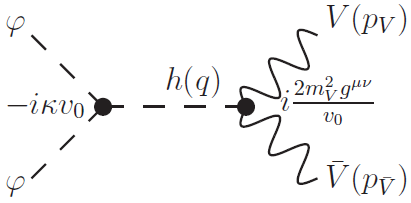}
	\vspace{3mm}
	\caption{Dark Matter to Standard Model Vector Bosons, $\varphi\varphi\rightarrow V\bar{V}$}
	\label{fig:tovv}
\end{figure}
The amplitude for the diagram in Figure \ref{fig:tovv}, for the s-channel DM annihilation to two vector bosons, where $m_V$ is the mass of the vector boson, is
\begin{equation}\label{eq:phiToVAmp}
i\mathcal{M}_{\varphi\varphi\rightarrow V\bar{V}}=(1)(-i\kappa v_0)(1)\frac{i}{q^2-m_h^2+im_h\Gamma_h}\varepsilon_\mu^*(p_V)\left(i\frac{2m_Vg^{\mu\nu}}{v_0}\right)\varepsilon_\nu^*(p_{\bar{V}})_.
\end{equation}
The CM cross-section is then
\begin{equation}\label{eq:phiToVCS}
\sigma_{\varphi\varphi\rightarrow V\bar{V}}=\frac{\delta_V}{16\pi s}\frac{\sqrt{1-4m_V^2/s}}{\sqrt{1-4m_\varphi^2/s}}\sum_{\rm{polarization}}|\mathcal{M}_{\varphi\varphi\rightarrow V\bar{V}}|^2_,
\end{equation}
where $\delta_V$ (which equals 1 for the $W$ boson and 1/2 for the $Z$ boson) is a symmetry factor. Evaluating the sum over polarizations,
\begin{multline}\label{eq:AmpPolSum}
\begin{split}
\sum_{\rm{polarization}}|\mathcal{M}_{\varphi\varphi\rightarrow V\bar{V}}|^2
&=\frac{4\kappa^2m_V^4}{(s-m_h^2)^2+m_h^2\Gamma_h^2}g^{\mu\nu}g^{\rho\sigma}\sum_{r,s}\varepsilon^{*r}_\mu(p_V)\varepsilon^{r}_\rho(p_V)\varepsilon^{*s}_\nu(p_{\bar{V}})\varepsilon^{s}_\sigma(p_{\bar{V}})\\
&=\frac{4\kappa^2m_V^4}{(s-m_h^2)^2+m_h^2\Gamma_h^2}\left(\delta^\nu_\nu-\frac{p_V^2}{m_V^2}-\frac{p_{\bar{V}}^2}{m_V^2}+\frac{(p_Vp_{\bar{V}})^2}{m_V^4}\right)\\
&=\frac{4\kappa^2m_V^4}{(s-m_h^2)^2+m_h^2\Gamma_h^2}\left(2+\frac{(s-2m_V^2)^2}{4m_V^4}\right)_.
\end{split}
\end{multline}
The cross-section is then
\begin{equation}\label{eq:phiToVCS2}
\sigma_{\varphi\varphi\rightarrow V\bar{V}}=\frac{\delta_V\kappa^2m_V^4}{4\pi s}\frac{\sqrt{1-4m_V^2/s}}{\sqrt{1-4m_\varphi^2/s}}\frac{(2+(s-2m_V^2)^2/4m_V^4)}{(s-m_h^2)^2+m_h^2\Gamma_h^2}_.
\end{equation}
Expanding the cross-section multiplied by $v$ in powers of $v$ we have
\begin{equation}\label{eq:phiToVCSexp}
\sigma_{\varphi\varphi\rightarrow V\bar{V}}v=\sigma_{\varphi\varphi\rightarrow V\bar{V}}^{(0)}+\sigma_{\varphi\varphi\rightarrow V\bar{V}}^{(2)}v^2+\mathcal{O}(v^4)_,
\end{equation}
where
\begin{equation}\label{eq:phiToVCSexp0}
\sigma_{\varphi\varphi\rightarrow V\bar{V}}^{(0)}\equiv\frac{\delta_V\kappa^2}{32\pi m_\varphi^2}\frac{(1-4r_V^2)^{1/2}(1-2r_V^2+8r_V^4)}{(1-r_h^2)^2+r_\Gamma^2}
\end{equation}
and
\begin{multline}\label{eq:phiToVCSexp2}
\sigma_{\varphi\varphi\rightarrow V\bar{V}}^{(2)}\equiv\frac{\delta_V\kappa^2(1-4r_V^2)^{-1/2}}{64\pi m_\varphi^2((1-r_h^2)^2+r_\Gamma^2)^2}\bigg (-1+8r_V^2-26r_V^4+56r_V^6+\cdot\cdot\cdot\\r_h^2(1-10r_V^2+36r_V^4-80r_V^6)+2r_V^2(1-5r_V^2+12r_V^4)(r_h^4+r_\Gamma^2)\bigg )_.
\end{multline}

\subsubsection{Dark Matter to Higgs Bosons, $\varphi\varphi\rightarrow hh$}

\begin{figure}[b!]
	\centering
	\includegraphics[width=0.21\textwidth]{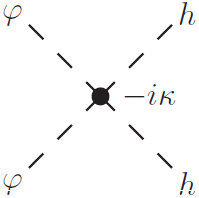}
	\caption{Dark Matter to Higgs Bosons, $\varphi\varphi\rightarrow hh$}
	\label{fig:tohh}
\end{figure}

Finally, the amplitude for the  diagram in Figure \ref{fig:tohh} is
\begin{equation}\label{eq:phiToHAmp}
i\mathcal{M}_{\varphi\varphi\rightarrow hh}=(1)(1)(-i\kappa)(1)(1)_.
\end{equation}
The above feynman diagram is the leading order DM annihilation process to two Higgs bosons. The $s$-channel diagram with a Higgs mediator and the $t$-and $u$-channel diagrams with a DM mediator being exchanged are sub-dominant to the four-vertex diagram. Thus, the $s$-channel amplitude is
\begin{equation}\label{eq:phiToHAmpS}
i\mathcal{M}^{s\rm{-channel}}_{\varphi\varphi\rightarrow hh}=(1)(-i\kappa v_0)(1)\frac{i}{s-m_h^2+im_h\Gamma_h}(1)\left(-i\frac{3m_h^2}{v_0}\right)(1)_,
\end{equation}
which is considerable only near resonance. But at threshold, $v=0$, and resonance (i.e. $2m_\varphi=m_h$) is kinematically forbidden since an on-shell Higgs boson cannot transmute to two Higgs bosons. Moreover, the $t$-channel amplitude (the $u$-channel is the same with the substitution $t\leftrightarrow u$) is
\begin{equation}\label{eq:phiToHAmpT}
i\mathcal{M}^{t\rm{-channel}}_{\varphi\varphi\rightarrow hh}=(1)(-i\kappa v_0)(1)\frac{i}{t-m_\varphi^2}(1)(-i\kappa v_0)(1)_,
\end{equation}
where $t$ (and $u$) is the Mandelstam variable, at threshold, $t=u=m_h^2-m_\varphi^2$. The $t$/$u$-channels are only considerable when $m_h\approx\sqrt{2}m_\varphi$, which is kinematically forbidden.

The CM cross-section is then
\begin{equation}\label{eq:phiToHCS}
\sigma_{\varphi\varphi\rightarrow hh}=\frac{1}{16\pi s}\frac{\sqrt{1-4m_h^2/s}}{\sqrt{1-4m_\varphi^2/s}}|\mathcal{M}_{\varphi\varphi\rightarrow hh}|^2=\frac{\kappa^2}{16\pi s}\frac{\sqrt{1-4m_h^2/s}}{\sqrt{1-4m_\varphi^2/s}}_.
\end{equation}
Expanding this cross-section multiplied by $v$ in powers of $v$ gives
\begin{equation}\label{eq:phiToHCSexp}
\sigma_{\varphi\varphi\rightarrow hh}v=\sigma_{\varphi\varphi\rightarrow hh}^{(0)}+\sigma_{\varphi\varphi\rightarrow hh}^{(2)}v^2+\mathcal{O}(v^4)_,
\end{equation}
where
\begin{equation}\label{eq:phiToHCSexp0}
\sigma_{\varphi\varphi\rightarrow hh}^{(0)}\equiv\frac{\kappa^2}{64\pi m_\varphi^2}(1-4r_h^2)^{1/2}
\end{equation}
and
\begin{equation}\label{eq:phiToHCSexp2}
\sigma_{\varphi\varphi\rightarrow hh}^{(2)}\equiv\frac{\kappa^2}{256\pi m_\varphi^2}\frac{(-1+6r_h^2)}{(1-4r_h^2)^{1/2}}_.
\end{equation}

\section{Relic Density for the Scalar Higgs Portal}

Now that we have derived analytic expressions for the leading contributions to the annihilation cross section we look to calculate the relic density of dark matter after freeze-out.  As previously, to calculate the relic density we undertake a Boltzmann analysis as outlined in Section \ref{SecAbundance}. Whereas previously we used simple parameterizations of the cross section, in working with a specific model there are a number of technicalities that one must take care of. 

The Boltzmann equation is typically written in terms of the thermally-averaged cross-section $\langle\sigma_{\rm A} v\rangle$ which is a product of the  annihilation cross section and relative dark matter velocity averaged over the equilibrium distribution functions. More specifically, this quantity is defined as \cite{Gondolo:1990dk}
\begin{equation}\label{eq:thermAvgGond}
\langle\sigma_{\rm A} v\rangle=\frac{1}{8m_\varphi^4TK_2^2(x)}\int^\infty_{4m_\varphi^2}\sigma_{\rm A}(s-4m_\varphi^2)\sqrt{s}K_1\left(\sqrt{s}/T\right)ds_,
\end{equation}
where we take $\sigma_{\rm A}$ to be
\begin{equation}\label{eq:annCSdef}
\sigma_{\rm A}=\sum_{\rm SM~particles}\sigma_{\varphi\varphi\rightarrow ii},
\end{equation}
where the sum runs over the index $ii$ of pairs of each type of Standard Model quark and lepton, the $W$ and $Z$ vector bosons and the Higgs boson. The mean speed of a particle at freeze-out ($x_{\rm F}\sim20$) is of order a third the speed of light, thus collecting terms up to order $v^2$ in the expansion of the annihilation cross-section provides a good approximation. Thermally averaging term-by-term gives \cite{Gondolo:1990dk}
\begin{equation}\label{eq:thermAvgExpGond}
\langle\sigma_{\rm A}v\rangle=\langle\sigma_{\rm A}^{(0)}+\sigma_{\rm A}^{(2)}v^2+\mathcal{O}(v^4)\rangle=\sigma_{\rm A}^{(0)}+(3/2)\sigma_{\rm A}^{(2)}x^{-1}+\mathcal{O}(x^{-2})_,
\end{equation}
where for $(j=0,2)$,
\begin{equation}
\sigma_{\rm A}^{(j)}\equiv\sum_{\rm SM~particles}\sigma_{\varphi\varphi\rightarrow ii}^{(j)},
\end{equation}

\begin{figure}[b!]
	\centering
	\includegraphics[width=0.9\textwidth]{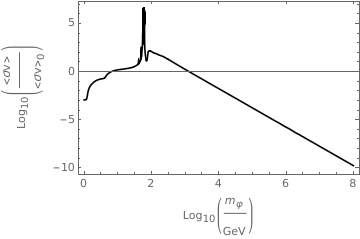}
	\caption[The total Higgs Portal annihilation cross-section.]{The total Higgs Portal annihilation cross-section in units of the standard DM relic density cross-section, $\langle\sigma v\rangle_0=3(10)^{-26}$cm$^3/$s, for a fixed value $\kappa=0.5$ of the mixed quartic $|H|^2|\phi|^2$, taking the Standard Model Higgs boson decay width to be $\Gamma_h^{\mathrm{SM}}=13~\mathrm{MeV}$.}
	\label{fig:TotalCS}
\end{figure}

\begin{figure}[h!]
	\centering
	\vspace{20mm}
	\includegraphics[width=\textwidth]{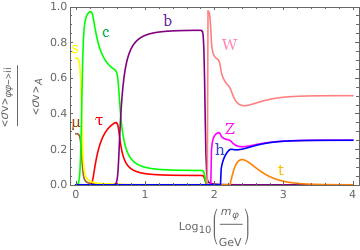}
	\caption[Fraction of annihilation cross-section  to Standard Model particles]{The fraction of the total annihilation cross-section at freeze-out going to various Standard Model particles for a fixed value $\kappa=0.1$ of the mixed quartic $|H|^2|\varphi|^2$ and taking $\Gamma_h^{\mathrm{SM}}=13~\mathrm{MeV}$. The curves indicate different Standard Model final state pairs: muons (brown), strange quarks (yellow), charm quarks (green), taus (red), bottom quarks (purple), $W$ bosons (pink), $Z$ bosons (magenta), Higgs Bosons (blue), and top quarks (orange).}
	\label{fig:FracAnnChann}
\end{figure}

Applying this to the partial cross sections derived in the previous section, we plot in Figure \ref{fig:FracAnnChann} the total $\varphi$ annihilation cross-section to Standard Model states via the Higgs portal (in units of the standard DM thermal cross-section: $\langle\sigma v\rangle_0=3(10)^{-26}$cm$^3/$s) for a fixed value $\kappa=0.5$ of the mixed quartic. Additionally, in Figure \ref{fig:TotalCS} we show the fraction of the total annihilation cross-section at freeze-out going to various Standard Model particles for a fixed value $\kappa=0.1$ of the mixed quartic $|H|^2|\varphi|^2$. In both cases we fix $\Gamma_h^{\mathrm{SM}}=13~\mathrm{MeV}$ which is consistent with current measurements \cite{PDGroup}. Note that our independent analytic derivations of the Higgs partial annihilation rates, and the results presented in  Figures \ref{fig:FracAnnChann} and  \ref{fig:TotalCS} are in good agreement with related studies of the Higgs portal in the literature e.g.~\cite{Cline:2012hg,Cline:2013gha,Feng:2014vea,Escudero:2016gzx}. 

\clearpage

\clearpage

With the appropriate thermally averaged cross section for the Higgs portal we can proceed to calculate the freeze-out temperature and subsequently the freeze-out abundance of dark matter.  Using the freeze-out condition \eqref{eq:FOcond} with the cross-section \eqref{eq:thermAvgExpGond}, we have that freeze-out for the Higgs Portal, $x_{\rm F}^{\rm HP}$, occurs for 
\begin{equation}\label{eq:higgsFO}
x_{\rm F}^{\rm HP}=\ln{\left[\frac{gm_\varphi^3\left(\sigma_{\rm A}^{(0)}+(3/2)\sigma_{\rm A}^{(2)}\left(x_{\rm F}^{\rm HP}\right)^{-1}\right)}{(2\pi)^{3/2}H_\star x_\star^{3/2}}\left((1-r)+\frac{rx_\star}{\left(x_{\rm F}^{\rm HP}\right)}\right)^{-1/2}\right]}_.
\end{equation}
Taking an initial guess of $\left(x_{\rm F}^{\rm HP}\right)^{(0)}=1$, $x_{\rm F}^{\rm HP}$ may be approximated as
\begin{equation}\label{eq:higgsFO1}
\left(x_{\rm F}^{\rm HP}\right)^{(1)}\equiv\ln{\left[\frac{gm_\varphi^3\left(\sigma_{\rm A}^{(0)}+(3/2)\sigma_{\rm A}^{(2)}\right)}{(2\pi)^{3/2}H_\star x_\star^{3/2}}\left((1-r)+rx_\star\right)^{-1/2}\right]}_.
\end{equation}
Notice that $\kappa^2$ may be factored out of the cross-section in \eqref{eq:higgsFO1} giving
\begin{equation}\label{eq:higgsFO1k}
\left(x_{\rm F}^{\rm HP}\right)^{(1)}=\ln{\kappa^2}+\left(x_{\rm F}^{\rm HP}\right)^{(1)}\big|_{\kappa=1},
\end{equation}
which provides a compact expression for deriving constraints on $\kappa$ and illustrates the freeze-out temperature is only logarithmically sensitive to changes in $\kappa$.


The only other relation that needs modification for the Higgs Portal is the abundance. Starting from eq.~\eqref{eq:BEy} and inserting the first two terms of \eqref{eq:thermAvgExpGond}, then following the line of argument used to derive \eqref{eq:Yinf}, the freeze-out abundance for the Higgs Portal is
\begin{equation}
\begin{aligned}\label{eq:higgsYfo}
Y_{\infty}^{\rm HP}\approx\bigg[\tilde{\lambda}(1-r)^{-1/2}\bigg(\sigma_{\rm A}^{(0)}\int_{x_{\rm F}}^{\infty}dx\left(1+\frac{r}{(1-r)}\frac{x_\star}{x}\right)^{-1/2}x^{-5/2}+\cdot\cdot\cdot\\(3/2)\sigma_{\rm A}^{(2)}\int_{x_{\rm F}}^{\infty}dx\left(1+\frac{r}{(1-r)}\frac{x_\star}{x}\right)^{-1/2}x^{-7/2}\bigg)\bigg]^{-1}_,
\end{aligned}
\end{equation}
where,
\begin{equation}\label{eq:lambdaTilde}
\tilde{\lambda}\equiv\frac{2\pi^2g_{*S}m_\varphi^3}{45H_\star x_\star^{3/2}}_.
\end{equation}
The analytic form for $Y_{\infty}^{\rm HP}$ is unwieldy, however it is interesting to note that by writing
$\tilde{a}\equiv rx_\star/(1-r)x_{\rm F}$, the integrals in \eqref{eq:higgsYfo} can be evaluated analytically,
\begin{equation}\label{eq:intHiggsYinf0}
\int_{x_{\rm F}}^{\infty}dx\left(1+\frac{r}{(1-r)}\frac{x_\star}{x}\right)^{-1/2}x^{-5/2}=\left(\frac{1-r}{rx_\star}\right)^{3/2}\left(\sqrt{\tilde{a}(1+\tilde{a})}-\arcsinh{\sqrt{\tilde{a}}}\right)
\end{equation}
and
\begin{equation}\label{eq:intHiggsYinf1}
\begin{aligned}
\int_{x_{\rm F}}^{\infty}dx&\left(1+\frac{r}{(1-r)}\frac{x_\star}{x}\right)^{-1/2}x^{-7/2}\\ &=\left(\frac{3}{4}\right)\left(\frac{1-r}{rx_\star}\right)^{5/2}\left(\sqrt{\tilde{a}(1+\tilde{a})}((2/3)\tilde{a}-1)+\arcsinh{\sqrt{\tilde{a}}}\right)_.
\end{aligned}
\end{equation}

Similar to the model independent  analysis (cf.~eq.~(\ref{eq:critDensZeta})) the freeze-out abundance is diluted as 
$\Omega_{\rm DM}^{\rm relic}h^2=\zeta\Omega_{\rm DM}^Fh^2$.  Conversely the to match the observed relic density one requires
\begin{equation}\label{eq:zetaChptHiggs}
\zeta=\frac{\Omega_\varphi^{\rm relic}h^2}{\Omega_\varphi^{\rm HP}h^2}=\frac{\rho_{\rm critical}\Omega_\varphi^{\rm relic}h^2}{s_0m_\varphi Y^{\rm HP}_\infty}
\end{equation}

Note that $\zeta$ is related to $T_{\rm RH}$ via eq.~(\ref{eq:zetaapprox}). In Figure \ref{fig:kContour7GeV} we display contours of the $T_{\rm RH}$  which give the correct relic density for scalar dark matter freezing out via the Higgs portal. This plot is overlaid with theoretical constraints, similar to those explored in Chapter \ref{Ch4}, which we discuss in the next section.

\section{Theoretical Constraints}

In Chapter \ref{Ch4} we studied the theoretical constraints on the matter dominated freeze-out scenario in a model independent manner, here we re-examine these requirements in the specific context of the scalar Higgs Portal. In the plots which appear at the end of this section we will use the full expression for $Y_{\infty}^{\rm HP}$, however to gain some intuition it will be insightful to make a minor approximation and derive some bounds on the various parameters analytically. 

\begin{figure}[h!]
	\centering
	\includegraphics[width=0.7\textwidth]{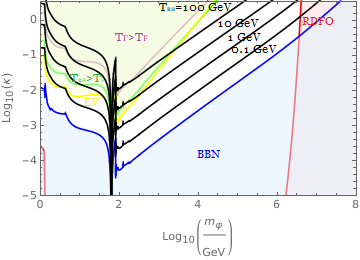}\\
	\vspace{3mm}
	\includegraphics[width=0.7\textwidth]{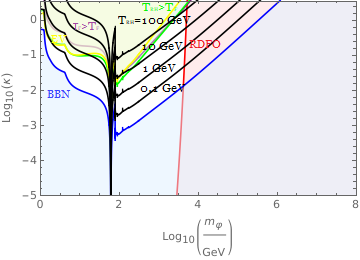}
	\caption[Constraints and relic density contours for $T_\star=10^7$ GeV and $T_\star=10^4$.]{Dominant constraints and relic density contours for $r=0.99$, $T_\star=10^7$ GeV (upper) and $T_\star=10^4$ (lower). Shaded regions excluded due to BBN constraint (blue), MD FO (red), decay before FO (purple), non-adiabicity (yellow), and no re-population (green). From bottom to top of graph, the black contours give the correct relic density for $T_{\rm RH}=0.1$ GeV, $1$ GeV, $10$ GeV, and $100$ GeV.}
	\label{fig:kContour7GeV}
\end{figure}

Specifically,  we consider a quantity $\hat Y$ in which we factor out the explicit $\kappa$ dependence from $Y_{\infty}^{\rm HP}$ as follows
\begin{equation}\label{eq:higgsYfok}
\hat Y=\frac{Y_{\infty}^{\rm HP}}{\kappa^{2}}\propto x_F^{\rm HP}.
\end{equation}
As noted in eq.~(\ref{eq:higgsFO1k}) the freeze-out (inverse) temperature $x_{\rm F}^{\rm HP}$ is only logarithmically sensitive to $\kappa$, thus the principal $\kappa$ dependence in $Y_{\infty}^{\rm HP}$ is due to the explicit factor. For the purpose of arriving at reasonable analytic bounds on this scenario we will treat $\hat Y$ as independent of $\kappa$, and neglect the logarithmic dependence due to $x_F^{\rm HP}$, which we fix at the characteristic value $\hat x_F\approx20$. The dependence of $x_F^{\rm HP}$ will be restored when we re-examine these constraints in Figure \ref{fig:kContour7GeV}.

Emulating the study in Chapter \ref{Ch4}, below we look at the leading theoretical constraints on this scenario, 
\begin{itemize}
	\vspace{-2mm}
	\item Freeze-out of dark matter before decay of $\Phi$ matter. 
	\vspace{-2mm}
	\item Freeze-out of dark matter during matter domination. 
	\vspace{-2mm}
	\item Sufficiently high reheat temperature from $\Phi$ Decays.
	\vspace{-2mm}
\end{itemize} 
In this case, however, we use these constraints to derive bounds on the portal coupling $\kappa$ which dresses the mixed quartic coupling $\frac{1}{2}\kappa\varphi^2 H^\dagger H$. We also check the constraints coming from unitarity and perturbativity here.

In Chapter \ref{Ch4} we derived that the theoretical requirement that dark matter freeze-out occurs before the decay of $\Phi$ can be expressed as a bound on $T_\star$ as follows
$T_\star\gtrsim\frac{x_{\rm F}^3}{(1-r)\hat{\gamma}}\frac{T_{\rm RH}^4}{m_\varphi^3}$, see eq.~\eqref{eq:c3b} and recall here $\hat\gamma$ is defined in eq.~(\ref{eq:gamhat}). In the context of the Higgs portal this constraint can be written as a constraint on the coupling
\begin{equation}\label{eq:higgsC3}
\kappa\lesssim\left[\frac{r^{4/3}\hat{\gamma}^{1/3}m_\varphi}{(1-r)\hat x_{\rm F} T_\star}\right]^{3/8}\left(\frac{s_0m_\varphi \hat Y}{\rho_{\rm critical}\Omega_\varphi^{\rm relic}h^2}\right)^{1/2}
\end{equation}
This constraint is shown in Figure \ref{fig:kContour7GeV} (with the full $\kappa$ dependence) as the purple curve.

Moreover, recall from Chapter \ref{Ch5} that if freeze-out occurs near $H\sim\Gamma_\Phi$ then there is loss of entropy conservation in the bath which significantly alters the physics. Thus to remain in the  matter dominated freeze-out regime we require that $T_{\rm F}\gtrsim T_{\rm EV}$,  where eq.~(\ref{eq:TentAltDef3}) gives the temperature of entropy violation $T_{\rm EV}$. Note that if this is satisfied then the instantaneous decay approximation for $\Phi$ decays is a good approximation.

The constraint $T_{\rm F}\gtrsim T_{\rm EV}$ can be expressed in terms of $\kappa$ as follows
\begin{equation}\label{eq:higgsCna}
\kappa^{8/5}\lesssim\left(\frac{5}{3\hat{\gamma}^{1/2}}\right)^{2/5}\left(\frac{r}{1-r}\right)^{6/5}\left(\frac{s_0m_\varphi \hat Y}{\rho_{\rm critical}\Omega_\varphi^{\rm relic}h^2}\right)^{4/5}\frac{m_\varphi}{T_\star\hat x_{\rm F}}
\end{equation}
This is shown in Figure \ref{fig:kContour7GeV}  as the yellow curve (including the full $\kappa$ dependence).

Next we constrain the parameter space to those parts in which dark matter freeze-out occurs during matter domination, comparing with eq.~\eqref{eq:c1b} in Chapter \ref{Ch4} this can be expressed as  $T_\star\gtrsim \frac{r}{(1-r)}\frac{m_\varphi}{3\beta x_{\rm F}}$ and in terms of $\kappa$ implies
\begin{equation}\label{eq:higgsC1}
\kappa^2\gtrsim\exp{\left[\frac{r}{(1-r)}\frac{m_\varphi}{3\beta T_\star}-\hat x_{\rm F}\right]}
\end{equation}
This is shown in Figure \ref{fig:kContour7GeV}  as the red curve (including the full $\kappa$ dependence).

Furthermore, there is the requirement that to reproduce the successes of early universe cosmology, the temperature after $\Phi$ decays is above the temperature of BBN $T_{\rm BBN}\sim 10$ MeV. Using \eqref{eq:zetaChptHiggs} and eq.~\eqref{eq:zetaapprox}, the constraint $T_{\rm RH}\gtrsim T_{\rm BBN}$ implies 
\begin{equation}\label{eq:higgsC6}
\kappa^2\gtrsim\frac{r}{1-r}\left(\frac{s_0m_\varphi \hat Y}{\rho_{\rm critical}\Omega_\varphi^{\rm relic}h^2}\right)\frac{T_{\rm BBN}}{T_\star}
\end{equation}
This is shown in Figure \ref{fig:kContour7GeV}  as the blue curve (including the full $\kappa$ dependence).

In addition to the reheat temperature being above $T_{\rm BBN}$ we also require that $T_{\rm RH}\lesssim T_{\rm F}$, to ensuring that the dark matter is not re-populated by the thermal bath as the expansion rate changes, as discussed in Section \ref{c3}.  Specifically, the requirement that  $T_{\rm RH}\lesssim T_{\rm F}$ may be expressed as follows
\begin{equation}\label{eq:higgsC7}
\kappa^2\lesssim\frac{r}{1-r}\left(\frac{s_0m_\varphi \hat Y}{\rho_{\rm critical}\Omega_\varphi^{\rm relic}h^2}\right)\frac{m_\varphi}{T_\star\hat x_{\rm F}}
\end{equation}
This is shown in Figure \ref{fig:kContour7GeV}  as the green curve (including the full $\kappa$ dependence).

It is also worth highlighting that the dark matter unitarity constraint, due to Griest and Kamionkowski \cite{Griest:1989wd}, is greatly relaxed in this scenario. Although the limit on plane wave unitary persists, namely that the maximum cross-section derived from unitarity consideration is $\langle\sigma v\rangle_{\rm max}=4\pi\sqrt{x_{F}}/\sqrt{6}m_\varphi^2$ \cite{Griest:1989wd}, since $\Omega_\varphi^{\rm relic}h^2\ll \Omega_\varphi^{\rm F}h^2$ due to the entropy injection, the freeze-out abundance can be much larger than the standard freeze-out abundance without overclosing the universe. While a  dark matter unitarity constraint still exists it does not effectively constrain either plot of  Figure \ref{fig:kContour7GeV}. Additionally, since we have treated this model perturbatively, there is a requirement of perturbitive couplings, thus we require that $\kappa\lesssim\sqrt{4\pi}$, and to illustrate this constraint in  Figure \ref{fig:kContour7GeV} we cut the $\kappa$ axis at $\kappa=\sqrt{4\pi}$.

We collect these constraints in Figure \ref{fig:kContour7GeV} for two values of the critical temperature $T_\star$ at which $\Phi$ becomes matter like, and where we fix $r=0.99$. Observe that the theoretical requirements effectively constrain the parameter space, but leave a modest range of parameters in which matter dominated dark matter freeze-out via the scalar Higgs portal can be successfully realized.  

\clearpage

\section{Experimental Constraints}

\begin{figure}[h]
	\centering
	\includegraphics[width=0.65\textwidth]{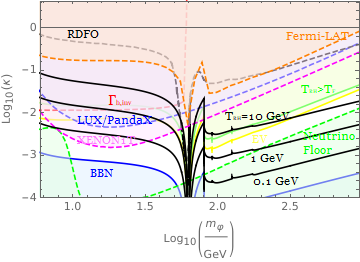}\\
	\includegraphics[width=0.65\textwidth]{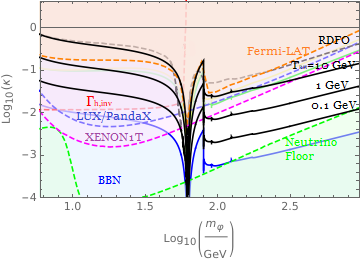}
	\caption[Experimental and theoretical  bounds for $T_\star=10^7$GeV and $T_\star=10^4$GeV.]{Experimental (dashed lines) and theoretical (solid lines) bounds for $r=0.99$, $T_\star=10^7$ GeV (upper) and $T_\star=10^4$ GeV (lower). For experimental constraints, shaded regions excluded from XENON1T \cite{Aprile:2018dbl} (above dashed magenta line), LUX\cite{Akerib:2016vxi} and PandaX\cite{Tan:2016zwf} (above dashed blue line), and Fermi-LAT\cite{Ackermann:2015zua} indirect detection experiment (above dashed orange line). The black dashed line is the standard RD FO scenario contour for matching the observed DM relic density\cite{Escudero:2016gzx}. For theoretical constraints, shaded regions excluded due to BBN constraint (above solid blue line), non-adiabicity (above solid yellow line), and no re-population (above solid green line). Green dashed line indicates the ``neutrino floor''. From bottom to top the black contours give the correct relic density for $T_{\rm RH}=0.1$ GeV, $1$ GeV, and $10$ GeV.}
	\label{fig:HiggsExpCon}
\end{figure}

In the previous section we argued that matter dominated freeze-out via the Higgs portal is theoretically viable in modest parameter regions, and we now turn to the experimental limits. First we consider direct detection experiments which rely on DM particles interacting with various nuclei \cite{PDGroup}. These experiments look for signals from the recoil that a nuclei would undergo in the event that it interacted with DM, thus providing considerable constraints on DM parameters \cite{Escudero:2016gzx}. The DM parameters that these experiments exclude for the Higgs model we study may be seen above the dashed magenta (XENON1T \cite{Aprile:2018dbl}) and blue (LUX\cite{Akerib:2016vxi} and PandaX\cite{Tan:2016zwf}) lines in Figure \ref{fig:HiggsExpCon}. To translate the cross-section bounds of these experiments into Higgs Portal terms, we used the common parameterization for spin-independent DM-nucleon scattering \cite{Djouadi:2011aa,Kanemura:2010sh},
\begin{equation}\label{eq:HPdmNcs}
\sigma_{\varphi-N}^{\rm SI}\simeq\frac{\kappa^2}{4\pi m_h^4}\frac{m_N^4f_N^2}{(m_\varphi+m_N)^2}
\end{equation}
where the nucleon mass is $m_N\approx1$ GeV, and the hadronic matrix element parameter $f_N=0.326$ \cite{Djouadi:2011aa,Young:2009zb}. Note that future experiments with improved sensitivity will need to confront the difficult task of distinguishing neutrino signals from DM signals in the region termed the ``neutrino floor,'' indicated as the green dashed line in Figure \ref{fig:HiggsExpCon}.

Indirect detection of DM experiments rely on detecting the products of DM annihilations that may occur, for instance, in the Milky Way galaxy center or near the Sun where DM concentrations are expected to be higher than average \cite{PDGroup}. For example, the Fermi-LAT collaboration's \cite{Ackermann:2015zua} data from observing DM annihilations into the $b\bar{b}$ channel moderately constrains the allowable parameter space for the Higgs portal \cite{Escudero:2016gzx}. In our model, the excluded parameters from this indirect experimental data is seen in Figure \ref{fig:HiggsExpCon} as the orange shaded region above the dashed orange line. Figure \ref{fig:HiggsExpCon} also shows that indirect experiments are currently less constraining than other experimental methods.

\begin{figure}[t!]
	\centering
	\includegraphics[width=0.5\textwidth]{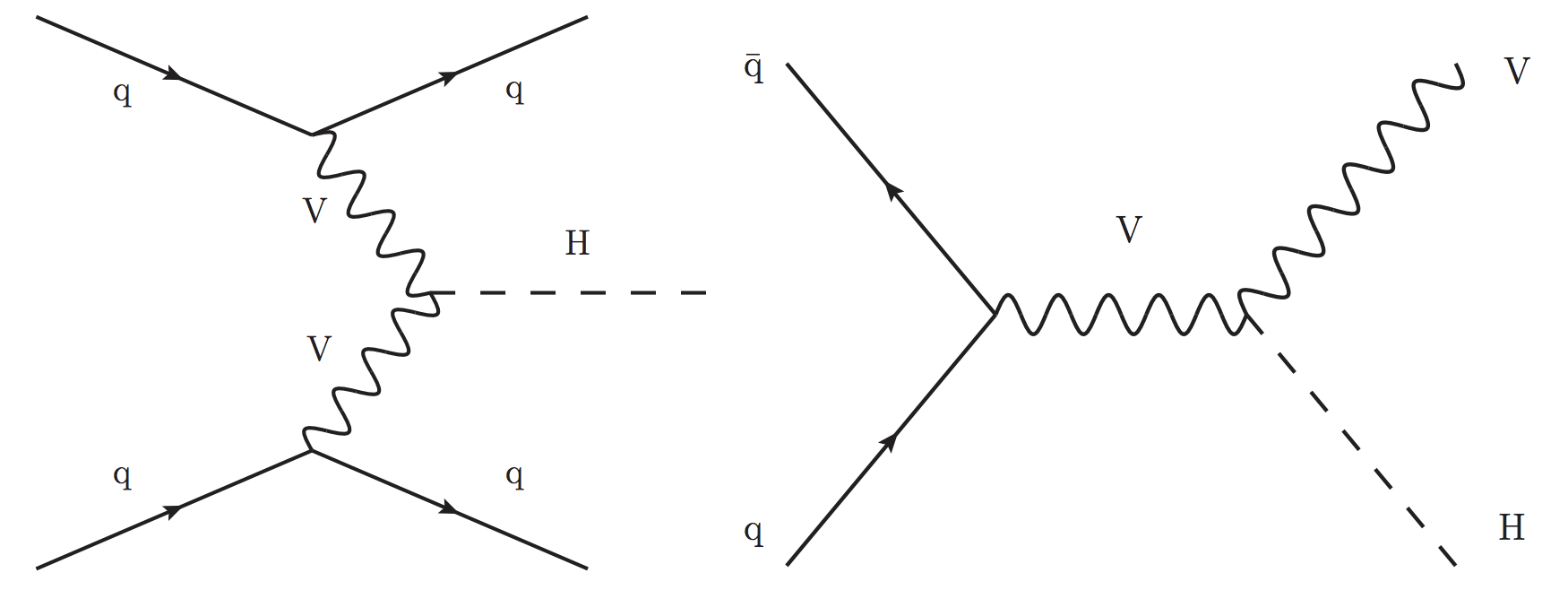}\includegraphics[width=0.3\textwidth]{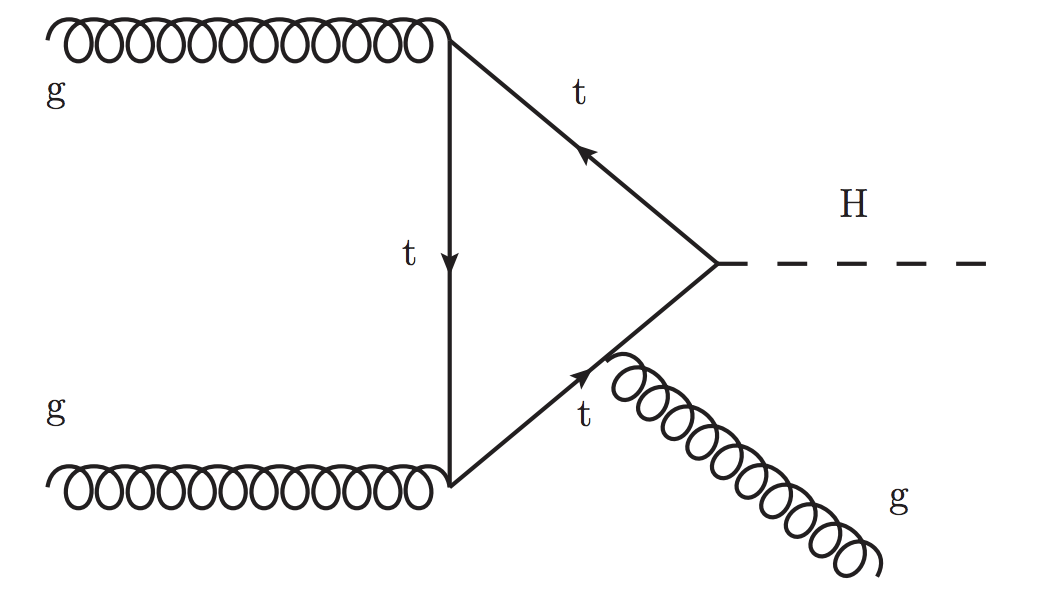}
	\caption[Three processes in the CMS search for invisible Higgs decays, from \cite{Khachatryan:2016whc}.]{Figure from V.~Khachatryan {\it et al.}~\cite{Khachatryan:2016whc}. The Figure illustrates the three production processes targeted in the CMS search for invisible Higgs boson decays.}
	\label{fig:44}
\end{figure}

Another experimental method for probing DM is through collider searches e.g.~\cite{Fox:2011fx,Rajaraman:2011wf,Fox:2011pm,Haisch:2012kf}. If DM interacts with the SM, one would expect that a fraction of the high energy particle collisions would produce DM. By measuring the energy of the products of particle collisions, it may be inferred if there are invisible states which carry away energy undetected through the experimental triggers. For the Higgs Portal, the Higgs width is the relevant parameter to probe. Specifically, emulating  \cite{Escudero:2016gzx}, we compare with the CMS search for invisible decays of the Higgs boson in $pp$ collisions at $\sqrt{s}$ = 7, 8, and 13 TeV \cite{Khachatryan:2016whc}, the processes targeted in this search are shown in Figure \ref{fig:44}. The current branching fraction of the Higgs width that may go to an invisible state is constrained by the Large Hadron Collider (LHC) searches as, $\mathcal{B}(h\rightarrow\mathrm{inv})\equiv\Gamma_{\mathrm{inv}}/(\Gamma_{\mathrm{inv}}+\Gamma_{\mathrm{SM}})<0.2$ with 90\% confidence level \cite{Khachatryan:2016whc}. Assuming that the invisible decays are due to DM  and using eq.~\eqref{eq:hDecayPhi}, this gives a constraint on Higgs portal DM as illustrated by region above the red dashed line in Figure \ref{fig:HiggsExpCon}.

In Figure \ref{fig:HiggsExpCon} we collect both the experimental and theoretical limits together; while this scenario is quite constrained in the low-mass/high-coupling regime shown in the figure, viable parameter space remains. Notably, radiation dominated freeze-out via the Higgs portal (without an entropy injection) is largely excluded apart from around the region of resonant annihilation as indicated by the black dashed line in Figure \ref{fig:HiggsExpCon} (also see e.g.~\cite{Escudero:2016gzx}). Thus it is interesting to observe that the classic Higgs portals returns as a possibility for providing the correct dark matter abundance, while avoiding constraints for models with an entropy injection such as  matter dominated dark matter freeze-out.

\newpage

%% file: PhDthesisV2.0chpt07Conclusion.tex
\chapter{~Conclusion}
\label{Ch7}

\vspace{-10mm}
{\em This chapter presents a summary and discussion of the results of this thesis.}
\vspace{2mm}

Dark matter poses one of the greatest and most exciting challenges for science today. Since the 1930s we have known that something was amiss with our description of the universe. As more data and theoretical work has accumulated, the $\Lambda$CDM model, of which dark matter is a critical component, has provided compelling explanations for the disparate phenomenon we observe in our universe. In this thesis we made the case for dark matter in Chapter \ref{Ch1}, and described the empirical searches for particle dark matter, which laid the ground for the validity of our investigation. Indeed it was shown how $\Lambda$CDM explains the galactic rotation curves, the baryon budget, the CMB power spectrum, large scale structure formation, and the Bullet Cluster in a coherent manner. With such powerful explanatory power, vigorous empirical effort has been made including efforts at direct detection, indirect detection, and collider searches.

It is vital that our empirical efforts are guided with the light of rigorous theoretical work. In Chapter \ref{Ch2} we laid out the plausible scenario that dark matter may have undergone freeze-out during an era of the universe in which matter was dominant. There we discussed the possibility that $T_{\rm MD}\gg T_F \gg T_\Gamma$ and $H\propto T^{\nicefrac{3}{2}}$ during freeze-out, as well as how the different dynamics and constituency of the universe may impact dark matter. The following two chapters were devoted to exploring this scenario rigorously and investigating its prominent constraints. In Chapter \ref{Ch5} we demonstrated that the instantaneous decay approximation we used for simplicity was well justified, and showed the regions of parameter space where it may break down. Finally in Chapter \ref{Ch6} the viability of a matter-dominated freeze-out scenario was well established with our presentation of how it re-opens wide the hitherto claimed, nearly ``closed'' Higgs Portal 

\begin{figure}[t!]
		\centering
		\includegraphics[width=0.49\textwidth,height=5cm]{Figs/AE-FigB.png}\includegraphics[width=0.49\textwidth,height=5.4cm]{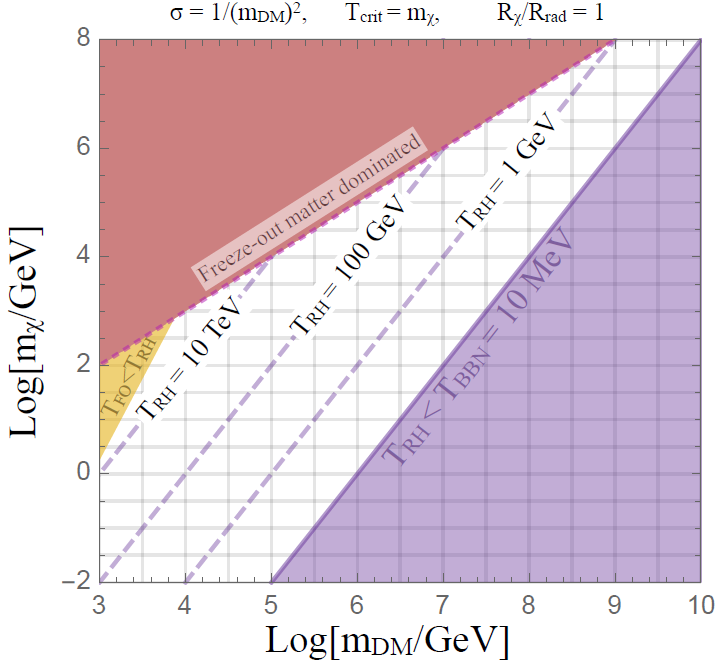}
		\caption[Figure from J.~Bramante and J.~Unwin JHEP 1702 (2017) 119]{Left panel is Figure \ref{fig:2} of this thesis. Right panel is figure from J.~Bramante and J.~Unwin JHEP 1702 (2017) 119 \cite{SuperHeavyBramante2017}.}
		\label{fig:4}
		\vspace{10mm}
	\end{figure}

This thesis initially grew out of considering what may lie in the red region of the right panel of Figure \ref{fig:4} from J.~Bramante and J.~Unwin's work \cite{SuperHeavyBramante2017} in which traditional freeze-out calculations broke-down. This thesis shows it is both a viable and rich region of parameter space, as it is elucidated in the left panel of Figure \ref{fig:4}, which is even capable of reviving hitherto excluded models of dark matter, such as the Higgs Portal. Future work will further investigate the rich possibilities that matter-dominated freeze-out creates, and studying this scenario is an essential step forward in our understanding of the range of possibilities for consistent and compelling models of dark matter.

%% file: PhDthesisV2.0bib.tex
\clearpage
\singlespace